\documentclass[
aps, prd, reprint, superscriptaddress, nofootinbib
]{revtex4-2}

\usepackage{graphicx}
\usepackage[usenames,dvipsnames]{xcolor}
\usepackage{xspace}
\usepackage{amsmath,amsfonts,amssymb,amsthm,bm}
\usepackage[utf8]{inputenc}
\usepackage[normalem]{ulem}
\usepackage{tensor}
\usepackage{array}
\usepackage{multirow}
\usepackage{colortbl}
\usepackage{booktabs}

\usepackage[force]{feynmp-auto}

\xdefinecolor{mylinkcolor}{rgb}{0,0,0.7}
\usepackage[
	breaklinks=true,
	colorlinks=true, filecolor=mylinkcolor, citecolor=mylinkcolor,
	linkcolor=mylinkcolor, urlcolor=mylinkcolor, menucolor=mylinkcolor,
]{hyperref}
\usepackage[depth=subsection]{bookmark}

\definecolor{pyBlue}{RGB}{31, 119, 180}
\definecolor{pyRed}{RGB}{214, 39, 40}
\definecolor{pyGreen}{RGB}{44, 160, 44}

\newcommand{\red}[1]{\textcolor{pyRed}{#1}}
\newcommand{\blue}[1]{\textcolor{pyBlue}{#1}}
\newcommand{\green}[1]{\textcolor{pyGreen}{#1}}

\renewcommand{\(}{\left(}
\renewcommand{\)}{\right)}
\renewcommand{\[}{\left[}
\renewcommand{\]}{\right]}
\renewcommand{\vec}[1]{\mathbf{#1}}

\def\mr{\mathrm}
\DeclareMathOperator{\Order}{\mathcal{O}}
\def\di{\mr d}

\allowdisplaybreaks

\begin{document}

\title{Post-Newtonian observables for aligned-spin binaries to sixth order in spin\\ from gravitational self-force and Compton amplitudes}

\author{Yilber Fabian Bautista}
\email{yilber-fabian.bautista-chivata@ipht.fr}
\affiliation{Institut de Physique Théorique, CEA, Université Paris–Saclay,
F–91191 Gif-sur-Yvette cedex, France}

\author{Mohammed Khalil}
\email{mkhalil@perimeterinstitute.ca}
\affiliation{Perimeter Institute for Theoretical Physics, 31 Caroline Street North, Waterloo, ON N2L 2Y5, Canada}

\author{Matteo Sergola}
\email{matteo.sergola@ipht.fr}
\affiliation{Institut de Physique Théorique, CEA, Université Paris–Saclay,
F–91191 Gif-sur-Yvette cedex, France}

\author{Chris Kavanagh}
\email{chris.kavanagh1@ucd.ie}
\affiliation{School of Mathematics \& Statistics, University College Dublin, Belfield, Dublin 4, Ireland, D04 V1W8}

\author{Justin Vines}
\affiliation{Mani L. Bhaumik Institute for Theoretical Physics, University of California at Los Angeles, Los Angeles, CA 90095, USA}

\begin{abstract}
Accurate modeling of compact binaries is essential for gravitational-wave detection and parameter estimation, with spin being an important effect to include in waveform models.
In this paper, we derive new post-Newtonian (PN) results for the conservative aligned-spin dynamics at next-to-next-to-leading order for the spin$^3$ and spin$^4$ contributions, in addition to the next-to-leading order (NLO) spin$^5$ and spin$^6$ contributions.
One approach we follow is  the Tutti Frutti method, which relates PN and gravitational self-force results through the redshift and spin-precession invariants, by making use of the simple dependence of the scattering angle on the symmetric mass ratio.
However, an ambiguity arises at the NLO spin$^5$ contribution, due to transcendental functions of the Kerr spin in the redshift; 
this is also the order at which Compton amplitudes calculations are affected by spurious poles.
Therefore, we follow an additional approach to determine the NLO spin$^5$ and spin$^6$ dynamics: using on-shell Compton amplitudes obtained from black hole perturbation theory.  
The Compton amplitude used in this work is composed of the unambiguous    tree-level far-zone part reported in [Phys. Rev. D 109, 084071 (2024)], as well as the full, non-interfering with the far-zone, $\ell=2$ partial wave contributions from the near zone, which are responsible for capturing Kerr finite-size effects. 
Other results in this paper include deriving the scattering angle of a spinning test body in a Kerr background from a parametrized worldline action, and computing the redshift and spin-precession invariants for eccentric orbits without an eccentricity expansion. 
\end{abstract}

\maketitle

\section{Introduction}

Several analytical approximation methods exist for modeling the dynamics of compact binary systems in general relativity: the post-Newtonian (PN) approximation~\cite{Blanchet:2013haa,Schafer:2018kuf,Levi:2015msa,Porto:2016pyg,Levi:2018nxp}, which is an expansion in small velocities and large separations; the post-Minkowskian (PM) approximation~\cite{Westpfahl:1979gu,Bel:1981be,Damour:2016gwp,Damour:2017zjx}, which is an expansion in large separations but for arbitrary velocities; and gravitational self-force (GSF)~\cite{Mino:1996nk,Quinn:1996am, Barack:2018yvs,Pound:2021qin}, which provides an expansion in the small mass ratio but is valid for arbitrary velocities in the strong-field regime.
Over the past two decades, remarkable advancements in the computation of  PN and PM observables  have been achieved using techniques inspired by quantum mechanics and particle physics, such as effective field theory (EFT)~\cite{Levi:2015msa,Porto:2016pyg,Levi:2018nxp,Foffa:2013gja,Foffa:2016rgu,Kalin:2019rwq,Blumlein:2020znm,Blumlein:2020pog,Blumlein:2021txe,Blumlein:2021txj,Kalin:2020fhe,Kalin:2020mvi,Dlapa:2021npj,Dlapa:2021vgp,Dlapa:2021vgp,Cho:2022syn,Dlapa:2022lmu,Dlapa:2024cje}, worldline quantum field theory~\cite{Mogull:2020sak,Jakobsen:2021smu,Jakobsen:2021lvp,Jakobsen:2021zvh,Jakobsen:2022fcj,Jakobsen:2023ndj,Jakobsen:2023hig,Driesse:2024xad}, and scattering amplitudes~\cite{Arkani-Hamed:2017jhn,Bjerrum-Bohr:2018xdl,Guevara:2018wpp,Kosower:2018adc,Cheung:2018wkq,Bautista:2019tdr,Guevara:2019fsj,Bern:2019nnu,Bjerrum-Bohr:2019kec,Bern:2019nnu,Bern:2020buy,Cristofoli:2020uzm,Bjerrum-Bohr:2021vuf,Herrmann:2021tct,Bern:2021dqo,DiVecchia:2021bdo,Bern:2021yeh,Bautista:2021wfy,Cristofoli:2021vyo,Herrmann:2021lqe,Travaglini:2022uwo,Bjerrum-Bohr:2022blt,Kosower:2022yvp,Bern:2024adl,Driesse:2024xad,Luna:2023uwd,Mougiakakos:2024nku}. Recently, there has also been work in applying and synergizing these methods with the self-force approach \cite{Cheung:2024byb,Cheung:2023lnj,Kosmopoulos:2023bwc}, and vice versa;
 i.e. using self-force for scattering calculations \cite{Barack:2023oqp,Long:2024ltn,Bini:2024icd}.
 
In the modern on-shell approach, 
compact objects are typically described as point particles interacting through the exchange and emission of gravitons, which resembles scattering problems in particle physics. 
This description is of course an effective one as the extended nature of the compact object manifests when the energies of the exchanged gravitons are large enough to resolve the object's finite-size structure. 
For instance, in a gravitational-wave scattering off a black hole (BH) scenario, the extended nature of the BH is elucidated starting at the fifth power of the ratio of the Schwarzschild radius to the wavelength of the graviton, i.e.~$\sim(G M/\lambda)^5$, when tidal deformations~\cite{Saketh:2023bul,Ivanov:2022qqt}, and tidal heating effects start to take place~\cite{Saketh:2022xjb,Chia:2024bwc}. 
To capture these effects, the effective point-particle description needs to be extended by, for example, incorporating non-minimal operators in an EFT Lagrangian, which renormalize UV divergences appearing from considering the compact objects as point particles, in addition to capturing the extended nature of such objects through the coefficients of those operators~\cite{Bern:2020uwk,Kalin:2020lmz,Ivanov:2024sds}.

A first-principle reconstruction of compact objects would naturally predict the values of the effective coefficients; this is however inaccessible at the moment. 
An EFT approach on the other hand uses a matching procedure to fix the value of such coefficients, but ignores the nature of how such values arise from a complete UV description. 
Although, an intermediate situation may arise in an EFT approach when a set of consistency rules, and perhaps hidden symmetries describing the compact objects, could emerge, providing alternative convenient ways to predict the values of some EFT coefficients.
Well-known examples appeared in the literature when considering the amplitudes that describe the linearized, isolated Kerr BH, where the uniqueness of its spin-multipole moments~\cite{Vines:2017hyw} can be recast in the simplest possible 3-point amplitude involving two massive spinning legs and one graviton~\cite{Guevara:2018wpp,Chung:2018kqs}.

Analogous proposals appeared at the level of the Compton amplitude, among them is the minimal coupling proposals for low orders in spin, including the double copy \cite{Guevara:2018wpp,Chung:2018kqs,Bautista:2019tdr,Aoude:2020onz,Bautista:2021inx,DeAngelis:2023lvf}, the  spin-shift symmetry\footnote{
Although this was first proposed at the level of the 2PM two-body amplitude, it can be shown to be a consequence of the  symmetry of the helicity-preserving gravitational Compton amplitude for low orders in spin. 
} \cite{Aoude:2022trd,Bern:2022kto,Aoude:2022trd,Aoude:2022thd,Haddad:2023ylx,Akpinar:2024meg},  spin-resummed predictions from amplitudes with massive gauge symmetries and improved high energy behavior~\cite{Chiodaroli:2021eug,Cangemi:2022bew,Chiodaroli:2021eug}, bootstrapped heavy-mass effective fields ansatz~\cite{Bjerrum-Bohr:2023jau,Brandhuber:2023hhl}, Neuman-Janis shifts and twistor dynamics~\cite{Alessio:2023kgf,Akhtar:2024mbg,Kim:2024grz}, worldline and dynamical spin-multipole moments~\cite{Ben-Shahar:2023djm,Scheopner:2023rzp},  superstring amplitudes~\cite{Cangemi:2022abk,Azevedo:2024rrf}, and Born amplitudes \cite{Correia:2024jgr}.
A proposal for  effective amplitudes obtained directly  from black hole perturbation theory (BHPT) computations appeared in Refs.~\cite{Bautista:2022wjf,Bautista:2021wfy}, with a recent connection to CFT~\cite{Bautista:2023sdf}; the latter reference also  discussed a separation of the tree-level Compton amplitude in an unambiguous far-zone contribution and a somehow more intriguing near-zone piece. 
These 3-point and Compton  amplitudes are building blocks for the binary BH problem at lower orders in the PM expansion, and as such, are of primordial interest. 

Identifying  symmetries and consistency conditions is more cumbersome when the BH's finite-size effects are present, the latter being  naturally captured by near-zone contributions in the Teukolsky solutions. 
Although the near-zone contributions in the Compton amplitude have a unique spin-multipole expansion (SME)---as we show in Sec.~\ref{sec:compton} by matching a unique spin-5 and spin-6 covariant contact term to the leading $\ell=2$ solutions of the Teukolsky equation in the near-zone region---it is ambiguous to separate a tree-level, point-particle contribution, from the BH finite-size  contributions. 
This is because higher-order-spin contributions get mixed with dynamical tidal deformations, as we also observe in this work. 
Hence, in this paper we follow a somewhat intermediate approach where the Compton amplitude used  to construct two-body observables is composed of the unambiguous tree-level far-zone  contribution, as well as the unambiguous, non-perturbative (in the Kerr spin) complete $\ell=2$ near-zone contribution, in turn capturing finite-size effects at first self-force order (1SF), and avoiding the ambiguities appearing when trying to extract a tree-level near-zone contribution. 
Analogous non-perturbative matching computations have been done in the absorptive sector, see for instance Refs.~\cite{Porto:2007qi,Saketh:2022xjb}. 

More precisely, in this paper, we study the imprints of the aforementioned Compton amplitude on the conservative 2PM scattering angle, and indirectly on the redshift~\cite{Detweiler:2008ft,Bini:2013rfa,Kavanagh:2015lva,Hopper:2015icj,Bini:2015bfb,Kavanagh:2016idg,Bini:2018zde,Bini:2019lcd,Bini:2020zqy,Munna:2022gio,Munna:2023wce} and spin-precession frequency~\cite{Dolan:2013roa,Bini:2014ica,Akcay:2016dku,Kavanagh:2017wot,Bini:2018ylh,Bini:2019lkm,Munna:2022xts}.
Our work then provides an alternative approach to obtaining gauge-invariant observables from the, also gauge-invariant, scattering amplitudes. 
In particular, we show how several terms in such observables, obtained from a traditional GSF approach involving the reconstruction of the regular piece of the off-shell metric perturbation, can be obtained purely from the scattering amplitudes in a Kerr BH gravitational-wave scattering process.  
The Compton amplitude then serves as an ``on-shell reconstruction'' of the perturbation avoiding any subtlety related to gauge choices.  
 
In addition to the scattering amplitudes approach, we also compute the aligned-spin dynamics using the Tutti Frutti method~\cite{Bini:2019nra,Bini:2020wpo,Siemonsen:2019dsu,Antonelli:2020ybz}, which takes advantage of the simple dependence of the PM-expanded scattering angle on the symmetric mass ratio~\cite{Damour:2019lcq} to relate PN results valid for generic mass ratios to GSF results at lower orders in the mass ratio.
The comparison of PN and GSF results is done through the redshift and spin-precession frequency; these gauge-invariant quantities are computed in the GSF literature by solving analytically the non-homogeneous Teukolsky or Regge–Wheeler-Zerilli equations in a PN expansion using, for example, the Mano-Suzuki-Takasugi (MST) method~\cite{Mano:1996mf,Mano:1996vt}, while in a PN calculation they can be computed from a Hamiltonian using the first law of binary mechanics~\cite{LeTiec:2011ab, Blanchet:2012at,LeTiec:2015kgg,Blanchet:2017rcn,Fujita:2016igj,Antonelli:2020ybz}.

Using this method, we derive the aligned-spin dynamics at the next-to-leading PN order (NLO) spin$^5$ and spin$^6$ for generic orbits. 
However, computing the redshift and spin-precession invariants for eccentric orbits in terms of frequencies requires more PN information at lower orders in spin that the NLO.
In particular, the NLO spin$^5$ and spin$^6$ contributions require information from the next-to-next-to-leading order (NNLO) spin$^3$ and spin$^4$ dynamics, which we derive in this paper for the first time.
However, while we obtain the full NNLO spin$^3$ contribution from existing GSF results, our result for the NNLO spin$^4$ contribution contains one unknown coefficient that requires currently unavailable GSF results.
We highlight the PN and spin orders  derived in this work  in Table~\ref{tab:PNspin}.

To perform the above calculations, we also needed to compute some results for a spinning test-body in a Kerr background. 
In particular, we computed the redshift and spin-precession frequency to sixth order in the Kerr spin for arbitrary eccentricities.
We also derived the scattering angle of a spinning test body, to fifth order in its spin and sixth order in the Kerr spin, starting from the parametrized worldline action presented in Ref.~\cite{Scheopner:2023rzp}, with generic Wilson coefficients. 
Then, we obtained constraints on the values of these coefficients by comparing the scattering angle with the one derived from a Compton amplitudes calculation, and with GSF results.

This paper is structured as follows:
\begin{itemize}
\item In Sec.~\ref{sec:boundObserv}, we compute the redshift and spin-precession frequency from a parametrized PN Hamiltonian, for both circular and eccentric orbits, after relating its coefficients to those of an ansatz for the scattering angle. 
Then, we compare our results with GSF results to solve for the unknowns.

\item In Sec.~\ref{sec:compton}, we use on-shell Compton amplitudes matched to BHPT to compute the 2PM scattering angle to sixth order in spin.

\item In Sec.~\ref{sec:NLOS5soln}, we combine results from both approaches and discuss three possibilities to resolve an ambiguity in determining the NLO spin$^5$ dynamics, in addition to confirming the unambiguous spin$^6$ solution. Summary of our findings is provided in Table~\ref{tab:summary_coeffs}.

\item In Sec.~\ref{sec:worldlineKerr}, we derive the scattering angle of a spinning test body from a worldline action with generic Wilson coefficients, and use GSF and Compton amplitudes results to constrain those coefficients.
\end{itemize}
We conclude in Sec.~\ref{sec:conclusions}, and provide in Appendix~\ref{app:zPsiTest} the test-body results for the redshift and spin-precession frequency to all orders in eccentricity, while Appendix~\ref{app_B} contains the full expressions of the BHPT-matched  Compton coefficients.
Most of our results are provided as electronic files in the Supplemental Material~\cite{ancMaterial}.

\subsection*{Notation}
We use the metric signature $(-,+,+,+)$, and geometric units in which $c=G=1$, but we write $c$ and $G$ explicitly for clarity in some PN and PM expansions.

For a binary with masses $m_1$ and $m_2$, we assume that $m_1 \geq m_2$, and define the following combinations of the masses:
\begin{gather}
M= m_1 + m_2, \qquad \mu = \frac{m_1m_2}{M}, \qquad \nu = \frac{\mu}{M}, \nonumber\\
q = \frac{m_2}{m_1}, \qquad \delta =\frac{m_1 - m_2}{M}.
\end{gather}

For an aligned-spin binary with spins $S_1$ and $S_2$, we define the spin lengths $a_{\mr i} = S_{\mr i} / m_{\mr i}$, where $\mr i = 1,2$. 
The spin length of a Kerr BH is denoted $a$, with the dimensionless spin $\chi$ defined as $\chi = a/m_1$, and we define the quantity $\kappa = \sqrt{1-\chi^2}\,$.

We also define the total energy $E = E_1+E_2$ of the binary and effective energy $E_\text{eff}$ via the energy map
\begin{subequations}
\label{gammaGamma}
\begin{align}
\Gamma &:= \frac{E}{M} = \sqrt{1 + 2\nu \left(\gamma - 1\right)}\,, \\
\gamma &:= \frac{E_\text{eff}}{\mu} = \frac{E^2 - m_1^2 - m_2^2}{2 m_1 m_2},
\end{align}
\end{subequations}
where the Lorentz factor $\gamma$ is related to the relative velocity $v$ by
\begin{equation}
\gamma = \frac{1}{\sqrt{1 - v^2}}.
\end{equation}

The azimuthal frequency is denoted $\Omega$ and the relativistic periastron advance is denoted $k$, from which we define the following variables:
\begin{subequations}
\begin{align}
x := (M \Omega)^{2/3}, \qquad \iota := \frac{3x}{k}, \\
y := (m_1 \Omega)^{2/3}, \qquad \lambda := \frac{3y}{k}.
\end{align}
\end{subequations}
The redshift and spin-precession frequency are denoted $z_{\mr i}$ and $\Omega_{S_{\mr i}}$, while the spin-precession invariant is defined as $\psi_{\mr i} := \Omega_{S_{\mr i}} / \Omega$.

The magnitude of the canonical orbital angular momentum is denoted $L$, and is related to the relative position $r$, radial momentum $p_r$, and total linear momentum $p$ via $p^2 = p_r^2 + L^2/r^2$. The Hamiltonian is denoted $\mathcal{H}$, and we define $u := M/r$.

In Sec. \ref{sec:compton}, related to wave perturbations of spin-weight $s$ off Kerr BH, we use the following notation:
\begin{equation}\label{eq:epsilon_param}
\epsilon=2G m_1 \omega,
\end{equation}
which is the PM wave scattering parameter, with $\omega$ the energy of the scattered wave.

\begin{table*}[th]
\setlength\extrarowheight{4pt}
\caption{This table summarizes the PN orders for the conservative dynamics at which each order in spin appears in the Hamiltonian, and the corresponding orders in $x$ (which are the same as $y$ and $u_p$) in the redshift $z$ and spin-precession invariant $\psi$.
We derive new PN results for the full NNLO S$^3$, NLO S$^5$, and NLO S$^6$ contributions (highlighted in green), in addition to the NNLO S$^4$ contribution (highlighted in orange) for which one  unknown coefficient remains at $\Order(\nu a_1^2a_2^2)$.
Note that the spin counting in this table follows the spin orders in the Hamiltonian, but since the spin-precession frequency involves a derivative with respect to spin, the spin orders in $\psi$ are one power lower than the Hamiltonian.}
\label{tab:PNspin}
\begin{ruledtabular}
\begin{tabular}{c|lll|lll|lll|lll|ll}
Spin~ & LO&$z$&$\psi$ & NLO&$z$&$\psi$ & NNLO&$z$&$\psi$ & N$^3$LO&$z$&$\psi$ & N$^4$LO&$z$ \\
\hline
S$^0$ &  0PN&$x$&--  & 1PN&$x^2$&--  &  2PN&$x^3$&--  & 3PN&$x^4$&-- & 4PN&$x^5$ \\
S$^1$ &  1.5PN&$x^{5/2}$&$x$  &  2.5PN&$x^{7/2}$&$x^2$  &  3.5PN&$x^{9/2}$&$x^3$  &  4.5PN&$x^{11/2}$&$x^4$ & & \\
S$^2$ &  2PN&$x^3$&$x^{3/2}$  &  3PN&$x^4$&$x^{5/2}$  & 4PN&$x^5$&$x^{7/2}$  &  5PN&$x^6$&$x^{9/2}$ & & \\
S$^3$ &  3.5PN&$x^{9/2}$&$x^3$  &  4.5PN&$x^4$&$x^4$  &  \cellcolor{green!25}5.5PN&$x^{13/2}$&$x^5$ & & & & & \\
S$^4$ &  4PN&$x^5$&$x^{7/2}$  &  5PN&$x^6$&$x^{9/2}$  &  \cellcolor{orange!25}6PN&$x^7$&$x^{11/2}$  & & & & & \\
S$^5$ &  5.5PN&$x^{13/2}$&$x^5$  &  \cellcolor{green!25}6.5PN&$x^{15/2}$&$x^6$ & & & & & & & & \\
S$^6$ & 6PN&$x^7$&$x^{11/2}$  &  \cellcolor{green!25}7PN&$x^8$&$x^{13/2}$ & & & & & & & & \\
\end{tabular}
\end{ruledtabular}
\end{table*}

\section{Bound-orbit PN observables from scattering and GSF results}
\label{sec:boundObserv}
In this section, we derive new PN results at the NNLO for the S$^3$ and S$^4$ conservative dynamics, in addition to the NLO S$^5$ and S$^6$ dynamics.
We follow the Tutti Frutti approach~\cite{Bini:2019nra,Bini:2020wpo,Antonelli:2020ybz} by taking advantage of the simple dependence of the PM-expanded scattering angle on the mass ratio, which makes it possible to obtain NNLO PN results valid for arbitrary mass ratios from GSF results at first order in the mass ratio.

We first obtain a Hamiltonian whose coefficients are matched to a scattering angle ansatz parametrized by unknown coefficients that satisfy the mass-ratio dependence of the scattering angle.
We then compute the redshift and spin-precession frequency from the Hamiltonian, for both circular and eccentric orbits in terms of the frequencies, and expand them to first order in the mass ratio.
Finally, we express these gauge-invariant quantities in terms of the Kerr geodesic variables used in GSF calculations, and compare them to 1SF results to determine the unknown coefficients. 
In Sec.~\ref{sec:compton}, we discuss an alternative approach to fixing the scattering angle's free coefficients at NLO S$^5$ and S$^6$ using information from Compton amplitudes and BHPT computations.

For circular orbits, computing the redshift and spin-precession frequency in terms of the orbital frequency up to NLO S$^n$ requires only the 1PN non-spinning contribution and the NLO S$^{m<n}$.
However, for eccentric orbits, two frequencies exist: the radial and azimuthal frequencies, which agree at Newtonian order but differ starting at 1PN order. 
Therefore, expressing the redshift and spin-precession invariants in terms of frequencies, which is needed for GSF comparisons (see e.g. Ref.~\cite{Akcay:2015pza}), requires including higher PN orders at lower orders in spin.
In particular, the NLO S$^5$ and S$^6$ contributions require including the NNLO S$^3$ and S$^4$, which in turn require the N$^3$LO S$^1$ and S$^2$ contributions, in addition to the 4PN non-spinning contribution.

In Table~\ref{tab:PNspin}, we include all the PN orders needed to compute the NLO S$^6$ redshift and spin-precession frequency for eccentric orbits, while indicating the powers in $x := (M \Omega)^{2/3}$ at each order.

\subsection{PN Hamiltonian and its relation to the scattering angle}

For the Hamiltonian, we use a ``quasi-isotropic'' gauge~\cite{Vines:2018gqi}, in which the aligned-spin Hamiltonian depends only on $p^2$ and $r$ but not explicitly on $L$ except as an overall factor for the odd-in-spin contributions.

We begin by writing down a general ansatz for the two-body Hamiltonian as 
\begin{equation}
\mathcal{H} = \sum_{n=0}^6 \mathcal{H}_{S^n}  +\mathcal{O}(S^7),
\end{equation}
where each $\mathcal{H}_{S^n}$ contains spin contributions of $\mathcal{O}(S^n)$.
We fix the gauge freedom in the 0PM part of the Hamiltonian using the effective-one-body gauge, as in Ref.~\cite{Vines:2018gqi}, which is given by
\begin{equation}
\label{gauge0PM}
\mathcal{H}_\text{0PM} = M \sqrt{1 + 2\nu \left(\sqrt{1 + \vec{p}^2/\mu^2} - 1\right)}\,.
\end{equation}

The non-spinning part of the Hamiltonian in this gauge was provided to 3PN  order in Ref.~\cite{Vines:2018gqi}. Schematically it reads
\begin{align}
\mathcal{H}_{S^0} &= Mc^2 + \left( \frac{\vec{p}^2}{2\mu^2}-G u\right) \nonumber\\
&\quad +\frac{1}{c^2}\left[- \frac{1}{8} (1+\nu) \frac{\vec{p}^4}{\mu^4} + \alpha_{11} G \frac{\vec{p}^2}{\mu^2}u +\alpha_{12} G^2 u^2 \right] \nonumber\\
&\quad 
+\frac{1}{c^4} \bigg[
\frac{1}{16} \left(\nu ^2+\nu +1\right) \frac{\vec{p}^6}{\mu^6} 
+ \alpha_{21} G \frac{\vec{p}^4}{\mu^4} u  \nonumber\\
&\qquad
 + \alpha_{22} G^2 \frac{\vec{p}^2}{\mu^2} u^2 + \alpha_{23} G^3 u^3 \bigg]
+ \Order(1/c^6),
\end{align}
where $u:=M/r$, the 0PM coefficients are the PN expansion of Eq.~\eqref{gauge0PM}, and we obtained the coefficients $\alpha_{ij}$ through 4PN order by canonically transforming the Hamiltonian of Ref.~\cite{Damour:2015isa}. We provide the Hamiltonian in the Supplemental Material.

The spin contributions can be organized for odd-in-spin terms as
\begin{align}\label{Hodd}
\mathcal{H}_{S^n}&= \frac{L}{c^{1+2n}}\frac{{\mathfrak{a}}^{(n)}}{r^{n+1}} \cdot \bigg[ \pmb{\alpha}_{11}^{(n)} G u \nonumber\\
&\qquad
    +\frac{1}{c^2}\left( \pmb{\alpha} _{31}^{(n)}
    G \frac{\vec{p}^2}{\mu^2} u + \pmb{\alpha} _{32}^{(n)} G^2 u^2 \right) \nonumber\\
&\qquad
+ \frac{1}{c^4}\left( \pmb{\alpha} _{51}^{(n)}
G \frac{\vec{p}^4}{\mu^4} u + \pmb{\alpha} _{52}^{(n)} G^2 \frac{\vec{p}^2}{\mu^2} u^2 + \pmb{\alpha} _{53}^{(n)} G^3 u^3 \right) \nonumber\\
&\qquad + \Order(1/c^6) 
\bigg],
\end{align} 
with $n=1,3,5$.
For even-in-spin terms, a convenient parametrization is
\begin{align}\label{Heven}
\mathcal{H}_{S^n} &= \frac{\mu \mathfrak{a}^{(n)}   }{c^{2n}\, r^n}  \cdot \bigg[ \pmb{\alpha}_{01}^{(n)} G u  \nonumber\\
&\qquad
+\frac{1}{c^2}  \left(
\pmb{\alpha}_{21} ^{(n)} G \frac{\vec{p}^2}{\mu^2} u + \pmb{\alpha}_{22}^{(n)} G^2 u^2\right)  \nonumber\\
&\qquad
+ \frac{1}{c^4}\left( \pmb{\alpha} _{41}^{(n)}
G \frac{\vec{p}^4}{\mu^2} u + \pmb{\alpha} _{42}^{(n)} G^2 \frac{\vec{p}^2}{\mu^2} u^2 + \pmb{\alpha} _{43}^{(n)} G^3 u^3 \right) \nonumber\\
&\qquad + \Order(1/c^6) 
\bigg],
\end{align}
with $n=2,4,6$.
As summarized in Table~\ref{tab:PNspin}, we include the N$^3$LO in $\mathcal{H}_{S^1}$ and $\mathcal{H}_{S^2}$, the NNLO in $\mathcal{H}_{S^3}$ and $\mathcal{H}_{S^4}$, and the NLO in $\mathcal{H}_{S^5}$ and $\mathcal{H}_{S^6}$.

Let us explain our notation in Eqs.~\eqref{Hodd} and \eqref{Heven}. 
To render the treatment of the spin sectors systematic, we introduced two $(n+1)$-vectors  
\begin{equation}\label{param}
\begin{split}
&{\mathfrak{a}}^{(n)}:=(a_1^n ,\, a_1^{n-1} a_2,\, a_1^{n-2} a_2^2, \cdots, \, a_2 ^n), \\
&\pmb{\alpha}_{ij}^{(n)}:=(\alpha^{n0}_{ij}, \alpha^{(n-1)1}_{ij}, \alpha^{(n-2)2}_{ij} , \cdots, \alpha^{0n}_{ij} ),
\end{split}
\end{equation}
whose number of entries increases with the spin order.  The symbol ${\mathfrak{a}}^{(n)}$   is   a vector where each entry is a monomial of the form $a_1^k a_2^l$   such that   $k+l=n$. ${\mathfrak{a}}^{(n)}$  is dotted into another vector, $\pmb{\alpha}_{ij}^{(n)}$, whose entries are coefficients that are matched to the scattering angle ansatz in the following subsection. 
The indices have the following meaning:  $\alpha^{lm}_{ij}$ is a coefficient which multiplies a contribution proportional to $(v/c)^{i} G^j a_1^l a_2^m$. 
Although we employ Eqs.~\eqref{Hodd} and \eqref{Heven} through $\Order(S^6)$, the same structure holds to arbitrarily high orders in spin. 

Next, we compute the scattering angle from the Hamiltonian written in terms of the $\{\alpha\}$ coefficients.
This is done by solving the equation $\mathcal{H}=E$ for the radial momentum  $p_r^2=\vec{p}^2-L^2/r^2$,  perturbatively in a PN expansion, and obtaining the scattering angle via~\cite{Damour:2016gwp}
\begin{equation}
\label{HamToAngle}
\theta(E,L,\{\alpha\}) =-\pi - 2 \int_{r_\text{min}}^{\infty} \di r\, \frac{\partial}{\partial L} p_r (E,L, r, \{\alpha\}),
\end{equation}
where $r_\text{min}$ is the turning point, obtained from the largest root of $p_r(E,L,r)=0$.

It is important to note that $L$ in the above equation is the magnitude of the canonical orbital angular momentum $L =|\bm{r} \times \vec{p}|$, corresponding to the Newton-Wigner spin-supplementary condition~\cite{pryce1948mass,newton1949localized}, which is needed when working in a Hamiltonian formalism.
However, in the scattering angle, it is more convenient to use the covariant angular momentum $L_\text{cov}$, corresponding  to the covariant (Tulczyjew-Dixon) spin-supplementary condition~\cite{Tulczyjew:1959,Dixon:1970zza,Dixon:1970zz}, since the impact parameter $b=\sqrt{-b^\mu b_\mu}$ (the covariant initial separation of the binary) can be defined as
\begin{equation}
b := \frac{L_\text{cov}}{|\vec{p}|}.
\end{equation}
The canonical and covariant angular momenta are related via the total angular momentum
\begin{align}
J & = L + m_1 a_1 + m_2 a_2, \nonumber\\
& = L_\text{cov} + E_1 a_1 + E_2 a_2,
\end{align}
which is valid to all orders in spin. 
Solving this equation for $L$, leads to~\cite{Vines:2017hyw,Vines:2018gqi}
\begin{equation}
L=\frac{\mu\gamma v}{\Gamma}b+M \frac{\Gamma-1}{2}\left[
a_1+a_2 -\frac{\delta}{\Gamma}(a_1-a_2)
\right],
\end{equation}
where $\Gamma$  and $\gamma = 1/\sqrt{1-v^2}$ are given by Eqs.~\eqref{gammaGamma}.

Thus, we can evaluate the scattering angle in Eq.~\eqref{HamToAngle} after replacing $L$ by $b$ such that
\begin{equation}
\theta(b,v,\{\alpha\})=-\pi + 2 \frac{M\Gamma}{\mu\gamma v} \int_0^{1/b} \frac{\di u}{u^2}\, \frac{\partial}{\partial b} p_r (E, b, u, \{\alpha\}).
\end{equation}
We note that this integral involves divergences which can be treated by taking the Hadamard partie finie at the unperturbed turning point~\cite{Damour:1988mr,Damour:2016gwp}.
Finally, we equate the angle obtained from the Hamiltonian to an ansatz for the angle that satisfies the mass-ratio dependence of the PM-expanded scattering angle, as explained in the following subsection.

\subsection{Scattering angle parametrization}
\label{sec:angleParam}
The PM expansion of the scattering angle $\theta$, in the center-of-mass frame scaled by the energy $\theta/\Gamma$, has a simple dependence on the mass ratio that was proven in Ref.~\cite{Damour:2019lcq} (see also Ref.~\cite{Vines:2018gqi}) and extended to spin in Ref.~\cite{Antonelli:2020ybz}.
The structure is such that at each $n$PM order for non-spinning binaries, $\theta/\Gamma$ is a polynomial in the symmetric mass ratio $\nu$ of degree $\lfloor (n-1)/2 \rfloor$.
When spin is included, the angle also depends on the antisymmetric mass ratio $\delta:= (m_1 - m_2)/ M $, because of the symmetry under the exchange of the two bodies' labels. (See Refs.~\cite{Damour:2019lcq,Antonelli:2020ybz,Khalil:2021fpm} for a more detailed discussion.)

We split the spin contributions to the scattering angle $\theta$ as
\begin{equation}
\theta = \sum_{n = 0}^{6} \theta_{S^n} + \Order(S^7).
\end{equation}
As discussed at the beginning of this section, computing the redshift and spin-precession invariants for eccentric orbits to NLO S$^5$ and S$^6$, in a form suitable for comparison with GSF results, requires information from the N$^3$LO S$^1$ and S$^2$ contributions, in addition to the NNLO S$^3$ and S$^4$ contributions. In the following paragraphs, we overview the knowledge of each of these pieces, some of which are fully known and some of which we give in parametrized form.

The 3PN non-spinning contribution $\theta_{S^0}$ is given to 3PN order by Eq.~(4.32a) of Ref.~\cite{Vines:2018gqi}, while the N$^3$LO (4.5PN) S$^1$ contribution was derived in Refs.~\cite{Antonelli:2020aeb,Antonelli:2020ybz,Mandal:2022nty,Kim:2022pou} and is given by Eq.~(7) of Ref.~\cite{Antonelli:2020aeb}.
The N$^3$LO (5PN) S$^2$ contribution was derived in Refs.~\cite{Kim:2022bwv,Kim:2021rfj,Mandal:2022ufb}, and is given by Eqs.~(6.17)-(6.20) of Ref.~\cite{Mandal:2022ufb}.
Thus, the scattering angle completely determines the N$^3$LO S$^1$ and S$^2$ contributions to the Hamiltonian.

In Ref.~\cite{Guevara:2018wpp} (see Eq.~(1.12) there), the authors obtained the fully-relativistic one-loop (2PM) scattering angle of two Kerr BHs with aligned spins $a_1$ and $a_2$. 
As explained there, this angle is to be trusted only through spin order $\mathcal{O}(a_1^{n}a_2^{m})$ where \emph{both} $n$ and $m$ are less than or equal 4. 
This is because higher spin terms are affected by spurious poles inside the Compton amplitudes which constitute the needed cuts.

In Sec.~\ref{sec:testScatter}, we derive the scattering angle of a spinning test body  to 6PM order and fifth order in spin, starting from the parametrized worldline action of Ref.~\cite{Scheopner:2023rzp} in terms of generic Wilson coefficients. 
Our result reduces to the 2PM angle of Ref.~\cite{Guevara:2018wpp} at fourth order in spin, after setting the Wilson coefficients to their values for BHs~\cite{Siemonsen:2019dsu}.

Therefore, because of the structure of the scattering angle discussed above, the test-body scattering angle at 2PM order yields the full NLO (4.5PN) S$^3$ and NLO (5PN) S$^4$ dynamics for aligned spins, which were later derived for generic spins in Refs.~\cite{Levi:2019kgk,Levi:2020lfn,Levi:2022rrq}.
We write the S$^3$ part of the scattering angle through NNLO as
\begin{equation}
\theta_{S^3} = \theta_{S^3}^\text{test} + \theta_{S^3}^\text{3PM,4.5PN} + \theta_{S^3}^\text{3PM,5.5PN},
\end{equation}
where $\theta_{S^3}^{\text{test}}$ is obtained from the cubic-in-spin part of the test-body scattering angle~\eqref{testAngle}, $\theta_{S^3}^\text{3PM,4.5PN}$ is the 4.5PN part of the 3PM contribution, while $\theta_{S^3}^\text{3PM,5.5PN}$ includes the 5.5PN part at 3PM order.
The term $\theta_{S^3}^\text{3PM,4.5PN}$ is known from the NLO S$^3$ Hamiltonian, which only includes 2PM terms at 4.5PN order, and we parametrize the unknown term $\theta_{S^3}^\text{3PM,5.5PN}$ as
\begin{align}
\label{scatterS3ansatz}
&\frac{\theta_{S^3}^\text{3PM,5.5PN}}{\Gamma} = \frac{G^3M^3}{b^6 v} \Big[(f_{30} + \delta g_{30} + \nu h_{30}) a_1^3  \nonumber\\
&\qquad
+ (f_{21} + \delta g_{21} + \nu h_{21}) a_1^2 a_2
+ (f_{30} - \delta g_{30} + \nu h_{30}) a_2^3 \nonumber\\
&\qquad
+ (f_{21} - \delta g_{21} + \nu h_{21}) a_1 a_2^2  \Big],
\end{align}
where $f_{ij},\,\,g_{ij}$ and $h_{ij}$ are some dimensionless coefficients independent of the mass, and we made use of the symmetry under the exchange of the two bodies' labels to reduce the number of unknowns in the above ansatz.

Similarly, we write the S$^4$ scattering angle as
\begin{equation}
\theta_{S^4} = \theta_{S^4}^\text{test} + \theta_{S^4}^\text{3PM,5PN} + \theta_{S^4}^\text{3PM,6PN},
\end{equation}
where $\theta_{S^4}^{\text{test}}$ is obtained from the quartic-in-spin part of Eq.~\eqref{testAngle} after setting the Wilson coefficients to zero, $\theta_{S^4}^\text{3PM,5PN}$ is the known 5PN part of the 3PM contribution, and we write the unknown contribution $\theta_{S^4}^\text{3PM,6PN}$ as
\begin{align}
\label{scatterS4ansatz}
&\frac{\theta_{S^4}^\text{3PM,6PN}}{\Gamma} = \frac{G^3M^3}{b^7 v^2} \Big[
(f_{40} + \delta g_{40} + \nu h_{40}) a_1^4  \nonumber\\
&\quad
+ (f_{31} + \delta g_{31} + \nu h_{31}) a_1^3 a_2
+ (f_{22} + \nu h_{22}) a_1^2 a_2^2 \nonumber\\
&\quad
+ (f_{31} - \delta g_{31} + \nu h_{31}) a_1 a_2^3 
+ (f_{40} - \delta g_{40} + \nu h_{40}) a_2^4 \Big].
\end{align}

At NLO S$^5$ and S$^6$, we parametrize the scattering angle as
\begin{subequations}
\begin{align}
\theta_{S^5} &= \theta_{S^5}^\text{test} = \bar\theta_{S^5}^\text{test}+\Delta\theta_{S^5}, \\
\theta_{S^6} &= \theta_{S^6}^\text{test} = \bar\theta_{S^6}^\text{test}+\Delta\theta_{S^6},
\end{align}
\end{subequations}
where the full NLO can in principle be obtained from the test-body scattering angle~\eqref{testAngle} derived in Sec.~\ref{sec:testScatter}.
However, since it includes Wilson coefficients whose values for BHs are unknown, we split the NLO contributions into terms of order $\Order(a_1^{\leq4} a_2^{\leq 4})$, denoted $\bar\theta_{S^5}^\text{test}$ and $\bar\theta_{S^6}^\text{test}$, and terms with higher orders in spin, which we parametrize as 
\begin{subequations}
\label{deltatheta56}
\begin{align}
\frac{\Delta\theta_{S^5}}{\Gamma} &=  
  \frac{\pi G^2 M^2}{b^7 v}   \left[
  (f_{50}+\delta\, g_{50}) a_1^5 + (f_{50}-\delta\, g_{50}) a_2^5 \right], \\
\frac{\Delta\theta_{S^6}}{\Gamma}  &=  
  \frac{\pi G^2 M^2}{b^8 v^2} \Big[
  (f_{60}+\delta\, g_{60}) a_1^6 + (f_{60}-\delta\, g_{60}) a_2^6 \nonumber\\
 &\quad
 + (f_{51}+\delta\, g_{51}) a_1^5 a_2 +  (f_{51}-\delta\, g_{51}) a_1 a_2^5 \Big].
\end{align}
\end{subequations}
We stress that the above parametrization for the scattering angle is completely fixed by parity invariance, dimensional analysis and a well-behaved small-velocity expansion.

We solve for the $f_{ij}$ and $g_{ij}$ unknowns in the NNLO S$^3$ and S$^4$ parts from the 3PM test-body scattering angle in Eq.~\eqref{testAngle}.
Comparison with the ansatzes in Eqs.~\eqref{scatterS3ansatz} and \eqref{scatterS4ansatz} yields the following values for the unknowns:
\begin{alignat}{2}
f_{30} &= -180, &\qquad g_{30} &= -100, \nonumber\\
f_{21} &= -540, &\qquad g_{21} &= -100, \nonumber\\
f_{40} &= 650, &\qquad g_{40} &= 400, \\
f_{31} &= 2600, &\qquad g_{31} &= 800, \nonumber\\
f_{22} &= 3900. && \nonumber
\end{alignat}
We solve for the remaining unknowns ($h_{30}, h_{21}, h_{40}, h_{31}$) using the 1SF redshift and spin-precession invariants, but $h_{22}$ requires 1SF results that are unavailable in the literature, namely the redshift (or spin-precession frequency) to quadratic (linear) order in the secondary spin and quadratic order in the Kerr spin.

We also use 1SF results to solve for the unknowns in the NLO S$^5$ and S$^6$ scattering angle, and in Sec.~\ref{subsec:constrains_wilson_coeffs}, we relate those unknowns to the Wilson coefficients in the test-body scattering angle~\eqref{testAngle}.
Additionally, in Sec.~\ref{sec:compton}, we solve for them from  gravitational-wave scattering.

\subsection{First law of binary mechanics}
Having related the coefficients of the Hamiltonian to the scattering angle ansatz, the next step is to compute the redshift and spin-precession invariants by making use of the first law of binary mechanics.

The first law was derived in Refs.~\cite{LeTiec:2011ab, Blanchet:2012at,LeTiec:2015kgg,Blanchet:2017rcn,Fujita:2016igj,Antonelli:2020ybz}, and it is known for spinning binaries in eccentric orbits to first order in spin.
It reads
\begin{equation}
\label{firstlaw}
\di E - \Omega_r \di I_r - \Omega \di L = \sum_{\mr i} ( z_{\mr i} \di m_{\mr i} + \Omega_{S_{\mr i}} \di S_{\mr i} ),
\end{equation}
where $\Omega_r$ and $\Omega$ are the radial and azimuthal frequencies, respectively, $I_r$ is the radial action, $z_{\mr i}$ is the redshift, $\Omega_{S_{\mr i}}$ is the spin-precession frequency, and the subscript $\mr i = 1,2$ labels the two bodies.

The first law implies that the redshift and spin-precession frequency can be computed by taking derivatives of the Hamiltonian (energy) while keeping $I_r$ and $L$ constant, i.e.,
\begin{equation}
z_{\mr i} = \left. \frac{\partial H}{\partial m_{\mr i}} \right|_{I_r,L}, \qquad
\Omega_{S_{\mr i}} = \left. \frac{\partial H}{\partial S_{\mr i}} \right|_{I_r,L}.
\end{equation}
However, it is difficult to express the Hamiltonian for spinning binaries in terms of the radial action, but we can still obtain a gauge-invariant quantity by taking an appropriate orbit average of the Hamiltonian derivatives, such as averaging over an eccentric orbit for aligned-spin binaries. (See, e.g., Sec.~IV.A of Ref.~\cite{Antonelli:2020ybz} for the derivation.)
In that case, we have
\begin{equation}
\label{zOmegaOrbAvg}
z_{\mr i} = \left\langle \frac{\partial H}{\partial m_{\mr i}} \right\rangle, \qquad
\Omega_{S_{\mr i}} = \left\langle \frac{\partial H}{\partial S_{\mr i}} \right\rangle.
\end{equation}
In Sec.~\ref{sec:zPsiEcc}, we explain how to perform this average.

For circular orbits, the calculation of the redshift and spin-precession frequency is much easier, since in that case, the first law can be written as
\begin{equation}
\di \mathfrak{M} \overset{\text{circ}}{=} \sum_{\mr i} ( z_{\mr i} \di m_{\mr i} + \Omega_{S_{\mr i}} \di S_{\mr i} ),
\end{equation}
where $\mathfrak{M}$ is the system's free energy defined as 
\begin{equation}
\label{freeEnergy}
\mathfrak{M}(\Omega) := E(\Omega)- \Omega\, L (\Omega),
\end{equation} 
leading to
\begin{equation}
\label{zOmegaCirc}
z_{\mr i}=\frac{\partial \mathfrak{M}(\Omega)}{\partial m_{\mr i}}, \qquad  
\Omega_{S_{\mr i}}=\frac{\partial \mathfrak{M}(\Omega)}{\partial S_{\mr i}}.
\end{equation}

It should be stressed that the first law in Eq.~\eqref{firstlaw} is only valid to first order in spin. Meaning that, when computing the redshift and spin-precession frequency of the lighter object, $m_2$, our results for $z_2$ and $\Omega_{S_2}$ are valid to linear order in $S_2$, but to all orders in $S_1$, since the primary object does not affect the derivatives $\partial /\partial m_2$ and $\partial / \partial S_2$ in Eq.~\eqref{zOmegaOrbAvg}.

In the following subsection, we first perform the calculation of the redshift and spin-precession frequency for circular orbits, which is enough to solve for all the unknowns in the scattering angle, except for $h_{22}$, which as explained above,  requires 1SF results unavailable in the literature.
Then, in Sec.~\ref{sec:zPsiEcc}, we perform the calculation for eccentric orbits to obtain additional constraints that confirm our results.\\

\subsection{Redshift and spin-precession invariants for circular orbits}
\label{sec:zPsiCirc}
To obtain $\mathfrak{M}(\Omega)$ for circular orbits, we solve Hamilton's equations
\begin{subequations}
\begin{gather}
\dot{p}_r = -\frac{\partial}{\partial r}\mathcal{H}(r,p_r = 0,L) = 0, \\
\dot{\phi} = \frac{\partial}{\partial L}\mathcal{H}(r,p_r = 0,L) = \Omega
\end{gather}
\end{subequations}
for $r(\Omega)$ and $L(\Omega)$, perturbatively in a PN expansion.
We then substitute the solution in $E=\mathcal{H}(r,p_r = 0,L)$ to obtain $E(\Omega)$, yielding $\mathfrak{M}(\Omega)$ via Eq.~\eqref{freeEnergy}.
The redshift $z_2$ and spin-precession frequency $\Omega_{S_2}$ are simply obtained by differentiating $\mathfrak{M}(\Omega)$ as in Eqs.~\eqref{zOmegaCirc}.
For convenience, we define the spin-precession invariant
\begin{equation}
\psi_2 := \frac{\Omega_{S_2}}{\Omega},
\end{equation}
and express $z_2$ and $\psi_2$ as functions of $x := (M\Omega)^{2/3}$. 

Next, we expand $z_2$ and $\psi_2$ to first order in the mass ratio $q := m_2/m_1$, by making use of the following variables:
\begin{subequations}
\begin{align}
y &:= (m_1 \Omega)^{2/3} = \frac{x}{(1 + q)^{2/3}}, \\
\chi &:= \frac{a_1}{m_1},
\end{align} 
\end{subequations}
setting $a_2 = 0$, and replacing $\delta$ and $\nu$ by
\begin{equation}
\delta = \frac{1 - q}{1 + q}, \qquad
\nu = \frac{q}{(1+q)^2}\,.
\end{equation}

Our result for the redshift reads
\begin{widetext}
\begin{subequations}
\label{UCirc}
\begin{align}
U_2 &:= \frac{1}{z_2} = U_2^{(0)} + q U_2^{(1)} + \Order(q^2), \\
U_2^{(0)} &= 1 +\frac{3}{2} y +\frac{27}{8} y^2 +\frac{135}{16} y^3 + \frac{2835}{128} y^4  + \Order(y^5) 
+ \chi \left[-2 y^{5/2}-9 y^{7/2}-\frac{135}{4} y^{9/2} + \Order(y^{11/2})\right]\nonumber\\
&\quad
+ \chi^2 \left[\frac{y^3}{2}+\frac{31}{12} y^4+\frac{255}{16} y^5 + \Order(y^6)\right]
+ \chi^3 \left[-\frac{79}{27} y^{11/2} -\frac{145}{6} y^{13/2} + \Order(y^{15/2})\right]
+ \chi^4 \left[\frac{3}{8} y^6 + \frac{4333}{1296} y^7 + \Order(y^8)\right] \nonumber\\
&\quad
+ \chi^5 \left[\left(\frac{8 f_{50}}{15}+\frac{8 g_{50}}{15}+14\right) y^{15/2} + \Order(y^{17/2})\right]
+ \chi^6 \left[\left(\frac{16 f_{60}}{35}+\frac{16 g_{60}}{35}-21\right) y^8 + \Order(y^9)\right] + \Order(\chi^7), \\
U_2^{(1)} &= -y -2 y^2 -5 y^3 + \left(\frac{41 \pi ^2}{32}-\frac{121}{3}\right) y^4
+ \chi \left[\frac{7 y^{5/2}}{3}+\frac{46 y^{7/2}}{3}+77 y^{9/2} + \Order(y^{11/2})\right] \nonumber\\
&\quad
+ \chi^2 \left[-y^3-\frac{86}{9} y^4-\frac{577}{9} y^5+\Order(y^6)\right]
+ \chi^3 \left[y^{9/2} +\frac{1526}{81} y^{11/2} +\left(\frac{3 h_{30}}{8}+\frac{72469}{648}\right)y^{13/2} + \Order(y^{15/2})\right]\nonumber\\
&\quad
+ \chi^4 \left[-2 y^6 + \left(\frac{5 h_{40}}{16}+\frac{506825}{7776}\right)y^7 + \Order(y^8) \right]
+ \chi^5 \left[\left(-\frac{16 f_{50}}{15}-\frac{16 g_{50}}{5}-47\right) y^{15/2} + \Order(y^{17/2})\right]\nonumber\\
&\quad
+ \chi^6  \left[\left(91-\frac{32 f_{60}}{21}-\frac{352 g_{60}}{105}\right) y^8 + \Order(y^9)\right] + \Order(\chi^7),
\end{align}
\end{subequations}
while for the spin-precession invariant, we obtain
\begin{subequations}
\label{PsiCirc}
\begin{align}
\psi_2 &= \psi_2^{(0)} + q \psi_2^{(1)} + \Order(q^2), \\
\psi_2^{(0)} &= \frac{3}{2} y +\frac{9}{8} y^2 + \frac{27}{16} y^3 + \Order(y^4)
+ \chi \left[-y^{3/2}-\frac{y^{5/2}}{2}-\frac{15}{8} y^{7/2} + \Order(y^{9/2})\right]
+ \chi^2 \left[-\frac{y^3}{2}+\frac{7}{12} y^4 + \frac{37}{16} y^5 + \Order(y^6)\right] \nonumber\\
&\quad
+ \chi^3 \left[-\frac{y^{9/2}}{2} -\frac{37}{108} y^{11/2} + \Order(y^{13/2})\right] 
+ \chi^4 \left[-\frac{3}{8} y^6 + \Order(y^7)\right] 
+ \chi^5 \left[\left(117-\frac{16 f_{51}}{35}-\frac{16 g_{51}}{35}\right) y^{13/2} + \Order(y^{15/2})\right] \nonumber\\
&\quad
+ \Order(\chi^6), \\
\psi_2^{(1)} &= y^2 - 3 y^3 + \Order(y^4)
+ \chi \left[y^{3/2}+0\times y^{5/2}+ \frac{16}{3} y^{7/2} + \Order(y^{9/2})\right] 
+ \chi^2 \left[-3 y^4 + \left(\frac{7409}{144}-\frac{3 h_{21}}{16}\right) y^5 + \Order(y^6)\right]\nonumber\\
&\quad
+ \chi^3 \left[y^{9/2} + \left(-\frac{5 h_{31}}{32}-\frac{1051}{16}\right) y^{11/2} + \Order(y^{13/2})\right] 
+ \chi^4 \left[0 \times y^6 + \Order(y^7)\right] \nonumber\\
&\quad
+ \chi^5 \left[\left(\frac{16 f_{51}}{21}+\frac{176 g_{51}}{105}-237\right) y^{13/2} + \Order(y^{15/2})\right] + \Order(\chi^6).
\end{align}
\end{subequations}
\end{widetext}
 
Let us pause for a moment and interpret the expressions above in terms of amplitudes.
The LO (NLO) PN contributions, corresponding to tree-level (one-loop) amplitude input appear in the redshift at order $y^{n+1}$ ($y^{n+2}$) for even powers of spin $\chi^n$, while for odd powers in spin $\chi^m$, they appear at orders $y^{m+3/2}$ ($y^{m+5/2}$).
For example, the tree-level, spin-squared scattering data is contained in the $\chi^2 y^3$ piece, and the one-loop S$^5$ information is in $\chi^5 y^{15/2}$. 
Similarly, in the spin-precession invariant, the LO (NLO) power counting goes as follows: for an even-in-spin term $\chi^n$ the LO (NLO) scattering data is in $y^{n+1}$ ($y^{n+2}$) while for an odd-in-spin term $\chi^m$ the LO (NLO) is $y^{m+1/2}$ ($y^{m+3/2}$). 
Table~\ref{tab:PNspin} summarizes the powers in $x$, or $y$, of the PN and spin orders considered in this paper.

Having computed the redshift and spin-precession invariants, we can compare them to GSF results to solve for the unknowns.
The test-mass limits (0SF parts) of these invariants are known to all PN orders, and are given by~\cite{Bini:2018ylh,Siemonsen:2019dsu}
\begin{subequations}
\label{UPsi0SF}
\begin{align}
U_\text{0SF}&= \left[1+\chi y^{3/2}-3 y (1-\chi y^{3/2})^{1/3}\right]^{-1/2} \nonumber\\
&\qquad\quad \times \left(1-\chi y^{3/2}\right)^{-1/2}, \\
\psi_\text{0SF} &= 1-\left(\frac{ 1+\chi y^{3/2}-3 y (1-\chi y^{3/2})^{1/3}}{1-\chi y^{3/2}}\right)^{1/2}.
\end{align}
\end{subequations}
In Appendix~\ref{app:zPsiTest}, we derive the 0SF redshift and spin-precession invariants for eccentric orbits.

The 1SF contribution to the redshift was derived in Ref.~\cite{Kavanagh:2016idg} for a non-spinning secondary in a Kerr background to all orders in the Kerr spin and to 8PN order,
while the spin-precession invariant was similarly derived in Ref.~\cite{Bini:2018ylh}.
The results of both references are included in the Black Hole Perturbation Toolkit~\cite{BHPToolkit}.

By comparing our results with GSF results, we obtain the following constraints at NNLO S$^3$ and S$^4$:
\begin{subequations}
\label{S3S4solns}
\begin{alignat}{2}
&\text{1SF } z: \quad && a^3 y^{13/2} \left(3 h_{30}-451\right) = 0, \\
&\text{1SF } \psi: \quad && a^2 y^5 \left(h_{21}-371\right) = 0, \\
&\text{1SF } z: \quad && a^4 y^7 \left(2 h_{40}+631\right) = 0, \\
&\text{1SF } \psi: \quad && a^3 y^{11/2} \left(h_{31}+446\right) = 0,
\end{alignat}
\end{subequations}
leading to the solution
\begin{subequations}
\begin{alignat}{2}
h_{30} &= \frac{451}{3}, &\qquad h_{21} &= 371, \\
h_{40} &= -\frac{631}{2}, &\qquad h_{31} &= -446.
\end{alignat}
\end{subequations}

The constraint at NLO S$^5$ from the 0SF redshift is
\begin{alignat}{2}\label{eq:0SFspin5constrain}
&\text{0SF } z: \quad && a^5 y^{15/2} \left(4 f_{50}+4 g_{50}+105\right) = 0,
\end{alignat}
leading to the following relation:
\begin{align}\label{S5solnTest}
f_{50} = -\frac{105}{4}-g_{50}.
\end{align}
The 1SF constraint from the redshift would be enough to determine the remaining unknowns, but at $\Order(S^5)$, the 1SF redshift depends on transcendental functions of spin, and it is not clear how to translate that nonperturbative information into a PN spin expansion; we discuss possible ways of resolving this issue in Sec.~\ref{sec:NLOS5soln}.
No constraints are available from the spin-precession frequency, since it would require a GSF calculation at fourth order of the secondary spin in a Schwarzschild background.

The NLO S$^6$ unknowns can be solved for with no ambiguity since no transcendental functions of spin appear at $\Order(y^8)$ in the redshift or at $\Order(y^{13/2})$ in the spin-precession invariant. 
Comparison with GSF results yields the following constraints:
\begin{subequations}
\begin{alignat}{2}
&\text{0SF } z: \quad && a^6 y^8 \left(16 f_{60}+16 g_{60}-735\right) = 0, \label{constCirc0SFz}\\
&\text{0SF } \psi: \quad && a^5 y^{13/2} \left(16 f_{51}+16 g_{51}-4095\right) = 0, \label{constCirc0SFpsi} \\
&\text{1SF } z: \quad && a^6 y^8 \left(-160 f_{60}-352 g_{60}+9555\right) = 0, \\
&\text{1SF } \psi: \quad && a^5 y^{13/2} \left(80 f_{51}+176 g_{51}-24885\right) = 0,
\end{alignat}
\end{subequations}
leading to the solution
\begin{equation}\label{S6Soln}
\begin{alignedat}{2}
f_{60} &= \frac{2205}{64}, &\qquad g_{60} &= \frac{735}{64}, \\
f_{51} &= 210, &\qquad g_{51} &= \frac{735}{16}.
\end{alignedat}
\end{equation}

With the solutions obtained above, the NNLO S$^3$ and S$^4$ contributions to the  scattering angle are given by
\begin{widetext}
\begin{subequations}
\begin{align}
\frac{\theta_{S^3}}{\Gamma} &= -\frac{4 G M a_+^3}{b^4 v} 
+ \frac{\pi G^2 M^2}{b^5 v^3} \left[-\frac{3}{2} \delta a_- a_+^2  -\frac{21}{2}  a_+^3 + \left(-\frac{3}{32} \delta a_-^3 -\frac{141}{32} \delta a_+^2 a_- +\frac{21}{32} a_+ a_-^2-\frac{357 a_+^3}{32}\right)v^2\right] \nonumber\\
&\quad + \frac{G^3 M^3}{b^6 v^5} \bigg\{
{-}20 \delta a_- a_+^2 -100 a_+^3
+ \left[\left(\frac{100 \nu }{3}-\frac{1400}{3}\right) a_+^3 -200 a_- a_+^2 \delta +\frac{200}{3} \nu a_-^2 a_+ \right] v^2 \nonumber\\
&\qquad
+ \left[-100 \delta a_- a_+^2 + \left(\frac{391 \nu }{3}-180\right)a_+^3 +20 \nu a_-^2 a_+ \right] v^4 + \Order(v^6)
\bigg\} + \Order(G^4), \\
\frac{\theta_{S^4}}{\Gamma} &= \frac{2G M}{b^5 v^2} a_+^4 \left(1 +v^2\right)
+ \frac{\pi G^2 M^2}{b^6 v^4} \left[
\frac{15}{4} a_+^4
+ \left(\frac{105}{16} \delta a_- a_+^3 +\frac{705}{32} a_+^4 -\frac{15}{32} a_-^2 a_+^2\right) v^2
\right] \nonumber\\
&\quad
+ \frac{G^3 M^3}{b^7 v^6} \bigg\{
30 a_+^4 
+ \left[200 \delta a_- a_+^3 +a_+^4 (550-10 \nu )-40 \nu a_-^2 a_+^2 \right] v^2 
+ \bigg[400 \delta a_+^3 a_- + \left(\frac{261 \nu }{16}-\frac{975}{4}\right) a_-^4\nonumber\\
&\qquad\quad
+ \left(\frac{975}{2}-\frac{1893 \nu }{8}\right) a_+^2 a_-^2
+ \left(\frac{1625}{4}-\frac{1523 \nu }{16}\right) a_+^4 
+ (h_{22} \nu +3900) a_1^2 a_2^2\bigg] v^4 + \Order(v^6)
\bigg\} + \Order(G^4),
\end{align}
where we defined $a_+ := a_1 + a_2$ and $a_- := a_1 - a_2$.
The NLO S$^5$ and S$^6$ contributions read
\begin{align}
\frac{\theta_{S^5}}{\Gamma} &= -\frac{4 GM a_+^5}{b^6 v} 
+ \frac{\pi G^2 M^2}{b^7 v^3} \bigg\{
{-}\frac{45}{16} \delta a_- a_+^4 -\frac{315}{16}  a_+^5 
+ \bigg[\frac{315 \delta a_-^5}{512}+\frac{1515}{256} \delta a_+^2 a_-^3 -\frac{3345}{512} \delta a_+^4 a_- +\frac{2625}{512} a_+ a_-^4\nonumber\\
&\qquad
+\frac{3045}{256} a_+^3 a_-^2 -\frac{8715 a_+^5}{512}
- a_1^5 \left(\frac{105}{4}+g_{50}(1-\delta)\right)
-a_2^5 \left(\frac{105}{4}+g_{50}(1+\delta)\right)\bigg] v^2 + \Order(v^4)
\bigg\}  + \Order(G^3),\\
\frac{\theta_{S^6}}{\Gamma} &= \frac{2G M}{b^7 v^2} a_+^6 \left(1 + v^2\right)
+ \frac{\pi G^2 M^2}{b^8 v^4} \left[
\frac{105}{16} a_+^6
+ \left(\frac{735}{64} \delta a_- a_+^5 +\frac{4515}{128} a_+^6 -\frac{105}{128} a_-^2 a_+^4\right) v^2 + \Order(v^4)
\right] + \Order(G^3),
\end{align}
\end{subequations}
\end{widetext}
where we left the coefficient $g_{50}$ in $\theta_{S^5}$ since there is more than one possibility of determining it from 1SF results, as explained in Sec.~\ref{sec:NLOS5soln}.

In the following subsection, we obtain additional constraints on the unknowns by comparing our results with eccentric-orbit GSF results. The solutions we obtain are consistent with the circular-orbit calculation.

\subsection{Redshift and spin-precession invariants for eccentric orbits}
\label{sec:zPsiEcc}
For eccentric orbits, the redshift and spin-precession invariants can be computed using Eq.~\eqref{zOmegaOrbAvg}, by taking derivatives of the Hamiltonian, orbit averaging the result and expressing it in terms of frequencies.
To perform such a calculation, we follow the method in Refs.~\cite{Bini:2019lcd,Bini:2019lkm}, by making use of the Keplerian parametrization for the conservative dynamics~\cite{darwin1959gravity}
\begin{equation}
\label{Keplerian}
r = \frac{M}{u_p\left(1+e \cos \zeta\right)},
\end{equation}
where $u_p$ is the inverse of the semi-latus rectum (scaled by the total mass), $e$ is the eccentricity, and $\zeta$ is the relativistic anomaly.

The radial and azimuthal periods can be calculated from the Hamiltonian using
\begin{subequations}
\begin{align}
T_r &:= \oint \di t = 2 \int_0^\pi \left(\frac{\partial \mathcal{H}}{\partial p_r}\right)^{-1} \frac{\di r}{\di\zeta} \di\zeta \,,  \\
T_\phi &:= \oint \di\phi = 2 \int_0^\pi \frac{\partial \mathcal{H}}{\partial L}\left(\frac{\partial \mathcal{H}}{\partial p_r}\right)^{-1} \frac{\di r}{\di\zeta} \di\zeta \,, 
\end{align}
\end{subequations}
from which we define the gauge-independent variables
\begin{equation}
\label{xiota}
x := (M \Omega)^{2/3}, \qquad \iota := \frac{3 x}{k},
\end{equation}
where $\Omega = T_\phi/T_r$ is the orbit-averaged azimuthal frequency, and $k$ is the fractional periastron advance computed from
\begin{equation}
k = \frac{T_\phi}{2\pi} - 1.
\end{equation}

Then, we perform the orbit averages as follows:
\begin{subequations}
\begin{align}
z_{\mr i} &= \left\langle \frac{\partial \mathcal{H}}{\partial m_{\mr i}} \right\rangle  = \frac{1}{T_r} \oint \frac{\partial \mathcal{H}}{\partial m_{\mr i}} \di t,  \\
\Omega_{S_{\mr i}} &= \left\langle \frac{\partial \mathcal{H}}{\partial S_{\mr i}} \right\rangle = \frac{1}{T_r} \oint \frac{\partial \mathcal{H}}{\partial S_{\mr i}} \di t,
\end{align}
\end{subequations}
where we express the integrands as functions of the Keplerian parameters $(u_p,e,\zeta)$, then evaluate the time integrals as
\begin{align} 
\langle f\rangle(u_p,e) = \frac{1}{T_r}  \int_{0}^{2 \pi}  f(u_p,e,\zeta)  \left( \frac{\partial \mathcal{H}}{\partial p_r} \right)^{-1} \frac{\di r}{\di \zeta} \, \di\zeta, 
\end{align}
for any function $f$.

The above steps yield the redshift and spin-precession invariants, $z_{\mr i}(u_p, e)$ and $\psi_{\mr i}(u_p, e)$, in terms of the gauge-dependent variables $u_p$ and $e$.
Therefore, we first invert the PN expansions for $x$ and $\iota$ from Eq.~\eqref{xiota} to obtain $u_p(x,\iota)$ and $e(x,\iota)$, which yield gauge-invariant expressions for $z_{\mr i}(x, \iota)$ and $\psi_{\mr i}(x, \iota)$.
Note that the denominator of $\iota$ in Eq.~\eqref{xiota} is of order 1PN, which effectively scales down the PN order such that, to obtain $z_{\mr i}$ and $\psi_{\mr i}$, we need to include higher PN orders at lower spin orders, as summarized in Table~\ref{tab:PNspin}.

The next step is to expand $z_2(x, \iota)$ and $\psi_2(x, \iota)$ (of the lighter object) to first order in the mass ratio $q = m_2/m_1$ and zeroth order in the secondary spin $a_2$, and we express the result in terms of the following variables:
\begin{subequations}\label{ylambda}
\begin{align}
y &:= (m_1 \Omega)^{2/3} = \frac{x}{(1+q)^{2/3}}, \\
\lambda &:=\frac{3y}{k}  = \frac{\iota}{(1+q)^{2/3}} \,.
\end{align}
\end{subequations}

To compare $z_2(y,\lambda)$ and $\psi_2(y,\lambda)$ with 1SF results, we need to write them in terms of the variables used in GSF calculations, which are the Kerr geodesic variables $(u_p, e)$.
These are technically in a different gauge from the one we use in the PN calculations, but we use the same notation for simplicity.
In Appendix~\ref{app:zPsiTest}, we show how to derive the relations $y(u_p, e)$ and $\lambda(u_p, e)$.
Typically, these relations are derived in the literature in an eccentricity expansion~\cite{Bini:2019lcd,Bini:2016dvs,Antonelli:2020ybz}, but we were able to obtain them, in addition to the redshift and spin-precession frequency, without any eccentricity expansions.

In the test-mass limit, our result for the (inverse) redshift $U_2 := 1/z_2 = U_2^{(0)} + q U_2^{(1)} + \Order(q^2)$ is given by 
\begin{align}
U_2^{(0)} &= [\dots]
+\chi^5 \varepsilon^3 u_p^{15/2} \bigg[
{-}\frac{33 \varepsilon ^7}{2}+\frac{903 \varepsilon ^6}{4}-513 \varepsilon ^5 \nonumber\\
&\qquad
-\frac{7443 \varepsilon ^4}{8}+\frac{3659 \varepsilon ^3}{2}+\frac{14457 \varepsilon ^2}{4}-84 \varepsilon -\frac{32921}{8} \nonumber\\
&\qquad
+ \left(\frac{77}{10}-\frac{\varepsilon ^6}{6}+\frac{7 \varepsilon ^4}{2}-\frac{21 \varepsilon ^2}{2}\right) \left(f_{50} + g_{50}\right)\bigg] \nonumber\\
&\quad
+ \chi^6 \varepsilon^3 u_p^8 \bigg[
{-}\frac{495 \varepsilon ^6}{16}+\frac{351 \varepsilon ^5}{2}-\frac{201 \varepsilon ^4}{16}-612 \varepsilon ^3\nonumber\\
&\qquad
+ \left(\frac{33}{5}-\frac{\varepsilon ^6}{7}+3 \varepsilon ^4-9 \varepsilon ^2\right) \left(f_{60} + g_{60}\right) \nonumber\\
&\qquad
-\frac{10485 \varepsilon ^2}{16}+\frac{17829}{16}
\bigg],
\end{align}
where the dots $[\dots]$ indicate lower PN contributions that are the same as the 0SF redshift derived in Appendix~\ref{app:zPsiTest} (cf. Eq.~\eqref{U0test}), and to ease the notation, we defined the variable
\begin{equation}
\varepsilon := \sqrt{1 - e^2}\,.
\end{equation}
Note that the 0SF eccentric-orbit redshift only depends on the sums $f_{50}+g_{50}$ and $f_{60}+g_{60}$, meaning that it cannot provide additional constraints on these unknowns than the circular-orbit comparison from the previous subsection; indeed,
comparing $U_2^{(0)}$ with the 0SF result from Eq.~\eqref{U0test} leads to the same circular-orbit constraint in Eq.~\eqref{constCirc0SFz}.

At first order in the mass ratio, our result for the redshift reads
\begin{widetext}
\small
\begin{align}\label{eq:red_shift_eccentric}
U_2^{(1)} &= -\varepsilon ^2 u_p -2 \varepsilon ^4 u_p^2
+\left(5 \varepsilon ^6-9 \varepsilon ^5+4 \varepsilon ^4-5 \varepsilon ^3\right) u_p^3
+ \left[\frac{27 \varepsilon ^7}{2}-4 \varepsilon ^8+16 \varepsilon ^6+\left(\frac{79}{6}-\frac{41 \pi ^2}{64}\right) \varepsilon ^5+16 \varepsilon ^4+\left(\frac{123 \pi ^2}{64}-95\right) \varepsilon ^3\right] u_p^4 \nonumber\\
&\quad
+ \chi \, \varepsilon^3 \bigg\{
(5-2 \varepsilon ) u_p^{5/2}
+\left(-8 \varepsilon ^3-5 \varepsilon ^2-12 \varepsilon +43\right) u_p^{7/2}
+\left(30 \varepsilon ^5-\frac{581 \varepsilon ^4}{8}-\frac{27 \varepsilon ^3}{2}-\frac{307 \varepsilon ^2}{2}-60 \varepsilon +\frac{2853}{8}\right) u_p^{9/2} \nonumber\\
&\qquad
+ \left[\frac{1447 \varepsilon ^6}{8}-32 \varepsilon ^7+68 \varepsilon ^5- \!\left(\frac{9773}{24}+\frac{569 \pi ^2}{256}\right)\! \varepsilon ^4+116 \varepsilon ^3+\!\left(\frac{2189 \pi ^2}{128}-\frac{58253}{24}\right)\! \varepsilon ^2-280 \varepsilon +\frac{231391}{72}-\frac{13355 \pi ^2}{768}\right]\! u_p^{11/2}\!
\bigg\} \nonumber\\
&\quad
+ \chi^2 \, \varepsilon^3 \bigg\{
(\varepsilon -2) u_p^3
+\left(\varepsilon ^3+29 \varepsilon ^2+12 \varepsilon -56\right) u_p^4
+\left(21 \varepsilon ^4-35 \varepsilon ^5+7 \varepsilon ^3+\frac{1011 \varepsilon ^2}{2}+80 \varepsilon -\frac{1363}{2}\right) u_p^5  
+ \bigg[121 \varepsilon ^7-\frac{3689 \varepsilon ^6}{8}\nonumber\\
&\qquad\quad
+\frac{153 \varepsilon ^5}{2}+\left(\frac{4403 \pi ^2}{4096}-\frac{4375}{24}\right) \varepsilon ^4-96 \varepsilon ^3+\left(\frac{54307}{8}-\frac{16767 \pi ^2}{2048}\right) \varepsilon ^2+448 \varepsilon +\frac{33875 \pi ^2}{4096}-\frac{175747}{24}\bigg] u_p^6
\bigg\} \nonumber\\
&\quad
+ \chi^3 \, \varepsilon^3 \bigg\{
\left(2 \varepsilon ^3-\frac{37 \varepsilon ^2}{2}-4 \varepsilon +\frac{47}{2}\right) u_p^{9/2}
+\left(12 \varepsilon ^5+\frac{309 \varepsilon ^4}{4}-27 \varepsilon ^3-588 \varepsilon ^2-40 \varepsilon +\frac{2463}{4}\right) u_p^{11/2} \nonumber\\
&\qquad
+\bigg[-130 \varepsilon ^7+\frac{4749 \varepsilon ^6}{16}-\frac{283 \varepsilon ^5}{4}-\frac{2875 \varepsilon ^3}{4}-280 \varepsilon +\varepsilon ^4 \left(\frac{45 h_{30}}{64}+\frac{106811}{64}\right)+\varepsilon ^2 \left(-\frac{105 h_{30}}{32}-\frac{307453}{32}\right) \nonumber\\
&\qquad\quad
+\frac{189 h_{30}}{64}+\frac{590739}{64}\bigg] u_p^{13/2}
\bigg\} 
+ \chi^4 \, \varepsilon^3 \bigg\{
\left(3 \varepsilon ^2-3\right) u_p^5
+\left(\varepsilon ^5-\frac{127 \varepsilon ^4}{2}+44 \varepsilon ^3+\frac{573 \varepsilon ^2}{2}-276\right) u_p^6\nonumber\\
&\qquad
+\left[50 \varepsilon ^7+\frac{339 \varepsilon ^6}{4}-197 \varepsilon ^5+1591 \varepsilon ^3+\varepsilon ^4 \left(\frac{75 h_{40}}{128}-\frac{506211}{256}\right)+\varepsilon ^2 \left(\frac{971975}{128}-\frac{175 h_{40}}{64}\right)+\frac{315 h_{40}}{128}-\frac{1843795}{256}\right] u_p^7
\bigg\} \nonumber\\
&\quad
+ \chi^5 \, \varepsilon^3 \bigg\{
\left(\frac{27 \varepsilon ^4}{2}-\frac{47 \varepsilon ^3}{2}-\frac{239 \varepsilon ^2}{4}+4 \varepsilon +\frac{263}{4}\right) u_p^{13/2}
+\bigg[\varepsilon ^6 \left(f_{50}+\frac{5 g_{50}}{3}-\frac{1697}{16}\right)-4 \varepsilon ^7+\frac{637 \varepsilon ^5}{2}-\frac{2763 \varepsilon ^3}{2}+56 \varepsilon \nonumber\\
&\qquad
+\varepsilon ^4 \left(-\frac{77 f_{50}}{3}-\frac{119 g_{50}}{3}+\frac{2783}{16}\right)+\varepsilon ^2 \left(91 f_{50}+133 g_{50}-\frac{18899}{16}\right)-77 f_{50}-\frac{539 g_{50}}{5}+\frac{29493}{16}\bigg] u_p^{15/2}
\bigg\} \nonumber\\
&\quad
+ \chi^6 \, \varepsilon^3 \bigg\{
\left(3 \varepsilon ^3+\frac{9 \varepsilon ^2}{2}-\frac{15}{2}\right) u_p^7 
+\bigg[\varepsilon ^6 \left(\frac{8 f_{60}}{7}+\frac{12 g_{60}}{7}-\frac{393}{16}\right) -\frac{327 \varepsilon ^5}{2}+\varepsilon ^4 \left(-28 f_{60}-40 g_{60}+\frac{20469}{16}\right)+570 \varepsilon ^3 \nonumber\\
&\qquad
+\varepsilon ^2 \left(96 f_{60}+132 g_{60}-\frac{59355}{16}\right)-\frac{396 f_{60}}{5}-\frac{528 g_{60}}{5}+\frac{40503}{16}\bigg] u_p^8
\bigg\}+\Order(\chi^7).
\end{align}
\normalsize
The 1SF contribution to the redshift for eccentric orbits in a Kerr background was derived in Ref.~\cite{Munna:2023wce} up to 8PN order, without an eccentricity expansion to 3.5PN order, expanded in eccentricity to $\Order(e^{16})$ through 6PN order, and to $\Order(e^{10})$ at higher PN orders.
Comparing Eq.~\eqref{eq:red_shift_eccentric} with the 1SF redshift from Ref.~\cite{Munna:2023wce} leads to several constraints at each order in $\varepsilon$, leading to the same circular-orbit solutions in Eqs.~\eqref{S3S4solns} and \eqref{S6Soln}.

For the spin-precession invariant, we obtain
\begin{align}
\psi_2^{(0)} &= [\dots] + \chi^5 u_p^{13/2} \left[
-\frac{537 \varepsilon ^6}{16}+\frac{13029 \varepsilon ^4}{16}-\frac{34887 \varepsilon ^2}{16}+\frac{24267}{16}
+ \left(\frac{\varepsilon ^6}{7}-3 \varepsilon ^4+9 \varepsilon ^2-\frac{33}{5}\right) \left(f_{51} + g_{51}\right)
\right],
\end{align}
where the dots represent the lower order PN contributions in Eq.~\eqref{psi0Test}.
The 0SF eccentric-orbit spin-precession invariant produces one constraint on the sum $f_{51} + g_{51}$, which is the same as the circular-orbit constraint in Eq.~\eqref{constCirc0SFpsi}.

Our result for the spin-precession invariant at first order in the mass ratio reads
\small
\begin{align}\label{psi1SFeccen}
\psi_2^{(1)} &= -u_p
+\left(\frac{13}{4}-\varepsilon ^2\right) u_p^2
+\left[-\frac{\varepsilon ^4}{2}+\left(\frac{123 \pi ^2}{256}-\frac{325}{16}\right) \varepsilon ^2-\frac{615 \pi ^2}{256}+67\right] u_p^3
+ \bigg[
\frac{628 \log u_p}{15}+\frac{31697 \pi ^2}{6144}-\frac{587831}{2880}+\frac{1256 \gamma_E }{15} \nonumber\\
&\qquad
+\frac{296 \log 2}{15} +\frac{729 \log 3}{5} 
+ e^2 \left(\frac{268 \log u_p}{5}-\frac{164123}{480}-\frac{23729 \pi ^2}{4096}+\frac{536 \gamma_E}{5}+\frac{11720 \log 2}{3}-\frac{10206 \log 3}{5}\right) \bigg] u_p^4 \nonumber\\
&\quad
+ \chi \bigg\{
{-}\frac{1}{2} u_p^{3/2}
+\left(\frac{\varepsilon ^2}{8}-\frac{21}{4}\right) u_p^{5/2}
+\left[\frac{43 \varepsilon ^4}{32}+\left(1+\frac{123 \pi ^2}{256}\right) \varepsilon ^2+\frac{81}{16}-\frac{615 \pi ^2}{256}\right] u_p^{7/2}
+ \bigg[\frac{628 \log u_p}{15}+\frac{52225 \pi ^2}{6144}-\frac{2580077}{5760} \nonumber\\
&\qquad
+\frac{1256 \gamma_E }{15}+\frac{296 \log 2}{15}+\frac{729 \log 3}{5} 
+ e^2 \!\left(\frac{268 \log u_p}{5}-\frac{274889}{640}-\frac{39529 \pi ^2}{4096}+\frac{536 \gamma_E }{5}+\frac{11720 \log 2}{3}-\frac{10206 \log 3}{5}\right)\!\bigg] u_p^{9/2}
\bigg\} \nonumber\\
&\quad
+ \chi^2 \bigg\{
{-}u_p^2
+\left(2 \varepsilon ^2+\frac{7}{4}\right) u_p^3
+\left[\frac{781}{16}-\frac{615 \pi ^2}{256}-\frac{5 \varepsilon ^4}{8}+\left(\frac{9}{2}+\frac{123 \pi ^2}{256}\right) \varepsilon ^2\right] u_p^4
+ \bigg[\frac{628 \log u_p}{15}+\frac{5155 \pi ^2}{1536}+\frac{159179}{2880}+\frac{1256 \gamma_E }{15}  \nonumber\\
&\qquad
 +\frac{296 \log 2}{15}+\frac{729 \log 3}{5} -\frac{3h_{21}}{16}
+ e^2 \!\left(\frac{268 \log u_p}{5}-\frac{15 h_{21}}{16}-\frac{9441}{160}-\frac{22037 \pi ^2}{2048}+\frac{536 \gamma_E }{5}+\frac{11720 \log 2}{3}-\frac{10206 \log 3}{5}\right)\!
\bigg] u_p^5
\bigg\} \nonumber\\
&\quad
+ \chi^3 \bigg\{
{-}u_p^{5/2}
+\left(\frac{5}{4}-\frac{\varepsilon ^2}{8}\right) u_p^{7/2}
+\left[\frac{477}{16}-\frac{615 \pi ^2}{256}-\frac{39 \varepsilon ^4}{16}+\left(\frac{123 \pi ^2}{256}-\frac{189}{8}\right) \varepsilon ^2\right] u_p^{9/2}
+ \bigg[\frac{628 \log u_p}{15}+\frac{5155 \pi ^2}{1536}-\frac{2922857}{5760} \nonumber\\
&\qquad
+\frac{1256 \gamma_E }{15}+\frac{296 \log 2}{15}+\frac{729 \log 3}{5} -\frac{5h_{31} }{32} 
+ e^2 \bigg(\frac{268 \log u_p}{5}-\frac{25 h_{31}}{32}-\frac{214929}{640}-\frac{22037 \pi ^2}{2048}+\frac{536 \gamma_E }{5}+\frac{11720 \log 2}{3} \nonumber\\
&\qquad\quad
-\frac{10206 \log 3}{5}\bigg)\bigg] u_p^{11/2}
\bigg\}
+ \chi^4 \bigg\{
{-}u_p^3
+\left(\frac{\varepsilon ^2}{4}+\frac{5}{4}\right) u_p^4
+\left[\frac{51 \varepsilon ^4}{16}+\left(\frac{59}{8}+\frac{123 \pi ^2}{256}\right) \varepsilon ^2+\frac{267}{16}-\frac{615 \pi ^2}{256}\right] u_p^5 \nonumber\\
&\qquad
+ \bigg[\frac{628 \log u_p}{15}+\frac{5155 \pi ^2}{1536}-\frac{22351}{2880}+\frac{1256 \gamma_E }{15}+\frac{296 \log 2}{15}+\frac{729 \log 3}{5}
+ e^2 \bigg(\frac{268 \log u_p}{5}-\frac{196037}{320}-\frac{22037 \pi ^2}{2048}+\frac{536 \gamma_E }{5} \nonumber\\
&\qquad\quad
+\frac{11720 \log 2}{3}-\frac{10206 \log 3}{5}\bigg)\!\bigg] u_p^6
\bigg\} 
+ \chi^5 \bigg\{
{-}u_p^{7/2}
+ \!\left(\frac{\varepsilon ^2}{4}+\frac{5}{4}\right)\! u_p^{9/2}
+\left[\frac{93}{4}-\frac{615 \pi ^2}{256}-\frac{35 \varepsilon ^4}{32}+ \!\left(\frac{123 \pi ^2}{256}-\frac{23}{16}\right)\! \varepsilon ^2\right] u_p^{11/2} \nonumber\\
&\qquad
+ \bigg[\frac{176 f_{51}}{35}+\frac{72 g_{50}}{5}+\frac{208 g_{51}}{35}+\frac{628 \log u_p}{15}+\frac{5155 \pi ^2}{1536}-\frac{8434547}{5760}+\frac{1256 \gamma_E }{15}+\frac{296 \log 2}{15}+\frac{729 \log 3}{5} \nonumber\\
&\qquad
+ e^2 \left(\frac{152 f_{51}}{7}+52 g_{50}+\frac{200 g_{51}}{7}+\frac{268 \log u_p}{5}-\frac{3452929}{640}-\frac{22037 \pi ^2}{2048}+\frac{536 \gamma_E }{5}+\frac{11720 \log 2}{3}-\frac{10206 \log 3}{5}\right)\bigg] u_p^{13/2}
\bigg\}\nonumber\\
&\quad
+\Order(\chi^6),
\end{align}
\end{widetext}
where $\gamma_E$ is the Euler gamma constant.
The PN contributions in $\psi_2^{(1)}$ written in terms of $\varepsilon$ have been computed without an eccentricity expansion. 
However, the highest PN order at each order in spin (N$^3$LO S$^1$, N$^3$LO S$^2$, NNLO S$^3$, NNLO S$^3$, NLO S$^5$, NLO S$^6$---see Table~\ref{tab:PNspin}) has been computed in an eccentricity expansion, because it receives contributions from the 4PN non-spinning part of the Hamiltonian, in which the nonlocal-in-time contribution is only known in an eccentricity expansion. 
We wrote the $\Order(e^2)$ expansion in the above equation, but in the Supplemental Material, we include the $\Order(e^6)$ expansion of those terms, since we used the 4PN non-spinning Hamiltonian of Ref.~\cite{Damour:2015isa}, whose nonlocal part was expanded to $\Order(e^6)$.

This mixing of the non-spinning and spin contributions from the Hamiltonian into the spin-precession frequency for eccentric orbits (when expressed in terms of frequencies) leads to unintuitive powers of $u_p$ at high orders in spin (cf. Table~\ref{tab:PNspin}). 
For example, the first PN order in $\psi_2^{(1)}$ at $\Order(\chi^5)$ is $u_p^{7/2}$, but the full information from the LO S$^6$ part of the Hamiltonian is in the $\chi^5 u_p^{11/2}$ term, and the NLO S$^6$ information is in the $\chi^5 u_p^{13/2}$ term.

Furthermore, note that the $e\to 0$ limit of the 1SF eccentric-orbit spin-precession invariant does not reduce to the circular-orbit result in Eq.~\eqref{PsiCirc}.
This is because the invariant was computed by fixing both the radial and azimuthal frequencies of the perturbed and unperturbed spacetime.
Thus, even if the orbit in the unperturbed spacetime is circular, the orbit in the perturbed spacetime is not necessarily circular. (See Sec.~III.B of Ref.~\cite{Akcay:2016dku} for an explanation of how to recover the circular-orbit spin-precession invariant from the eccentric-orbit result.)

The 1SF spin-precession invariant has not been derived in the literature beyond quadratic order in the Kerr spin~\cite{Bini:2019lkm}. 
Therefore, we can only solve for the NNLO S$^3$ coefficient $h_{21}$ using existing GSF results; the eccentric-orbit comparison leads to the same solution $h_{21} = 371$ that was obtained from the circular-orbit calculation.

\section{Scattering angle from Compton  Amplitude}\label{sec:compton}

In order to compute the green slots in the NLO column  of Table~\ref{tab:PNspin} and provide a match to the scattering angle coefficients in Eq.~\eqref{deltatheta56}, in this section we follow a scattering amplitude approach to the problem. The scattering angle receives contributions from the  1-loop triangle leading singularity \cite{Cachazo:2017jef}, which uses as main ingredient the gravitational Compton amplitude, which we extract from BHPT solutions. 
We start this section by reviewing the Compton matching to Teukolsky solutions, and then show how such a matching completely fixes the scattering angle free coefficients.  We leave to Sec. \ref{sec:NLOS5soln} the comparison of the results obtained in this section  with the ones obtained from the  GSF  approach discussed in the previous section.

\subsection{Gravitational Compton Amplitude}
For perturbations of spin-weight  $s$  off a Kerr BH whose spin parameter is $\chi=a_1/m_1  \equiv a/m$, Compton amplitudes are well known to be obtained from  infinite sums of partial-wave amplitudes. 
Denoting  $\gamma$ as the angle formed between the direction of the incoming wave and the orientation of the BH's spin, and $\vartheta$ and $\varphi$ as the polar and  azimuthal angles defining a direction on the celestial sphere, this sum reads
\cite{Dolan:2008kf}
\begin{equation}
{}_s\mathcal{A} {=}\sum_{\ell,m}\left[{}_{-2}S_{\ell m}\left(\gamma,0;\frac{\epsilon \chi}{2}\right){}_{-2}S_{\ell m}\left(\vartheta',\varphi;\frac{\epsilon \chi}{2}\right) {}_s\mathcal{ A}_{\ell,m} \right]\,.\label{Eq:fKerrGeneric}
\end{equation}
 For the helicity-preserving amplitude, the angle $\vartheta'=\vartheta$, and the 
the partial wave  modes are obtained  via
\begin{equation}
\label{eq:amplitude_modes}
{}_s\mathcal{A}_{\ell m} |_{\text{HP}}=\frac{2\pi}{i\omega} \sum_{P=\pm1} \left({}_s\eta_{\ell m} e^{2 i (\red{{}_s \delta_{\ell m}^{P,\rm FZ}}+\blue{{}_s \delta_{\ell m}^{ \rm NZ})}} -1\right),
\end{equation}
here $\omega$ is the energy of the wave perturbation. For the helicity-reversing scenario,  the angle $\vartheta'=\pi-\vartheta$, and the amplitude modes  are  instead
\begin{equation}
\label{eq:amplitude_modes HR}
{}_s\mathcal{A}_{\ell m} |_{\text{HR}}=\frac{2\pi}{i\omega} \sum_{P=\pm1} P(-1)^\ell\left({}_s\eta_{\ell m} e^{2 i (\red{{}_s \delta_{\ell m}^{P,\rm FZ}}+\blue{{}_s \delta_{\ell m}^{ \rm NZ})}} -1\right).
\end{equation}

Here, $P$  is a parity label which appears explicit in the sum factors in Eq.~\eqref{eq:amplitude_modes HR},  or implicit in the phase-shift for the modes in both Eqs.~\eqref{eq:amplitude_modes} and~\eqref{eq:amplitude_modes HR}  due  to the  imaginary part of the   Teukolsky-Starobinski constant\footnote{ Recall the Teukolsky-Starobinski constant is complex only for perturbations of spin-weight $s=-2$. This is the reason why the the helicity-reversing modes \eqref{eq:amplitude_modes HR} vanish  for non-gravitational  perturbations.}. 
The sum over $P$ is understood as the change from the    parity to the helicity basis. 
Notice  we have  made explicit the separation of the near  and far-zone   contributions to the scattering phase-shift \cite{Ivanov:2022qqt,Bautista:2023sdf}, as we are interested in tracking the different imprints these components have  on the  two-body observables\footnote{The near/far-zone separation here is a different separation compared to other splitting proposed in the literature, where the perturbation is divided not in two but in three  regions: the far-zone $(r\gg r_+)$ which agrees with the one here, the near-horizon region $((r-r_+)\omega \ll1)$, and the near-zone which is the overlap between the near-horizon and the far-zone ( $(r-r_+)\omega \ll1$ and $r\gg r_+$ ). In our splitting, our near-zone compresses both the near horizon and the near-zone used  for instance in Ref.~\cite{Fucito:2023afe}. }. 
 The  conservative far- and near-zone contributions to the phase-shift read respectively   \cite{Bautista:2023sdf}:
\begin{align}\label{eq:phase-far}
   \red{ {}_s \delta_{\ell m}^{P,\rm FZ}} &{=}  \underbrace{ \frac{1}{2} {\rm Im} \[\partial_{M_3} F \] {-} \frac{1{-} \kappa}{2} \epsilon+\frac{1}{2}{\rm Arg}[A_s^P]}_{\rm rational}   {+} \underbrace{\epsilon \log(2 |\epsilon|)}_{\rm tail} \nonumber\\ &\quad
   {+}\underbrace{\frac{1}{2} {\rm Arg} \[ \frac{\Gamma(\frac{1}{2} {-}\bar{\nu}  {-} M_3)}{\Gamma(\frac{1}{2} {-} \bar{\nu}  {+} M_3)}\]{+} \frac{\pi}{2}\(\ell {+} \frac{1}{2} {+} \bar{\nu} \)}_{\rm transcendental}\,,
\end{align}
and
\begin{equation}\label{eq:phase-near}
   \blue{ {}_s \delta_{\ell m}^{ \rm NZ}} = \frac{1}{2} {\rm Arg} \left[ \frac{1 + e^{-i \pi \bar{\nu} } \mathcal{K}}{1 + e^{i\pi\bar{\nu} } \frac{\cos(\pi(M_3 -\bar{\nu} ))}{\cos(\pi(M_3+\bar{\nu} ))} \mathcal{K}} \right].
\end{equation} 

Here, we have used the dictionary relating gauge theory to Kerr BH parameters   \cite{Aminov:2020yma}
\begin{equation}\label{eq:dict}
\begin{split}
M_1 &= i \frac{m\chi-\epsilon}{\kappa} \,, \qquad M_2 = -s - i \epsilon \, ,   \\
M_3 &=  i \epsilon-s \,, \qquad \Lambda = -2i \epsilon\kappa \,, \\
\mathsf{u} &= -\lambda - s(s+1) + \epsilon (  i s\kappa- m \chi) 
 + \epsilon^2 (2 + \kappa ) \,,
\end{split}
\end{equation}
with  $\lambda$ and $m$ the  spheroidal and  azimuthal eigenvalues respectively, and we recall that $\kappa := \sqrt{1 - \chi^2}$.  We use the     Matone relation \cite{Flume:2004rp,Matone:1995rx}  to solve for the  partial wave ``shifted-renormalized angular momentum'' parameter\footnote{In the notation of Ref.~\cite{Bautista:2023sdf}, here $\bar{\nu}=\alpha$ in Eq. (15) of Ref.~\cite{Bautista:2023sdf}.}, $\bar{\nu}$, given by 
\begin{align*}
\end{align*}
\begin{equation}\label{eq:Matone}
\mathsf{u}=\frac{1}{4}-\bar{\nu}^2+\Lambda\partial_\Lambda F(M_1,M_2,M_3,\bar{\nu},\Lambda)\,,
\end{equation}
where $F$ is the so called  Nekrasov-Shatashvili Function. An explicit evaluation of $F$   up to $\mathcal{O}(\Lambda^9)$ was presented in Ref.~\cite{Bautista:2023sdf}, and we make use of it in the discussion  below.

The amplitude modes in Eqs.~\eqref{eq:amplitude_modes} and \eqref{eq:amplitude_modes HR} contain additionally  absorptive contributions   enclosed in the absorption probability $1-({}_s\eta_{\ell m})^2$, which we discard  in  this work. The conservative amplitude is then defined as the  piece that  conserves unitarity.

The function $\mathcal{K}$ entering in the near-zone contributions has the explicit form 
\begin{widetext}
\begin{equation}\label{eq:Kappa}
    \mathcal{K} = |\Lambda|^{-2 \bar{\nu} } \frac{ \Gamma(2 \bar{\nu} ) \Gamma(2 \bar{\nu}  +1) \Gamma\left(M_3-\bar{\nu} +\frac{1}{2}\right) \Gamma\left(M_2-\bar{\nu}  +\frac{1}{2}\right) \Gamma\left(M_1-\bar{\nu}  +\frac{1}{2}\right)}{ \Gamma(-2 \bar{\nu} ) \Gamma(1-2\bar{\nu} ) \Gamma\left(M_3+\bar{\nu} +\frac{1}{2}\right) \Gamma\left(M_2+\bar{\nu}  +\frac{1}{2}\right) \Gamma\left(M_1+\bar{\nu}  +\frac{1}{2}\right)} e^{\partial_{\bar{\nu}}  F}\,,
\end{equation}
\end{widetext}
and goes by the name of   tidal response function, as it captures the tidal deformability of the BH due to the wave perturbations.

The gravitational Compton amplitude is computed  for  wave perturbations of spin-weight $s=-2$. In Ref.~\cite{Bautista:2023sdf}, the tree-level contributions to the far-zone amplitude were presented up to eighth order in the Kerr spin, for both the helicity-preserving and the  helicity-reversing scenarios. In this work, we are interested in introducing the leading $\ell=2$ near-zone contributions to the scattering amplitude,  up to sixth order in the Kerr spin. 
We however extract only the piece that does not  present interference between the near and far-zones\footnote{ Interference is used here to refer to the mixing between the near- and far-zone contributions when the exponential of the phase-shift is 
 PM expanded.  }, where interference effects are  manifested  starting at order $a^6$. The first interference piece is purely imaginary at the level of the amplitude. (As an illustration of this in the helicity-preserving amplitude,  see the green pieces in Eq.~\eqref{eq:coeffsnear} which appear, as mentioned, starting at $\mathcal{O}(a^6$) and are anyways  expected be canceled  with non-Compton contributions in the two-body observables \cite{Kosower:2018adc}.) 
 
\begin{figure}[h]
 \begin{align*}\nonumber
 \begin{fmffile}{a3a4}
  \parbox{22pt}{
  \begin{fmfgraph*}(65,65)
    \fmfleft{i2,i1}
    \fmfright{o2,o1}
    \fmftop{t}
    \fmf{phantom,tension=0.2}{i1,v1,i2}
    \fmf{phantom}{o1,v2,o2}
    \fmf{phantom,tension=0.3}{v1,v2}
    \fmffreeze
    \fmf{plain,width=0.7
    ,foreground=(0,,0,,0)}{g,i1}
    \fmf{plain,width=0.7,tension=2.8}{g,v1}
    \fmf{plain,width=0.7}{i2,v1}
    \fmf{photon,width=0.7,foreground=(1,,0.1,,0.1),tension=2.8}{o1,v1}
    \fmf{photon,width=0.7,foreground=(1,,0.1,,0.1)}{o2,v1}
  \fmfv{decor.shape=circle,decor.filled=12,decor.size=.35w}{v1}
  \end{fmfgraph*}}_{\text{ tree-FZ}}
\hspace{1cm} 
 + \,\,\,
 \parbox{22pt}{
  \begin{fmfgraph*}(65,65)
    \fmfleft{i2,i1}
    \fmfright{o2,o1}
    \fmftop{t}
    \fmf{phantom,tension=0.2}{i1,v1,i2}
    \fmf{phantom}{o1,v2,o2}
    \fmf{phantom,tension=0.3}{v1,v2}
    \fmffreeze
    \fmf{plain,width=0.7
    ,foreground=(0,,0,,0)}{g,i1}
    \fmf{plain,width=0.7,tension=2.8}{g,v1}
    \fmf{plain,width=0.7}{i2,v1}
    \fmf{photon,width=0.7,foreground=(0.035,,0.168,,0.623),tension=2.8}{o1,v1}
    \fmf{photon,width=0.7,foreground=(0.035,,0.168,,0.623)}{o2,v1}
   \fmfv{decor.shape=square,decor.filled=12,decor.size=.15w}{v1}
  \end{fmfgraph*}}_{\ell=2\,\text{ NZ}}
   \end{fmffile}\qquad+\quad...
\end{align*}
\caption{Schematic representation of the different contributions to the Compton amplitude  used in this paper.  }
\label{eq:A4-figure}
\end{figure}
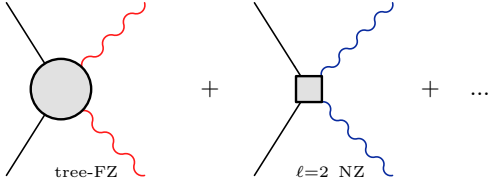

As discussed in Ref.~\cite{Bautista:2023sdf}, the extraction of  a  spin polynomial piece that features an explicit tree-level scaling is ambiguous for the near-zone contribution because of the appearance of irrational functions of the Kerr spin parameter. In this 
work, we take a different philosophy and match instead   the full $\ell=2$ near-zone contribution to a contact term in the amplitude, at the spin orders outlined above.  This matching  captures of course the tree-level part of the near-zone contributions but has the advantage of encapsulating additional 
non-perturbative-in-$\chi$ Kerr finite-size effects.  
In summary, the Compton amplitude we present  here is composed of the  unambiguous  tree-level contributions from the far-zone, plus    the full, non-interfering $\ell=2$ contributions from the near-zone, and   up to $\mathcal{O}((a\omega)^6)$. Schematically this is shown in Fig.~\ref{eq:A4-figure}.

The $\ell\to2$  near-zone phase-shift is easily obtained  using Eqs.~\eqref{eq:phase-near}, \eqref{eq:Kappa}, \eqref{eq:dict} and  \eqref{eq:Matone}. In a PM expansion, it has the schematic form:
\begin{align}
\label{eq:deltaNZl2_squem}
&\blue{{}_{-2}\delta^{\text{NZ}}_{\ell m}|_{\ell\to2}} = \,\sigma^{(6)}_{2,m}(\chi,\log\epsilon) \epsilon^6+ \sigma^{(7)}_{2,m}(\chi,\log\epsilon) \epsilon^7 \nonumber\\
&\qquad
+\frac{1}{\ell{-}2}\Big[\tilde{\sigma}^{(6)}_{2,m}(\chi) \epsilon^6 +   \tilde{\sigma}^{(7)}_{2,m}(\chi) \epsilon^7\Big]+\mathcal{O}(\epsilon^8)\,,
\end{align}
where the PM-wave parameter $\epsilon$, is defined in Eq. \eqref{eq:epsilon_param}.
The explicit expression for the  $\sigma_{i,j}^{(k)}$-coefficients can be obtained from  Eq. \eqref{eq:deltaNZl2}. These coefficients  are particular functions of $\log\kappa$, and   combinations of the polygamma functions of the form
\begin{equation}\label{eq:polygamma_conventions}
\begin{split}
\psi^{(n,m)}(\chi)&= \psi ^{(n)}\left(\frac{i m\chi}{\kappa}\right)+\psi ^{(n)}\left(-\frac{i m\chi}{\kappa}\right),\\
\tilde{\psi}^{(n,m)}(\chi)&= 
i\left[ \psi ^{(n)}\left(\frac{i m\chi}{\kappa}\right)-\psi ^{(n)}\left(-\frac{i m\chi}{\kappa}\right)\right],
\end{split}
\end{equation}
where  $\psi^{(n)}(z) = \frac{d^{n+1}}{dz^{n+1}}\log\Gamma(z)$, and we also explicitly highlight the presence of near-zone logarithmic tails $\sim \log \epsilon$.  

The divergence in Eq.~\eqref{eq:deltaNZl2_squem} in the $\ell\to 2$ limit 
is spurious  and 
cancels with the analog far-zone contribution; 
we safely ignore it in the following\footnote{ From a point-particle computation, this kind of poles appear as UV-divergent terms in the scattering amplitude that are removed using for instance a minimal subtraction scheme \cite{Ivanov:2024sds}.  The near + far-zone   BHPT result  is  finite as the BH  finite size acts as UV regulator. Since here we are dropping the $\ell\to2$ divergent terms in the near-zone,  in order to be consistent when computing the Compton amplitude  at higher loop (higher $\epsilon$) orders, the UV divergences appearing in the point-particle computation needs to be canceled using the 
same  prescription; that is, by matching the UV-divergent term exactly to the  divergent $\ell\to2$ piece in the far-zone amplitude (which comes with the opposite sign compared to the   near-zone one), and dropping it from the final result. 
}.
The piece of the Compton amplitude we use in this paper is then obtained  by considering the following terms in the  expansion of the phase-shift
\begin{widetext}
\begin{equation}\label{eq:expphase}
e^{2 i (\red{{}_s \delta_{\ell m}^{P,\rm FZ}}+\blue{{}_s \delta_{\ell m}^{ \rm NZ})}} = 
e^{i\Phi}\Big(1 + \red{\sum_{i=0} \beta^{P,\text{FZ}}_{\ell m,i}\chi^i\epsilon^{i+1}} + \blue{  \sum_{i=5,6} 
 \beta_{2 m,i}^{\text{NZ}}(\chi) \chi ^i\epsilon^{i+1} }  +  \blue{  \sum_{i=5,6} 
 \beta_{2 m,i}^{\log, \text{NZ}}(\chi) \chi ^i\epsilon^{i+1} \log\epsilon } +\cdots\Big)\,,
\end{equation}
\end{widetext}
where the dots ($\cdots$)  stand for non-tree terms ignored  in the far-zone, interference terms between the near and far-zones, and   higher $\epsilon$ and higher $\ell$ contributions in  both the near and the far-zones. We have also factored out the Newtonian phase  $\Phi=-\frac{1}{2}\epsilon+\epsilon\log(2\epsilon)$, appearing in the far-zone phase-shift \eqref{eq:phase-far};  we drop  it from the matched amplitude presented below. 

Notice that for the far-zone terms in Eq.~\eqref{eq:expphase}, we made explicit the tree-level scaling of the spin multipole expansion (SME),   and the 
coefficients $\beta_{\ell m ,i}^{P,\text{FZ}}$ are then understood  to be  pure real numbers. These were   reported  explicitly in Ref.~\cite{Bautista:2023sdf}; therefore, we do not include them here---there is  still a  parity label which have to be summed over as indicated in Eq.~\eqref{eq:amplitude_modes}.  
For the near-zone on the other hand, the  coefficients   $\beta_{\ell m }^{\text{NZ}}(\chi)$ are now functions of the Kerr spin $\chi$, which we keep in the physical region $\chi\le1$. We remark that in order to keep the PM expansion (expansion in $\epsilon$) and the SME (expansion in $a\omega$) independent, the Kerr parameter $\chi$ entering in this coefficients needs to be interpreted purely as a number. Therefore, an additional expansion in small $\chi$, will not change the PM scaling of the spin operators accompanied by such coefficients; we expand on this  below.  The explicit form of the   $\beta_{\ell m }^{\text{NZ}}(\chi)$ coefficients can be easily obtained by expanding the exponential of the near-zone phase-shift given in Eq.~\eqref{eq:deltaNZl2} up to the desired order. Analogously, the coefficients for the  logarithmic tail in the near-zone,  $\beta_{\ell m }^{\log, \text{NZ}}(\chi)$ are obtained from the $\log(\epsilon)$ contribution\footnote{In an EFT description of Kerr BHs, these terms are of physical interest as they capture the  renormalization group running of the tidal coefficients in the effective model \cite{Ivanov:2024sds,Saketh:2023bul}. The constant in the $\omega$ term can still have a tidal interpretation but it is   scheme dependent. 
The BHPT  matching  selects then a  preferred renormalization group scheme in such a way  the Compton  free coefficients in  Eq.~\eqref{eq:ansatz} no longer present renormalization group running.   }.  

We are now ready to match the amplitudes obtained by the pieces selected in  Eqs.~\eqref{eq:expphase} and \eqref{eq:amplitude_modes HR}, using Eqs.~\eqref{eq:amplitude_modes} and \eqref{Eq:fKerrGeneric}, to  covariant ansatzes. We start by postulating that for conservative scattering,  at any  order in the PM expansion  and  the  SME, keeping these as independent expansions,  the Compton amplitude, for both the helicity-preserving and helicity-reversing cases,  can be expanded \textit{entirely} in the spin basis 
\begin{equation}\label{eq:spin_basis}
\{k_2{\cdot} a,k_3{\cdot} a,w{\cdot} a\}\,,
\end{equation}
where $k_2^\mu$ ($k_3^\mu$) is the momentum of the incoming (outgoing) wave, and the vector $w^\mu$ constructed from the  spinors of the massless momenta, 
$w^\mu = p_1\cdot k_2 \frac{[3|\sigma^\mu|2\rangle}{[3|p_1|2\rangle}$.  The kinematic information in the   amplitudes will be on the other hand captured by  (in general complicated) functions of  the optical parameter, $\xi= \sin^2(\theta/2)$.

Let us start with the helicity-preserving scenario. 
Amazingly,  the terms extracted in Eq.~\eqref{eq:expphase}   can still be captured with the aid of the generic ansatz presented in Ref.~\cite{Bautista:2022wjf}:
\begin{equation}\label{eq:ansatzspin}
   {}_{-2} A |_{\text{HP}}= A^{(0)} |_{\text{HP}}\left[e^{-(2w{-}k_3{-}k_2)\cdot a} {+} P_\xi (k_2{\cdot} a,k_3 \cdot a, w{\cdot} a) \right],
\end{equation}
where we recall that the contact term deforming the  exponential is\footnote{Note some sign differences as compared to the original form written in Ref.~\cite{Bautista:2022wjf} due to the metric signature conventions used in this paper: One can recover the original ansatz simply by replacing $\xi\to-\xi$ and dot products of the form $p\cdot q\to -p\cdot q$. } 
\begin{widetext}
\begin{align}\label{eq:ansatz}
    P_\xi =& \sum_{m=0}^2 (-\xi)^{m-1} (w{\cdot} a)^{4-2m}(w{\cdot} a- k_2{\cdot} a)^m (w{\cdot} a - k_3 \cdot a)^m r^{(m)}_{|a|}(k_2{\cdot} a, k_3 \cdot a, w{\cdot} a)  \\ 
   & +\sum_{m=0}^{\infty} \left[ \frac{(w{\cdot} a)^{2m+6}}{(-\xi)^{m+2}} p^{(m)}_{|a|}(k_2{\cdot} a, k_3 \cdot a, w{\cdot} a)
     + (-\xi)^{m+2} (w{\cdot} a- k_2{\cdot} a)^{m+3}(w{\cdot} a - k_3 \cdot a)^{m+3}\,q^{(m)}_{|a|}(k_2{\cdot} a, k_3 \cdot a, w{\cdot} a) \right]\nonumber
\end{align}
\end{widetext}
and  up to S$^6$, the polynomials $r_{|a|},p_{|a|},q_{|a|}$ are
\begin{equation}\label{eq:ansatzr}
   \begin{split}r_{|a|}^{(m)}  =&\, c_{1}^{(m)}{-}c_{2}^{(m)}(k_{2}{+}k_{3}){\cdot}a{-}c_{3}^{(m)}w{\cdot}a{+}c_{4}^{(m)}|a|\omega\\
 & +c_{5}^{(m)}(w{\cdot}a-k_{2}{\cdot}a)(w{\cdot}a-k_{3}{\cdot}a)\\
 & +c_{6}^{(m)}(2w{\cdot}a-k_{2}{\cdot}a-k_{3}{\cdot}a)w{\cdot}a\\
 & +c_{7}^{(m)}(2w{\cdot}a-k_{2}{\cdot}a-k_{3}{\cdot}a)^{2}\\
 &+c_{8}^{(m)}(w{\cdot}a)^{2} -c_{9}^{(m)}(k_{2}{\cdot}a+k_{3}{\cdot}a)|a|\omega\\
&-c_{10}^{(m)}w{\cdot}a|a|\omega+\mathcal{O}(a^{3}),
\end{split}
\end{equation}
\begin{equation}
    \label{eq:ansatzd}p_{|a|}^{(m)}=d_{1}^{(m)}+\mathcal{O}(a)\,,\qquad q_{|a|}^{(m)}=f_{1}^{(m)}+\mathcal{O}(a)\,.
\end{equation}
The difference with the tree-level matching presented in Refs.~\cite{Bautista:2022wjf,Bautista:2023sdf} is that now the coefficients accompanying each  spin-multipole moment   $c_i^{(j)}(\chi)$ can now be  functions of the Kerr spin parameter $\chi$, which we interpret as a number, as mentioned above. 

In Ref.~\cite{Bautista:2022wjf}, the operator $|a|\omega = \sqrt{a^2}\omega$  was interpreted as being responsible for capturing    dissipative  (${}_s\eta_{\ell m} \ne 1 $)
 contributions  in  the super-extremal region of the Compton amplitude. 
Since in this work we are concerned with conservative (${}_s\eta_{\ell m} = 1 $) contributions only, it would be natural for us to drop these contributions in the ansatz since  $|a|\omega$ is not an element of the spin basis \eqref{eq:spin_basis}.  These terms are nevertheless still useful in a conservative Compton ansatz. 
We can rewrite  $|a|\omega = \chi\epsilon/2$. The  factor $\chi$ can then be reabsorbed inside the free coefficients, and the resulting spin-multipole operator   is interpreted as contributing one order lower in  the SME but one order higher in the PM expansion. Terms obtained in this way are not relevant for the computation of the two-body observables at the  PN orders we consider in this paper, as outlined in Table \ref{tab:PNspin}, but are still illustrative  to keep at the level of the Compton amplitude.

In order to match this ansatz to  both the near and far-zone contributions from the BHPT computation,  it  is also natural to split  the free coefficients as follows:
\begin{equation}\label{eq:coeffs_decomp}
c_i^{(m)}\to \blue{c_i^{\text{NZ}, (m)}(\chi)}+ \blue{c_i^{\log,\text{NZ}, (m)}(\chi) \log\epsilon} +\red{c_i^{\text{FZ},(m)}}\,,
\end{equation}
and analogously for the $d_i^{(j)}$ and $f_i^{j}$ coefficients. 
After the matching computation, the far-zone coefficients agree with those reported in Table 1 of Ref.~\cite{Bautista:2022wjf} for $\alpha=1$ and $\eta=0$ (recall we are considering the conservative contributions only), which  reduce the amplitude to agree with the  far-zone amplitude reported in Ref.~\cite{Bautista:2023szu}. 
We included the far-zone  solutions in  Eq.~\eqref{eq:far_Zone_solutions} for the reader's convenience.
The near-zone contributions   are reported in Eq.~\eqref{eq:coeffsnear}, and as expected, these have a  non-trivial dependence on the Kerr spin parameter $\chi$. 

Let us remark that although $\chi$ appears in the denominator of these coefficients, the near-zone $\ell=2$ contribution cannot be captured by a lower spin multipole moment  ansatz for the helicity-preserving  Compton amplitude as the angular dependence in the amplitude   is unique for a given harmonic.  
This allows us to conclude that the matched contact terms in Eq.~\eqref{eq:ansatz}, with  coefficients produced  via Eq.~\eqref{eq:coeffs_decomp} with explicit solution as obtained by replacing  
Eqs.~\eqref{eq:far_Zone_solutions} and  \eqref{eq:coeffsnear}, 
correspond indeed to  $\mathcal{O}(a\omega)^{5,6}$   spin multipole moments. More remarkable is the fact that  the construction of the two-body  observables through the Compton will naturally inherit this SME;  this is a feature  that is not manifested for two-body observables constructed via  GSF methods. 
For instance, if one takes directly the expressions for the 1SF observables reported for instance in 
Refs.~\cite{Munna:2022gio, Kavanagh:2015lva}, the SME is totally obscured, and the  only option one massage such observables is to do a  small-spin expansion, as we discuss in Sec. \ref{sec:NLOS5soln}, which is not equivalent to the SME presented in this section.  

Let us finally comment on the non-vanishing $|a|\omega$ terms which, although they do not play a role in our two-body analysis, it is  still illustrative to understand their origin in the wave scattering computation. 
These are terms accompanied by  the coefficients $c_8^{(i)}$. To obtain a non-vanishing contribution,  we have  purposely included an interference contribution given by a term of the form  $\beta_{2,m,0}^{P,FZ}\times \beta_{2,m,6}^{NZ}\times \chi^6\epsilon^{8}$. This gives a term contributing at sixth order in the spin-multipole expansion but at one loop in the PM expansion. The contribution of this  interference term is given by the green terms in Eq.~\eqref{eq:coeffsnear}. The remaining  (real) contributions in the $c_8^{(i)}$ coefficients come from 1-loop contributions in the near-zone phase-shift \eqref{eq:deltaNZl2}, more precisely from those terms proportional to $\pi$.

We now turn to the  helicity-reversing amplitude. In this case,  in the non-interfering  scenario  the non-zero contributions to the modes 
\eqref{eq:amplitude_modes HR} come purely from the far zone due to the overall parity factor $P$ inside the sums. 
The covariant amplitude obtained  by matching the non-interfering far-zone contributions is    the well-known exponential 
\begin{equation}
\label{eq:helicityreversing}
{}_{-2}A|_{\text{HR}} = A^{(0)}|_{\text{HR}} \times e^{-(k_3-k_2)\cdot a}\,,
\end{equation}
which was checked up to eighth order in the spin multipole expansion in Ref.~\cite{Bautista:2023sdf}. This  therefore sets to zero any deformation of the helicity-reversing Britto-Cachazo-Feng-Witten (BCFW) exponential, up to that spin order (see for instance Ref.~\cite{Haddad:2023ylx} for an explicit Compton ansatz for the helicity-reversing amplitude).

\subsection{Scattering angle}
Having discussed the Compton amplitude, we  
 now consider the scattering of two Kerr BHs with spins $a_1$ and $a_2$, aligned in the direction of the system's angular momentum. 
Compton contributions to the aligned-spin scattering angle are  well known to be controlled by the triangle leading singularity \cite{Guevara:2017csg}, where as shown in  Ref.~\cite{Guevara:2017csg,Bautista:2023szu},
on the support of the Teukolsky solutions, the helicity-reversing exponential \eqref{eq:helicityreversing} does not contribute to the scattering angle, leaving  as  only contribution that from the  helicity-preserving amplitude. 
In the following, we also subtract the near-zone logarithmic tails from the  Compton amplitude as these deserve a deeper understanding, especially considering that $\log\epsilon$-tails can also be obtained from the far zone.  

Using the ansatz \eqref{eq:ansatz}, the scattering  angle up to $\mathcal{O}(a^6)$ was computed in Ref.~\cite{Bautista:2023szu} for generic values of the free coefficients. Let us not import the total angle from that reference but instead consider its non-relativistic expansion as we are interested in making contact with the Hamiltonian formulation discussed above. In the PN expansion,  we  identify the map from the  coefficients in Eq.~\eqref{deltatheta56}   to those from the helicity-preserving  Compton ansatz \eqref{eq:ansatz}:
\begin{align}\label{eq:angle_compton}
f_{50}&= -\frac{3}{128}  (60 c_2^{(1)}+240 c_2^{(2)}-45 c_3^{(0)}+692)\,, \nonumber\\
g_{50}&= \frac{3}{128} (60 c_2^{(1)}+240 c_2^{(2)}-45 c_3^{(0)}-428)\,, \nonumber\\
f_{60}&=- \frac{25}{256}  (27 c_5^{(0)}-27 c_5^{(1)}+45 c_5^{(2)}+45 c_6^{(0)}-45 c_6^{(1)} \nonumber\\
&\quad +45 c_6^{(2)}+216 c_7^{(0)}-180 c_7^{(1)}+252 c_7^{(2)}-260)\,, \nonumber\\
g_{60}&= \frac{35}{256} (27 c_5^{(0)}-27 c_5^{(1)}+45 c_5^{(2)}+45 c_6^{(0)}-45 c_6^{(1)} \nonumber\\
&\quad +45 c_6^{(2)}+216 c_7^{(0)}-180 c_7^{(1)}+252 c_7^{(2)}+76)\,, \nonumber\\
f_{51}&= \frac{105}{64} (3 c_2^{(0)}+6 c_2^{(2)}+128)\,, \nonumber\\
g_{51}&= -\frac{105}{64}  (3 c_2^{(0)}+6 c_2^{(2)}-28)\,,
\end{align}
 where the $c_i^{(j)}$ coefficients are understood to be decomposed as in Eq.~\eqref{eq:coeffs_decomp} and discarding the near-zone logarithmic tails.  Out of the non vanishing contributions, the $c_3^{(i)}$ and $c_6^{(i)}$ ones are subject  of further analysis below as these are the ones contributing to the gauge-invariant observables at the  PN orders indicated in Table~\ref{tab:PNspin}.
Let us finally observe  that although the body associated to the Compton amplitude in the triangle diagram is the small body, the information about the larger mass enters through a symmetrization of the triangle, i.e. when the Compton effects are attached to the heavier BH.

\textit{Non-aligned spin:}
Before we discuss the comparison of our findings with the GSF literature, let us take a moment to discuss the  non-aligned spin scenario. In Ref.~\cite{Bautista:2023szu}, observables for  non-aligned spins   with generic $c_i^{(j)}$ deformations of the Compton amplitude  where computed from the eikonal phase at 2PM order using the formula derived in Refs.~\cite{Bern:2020buy,Kosmopoulos:2021zoq}. We remark  such results can be evaluated  on the non-perturbative in $\chi$ Teukolsky solutions presented here,   via the splitting \eqref{eq:coeffs_decomp} and the  explicit solution as obtained by replacing  
Eqs.~\eqref{eq:far_Zone_solutions} and  \eqref{eq:coeffsnear}; i.e. by 
allowing the $c_i^{(j)}$ coefficients to be functions of the Kerr spin parameter $\chi$, as the effect of the  derivatives on such $c_i^{(j)}(\chi)$ coefficients is effectively zero when using the eikonal formula, at least up to 2PM order. 

\section{NLO \texorpdfstring{S$^{5}$ and S$^6$}{S5 and S6}: Comparison between GSF and Compton/Teukolsky solutions}\label{sec:NLOS5soln}

\subsection{Solving for the NLO \texorpdfstring{S$^5$}{S5} unknowns}

The NLO S$^5$ part of the scattering angle~\eqref{deltatheta56} contains two unknown coefficients, $f_{50}$ and $g_{50}$.
From the 0SF redshift, we obtained the relation $f_{50} = -105/4 - g_{50}$ (see Eq.~\eqref{S5solnTest}). 
We could, in principle, solve for the remaining unknown from the 1SF redshift, if not for transcendental functions of the Kerr spin that appear at $\Order(\chi^5 y^{15/2})$.
These functions are likely due to the tidal deformations or other finite-size effects of the Kerr BH, and they make it difficult to separate the NLO S$^5$ PN contribution~\cite{Bautista:2023sdf}.

To illustrate this issue, we consider the $\Order(u_p^{15/2} e^2)$ piece of the 1SF redshift (noting that  the same conclusions in this subsection can be obtained from the circular-orbit part of the redshift or from higher orders in eccentricity), which is given by~\cite{Munna:2023wce}
\begin{align}\label{eq:up15o2e2}
&U^{(1)}|_{u_p^{15/2}e^2 } = \frac{18317 \chi ^5}{5} + \frac{96}{5} \left(3 \chi ^3 + \chi \right) [\psi^{(0,2)}(\chi) +  \log\kappa^2] \nonumber\\
&+\!\chi^3 \!\left[\frac{576 \log u_p}{5}{+}\frac{20600396}{225}{-}\frac{37445 \pi ^2}{768}{+}\frac{576 \gamma_E }{5}{+}\frac{576 \log 2}{5}\right] \nonumber\\
&+\!\chi \!\left[\frac{21874 \log u_p}{35} +\frac{42404 \gamma_E }{35}+\frac{932332 \log 2}{105}+\frac{2430 \log 3}{7} \right. \nonumber\\
&\quad \left. 
-\frac{86233969 \pi ^2}{6144}+\frac{10843142833}{44100} \right]+\frac{2687231 \pi }{4410}\,,
\end{align}
where the polygamma functions $\psi^{(n,k)}(\chi)$ were defined in Eq.~\eqref{eq:polygamma_conventions}.

We consider three possibilities to use the GSF and Compton amplitudes results to solve for the $f_{50}$ and $g_{50}$ coefficients: 
1) match the polynomial $\chi^5$ piece only, 2) expand the polygamma functions in spin, or 3) match nonperturbatively in spin by treating $f_{50}$ and $g_{50}$ as effective functions of spin.
We proceed to discuss these three possibilities.

\subsubsection{Polynomial matching}
One possible way to  determine the $f_{50}$ and $g_{50}$ coefficients is by simply equating  the spin powers in our computation of the redshift in Eq.~\eqref{eq:red_shift_eccentric}, restricting to the desired PN and eccentricity order, to the 1SF result in Eq.~\eqref{eq:up15o2e2}. 
This reduces to considering only the first term in Eq.~\eqref{eq:up15o2e2} and discarding any other irrational contributions, yielding the constraint
\begin{equation}\label{constPrescription1}
a^5 u_p^{15/2}e^2 \left(200 f_{50}+296 g_{50}+6213\right) = 0.
\end{equation}
Combined with the 0SF constraint~\eqref{eq:0SFspin5constrain}, one obtains
\begin{equation}\label{eq:coeffssoin5_prescription1}
f_{50} = -\frac{519}{32}\,, \qquad g_{50} = -\frac{321}{32}\,.
\end{equation}
This solution for $f_{50}$ and $g_{50}$ as rational numbers is consistent with the expectation that LO and NLO PN contributions only contain rational numbers, as can be seen from the lower spin orders in Eq.~\eqref{UCirc} for example.
In addition, both coefficients appear in the 0SF redshift (cf. Eq.~\eqref{U0test}), which does not contain irrational numbers at any order.

Let us remark that the solution in Eq.~\eqref{eq:coeffssoin5_prescription1} is automatically satisfied if one uses the Compton amplitude reported in Refs.~\cite{Bautista:2022wjf,Cangemi:2023bpe} to compute the 2PM angle, after removing the polygamma contributions (labeled with the $\alpha$ parameter), or simply by using the polynomial piece of the near+far-zone coefficients \eqref{eq:far_Zone_solutions} and \eqref{eq:coeffsnear} in the first and second line of Eq.~\eqref{eq:angle_compton}, using the splitting \eqref{eq:coeffs_decomp}.

\subsubsection{Small-spin expansion}
Another possibility is to perform a small-spin expansion of the log and polygamma functions in Eq.~\eqref{eq:up15o2e2}. 
This is motivated by the fact that GSF results of Ref.~\cite{Munna:2023wce} only assume that the spin value of the black hole is sub-extremal and as such should be valid in the Schwarzschild limit. 
In particular, one finds they agree  explicitly with the $\Order(\chi\, e^4)$ redshift derived in Refs.~\cite{Bini:2016dvs,Bini:2019lcd}, which suggests that this expansion produces sensible results.

The small-spin expansion of $\psi^{(0,2)}(\chi)$ and $\log \kappa^2$ from Eq.~\eqref{eq:up15o2e2} is given by:
\begin{subequations}
\begin{align}
\log \kappa^2 &= -\chi^2 - \frac{\chi^4}{2} - \frac{\chi^6}{3} + \Order(\chi^8), \\
\psi^{(0,2)}(\chi) &= -2 \gamma_E + 8 \chi^2 \zeta(3) + 8 \chi^4 [\zeta(3) - 4 \zeta(5)] \nonumber\\
& + 8 \chi^6 [\zeta (3)-8 \zeta (5)+16 \zeta (7)] + \Order(\chi^8),
\end{align}
\end{subequations}
where $\zeta(z)$ is the Riemann zeta function.
With these expansions, the first line of Eq.~\eqref{eq:up15o2e2} becomes
\begin{align}
&U^{(1)}|_{u_p^{15/2}e^2} = -\frac{192 \gamma_E}{5}  \chi
+ \chi^3 \left[\frac{768 \zeta (3)}{5}-\frac{576 \gamma_E}{5}\right] \nonumber\\
&\quad
+ \chi^5 \left[\frac{3072 \zeta (3)}{5}-\frac{3072 \zeta (5)}{5}+\frac{18029}{5}\right]\nonumber\\
&\quad
+ \chi^7 \left[\frac{3072 \zeta (3)}{5}-3072 \zeta (5)+\frac{12288 \zeta (7)}{5}-\frac{144}{5}\right] \nonumber\\
&\quad + \Order(\chi^9) + [\dots]\,,
\end{align}
where the dots represent the last three lines of Eq.~\eqref{eq:up15o2e2}.

Matching the $\Order(\chi^5)$ part of the above equation to our PN calculation from Eq.~\eqref{eq:red_shift_eccentric} yields the following constraint:
\begin{align}
&a^5 u_p^{15/2} e^2 \bigg[\frac{3806}{5}+\frac{80 f_{50}}{3}+\frac{592 g_{50}}{15} \nonumber\\
&\qquad\qquad
+ \frac{3072}{5}\zeta (3) - \frac{3072}{5} \zeta (5)\bigg] = 0,
\end{align}
which together with Eq.~\eqref{eq:0SFspin5constrain} lead to the solution
\begin{equation}
\begin{aligned}
f_{50} &= 48 \zeta (3)-48 \zeta (5) -\frac{21}{4} - \frac{519}{32} \,, \\
g_{50} &= -48 \zeta (3)+48 \zeta (5)   +\frac{21}{4}-\frac{321}{32}\,.
\end{aligned}
\end{equation}
We recognize the first two terms in each line as  polygamma contributions, the third terms  comes from the $\log\kappa$, and the last term correspond to  the polynomial contribution. Notice the same conclusion can be obtained by using directly the Compton solutions given in Eq.~\eqref{eq:angle_compton}, by replacing the splitting  \eqref{eq:coeffs_decomp}, together with the explicit  near-zone and the far-zone solutions, \eqref{eq:coeffsnear} and \eqref{eq:far_Zone_solutions}, followed by the small-spin expansion, and extracting the $\chi$-independent piece of such an expansion. 

By inspection, the small-spin expansion presents  several features: 1) the polygamma contributions always come accompanied by a transcendental number, which suggests those to be loop contributions at the level of the Compton amplitude; 
2) at $\Order(\chi^7)$, the expansion produces a pre-leading PN contribution at $\Order(u_p^{15/2})$, even though the leading PN order starts at $\Order(u_p^{17/2})$, as can be extrapolated from Table~\ref{tab:PNspin}; 
and 3) the $\Order(\chi^5)$ piece could be interpreted as the polynomial term contributing to the tree-level Compton amplitude at fifth order in the SME, which is similar to what Ref.~\cite{Bautista:2022wjf} considered but in that reference, the opposite $\chi\gg1$ limit was taken.

\subsubsection{Non-Perturbative spin matching}
Since the polynomial matching is ambiguous due to polygamma identities, a  third possibility is to do a non-perturbative in $\chi$ matching following  the absorptive computations performed in Refs.~\cite{Porto:2007qi,Saketh:2022xjb}, such that the  scattering angle coefficients $f_{50}$ and $g_{50}$ can be interpreted as functions of $\chi$. 
We   stress however  that  this matching cannot be done by direct  comparison of  Eq.~\eqref{eq:up15o2e2}  to Eq.~\eqref{eq:red_shift_eccentric} since the former contains lower-in-spin contributions at higher PM orders that mix with the NLO S$^5$ contributions we are after. It is here where the Compton matching of the previous section provides clarity, as the comparison to BHPT results has been obtained in a SME.  Indeed, by replacing  the non-perturbative  solutions from Eqs.~\eqref{eq:far_Zone_solutions} and \eqref{eq:coeffsnear}  into Eqs.~\eqref{eq:angle_compton} via \eqref{eq:coeffs_decomp}, we obtain

\begin{equation}\label{eq:sol3f50g50}
\begin{split}
f_{50}&= \frac{15}{128} (9 c_{3}^{\text{NZ},(0)}-100), \\
g_{50}&=-\frac{15}{128}  (9 c_{3}^{\text{NZ},(0)}+124),
\end{split}
\end{equation}
where  $c_{3}^{\text{NZ},(0)}$ is given in  Eqs.~\eqref{eq:coeffsnear}. 
These results  perfectly match  the polygamma and $\log\kappa$ contributions in \eqref{eq:up15o2e2}. Additionally, these captures some of the $\chi^3$ and $\chi^1$ pieces in Eq.~\eqref{eq:up15o2e2}  that come from actual $\mathcal{O}((a\omega)^5)$ spin multipole operators in the Compton amplitude. The difference when replacing Eq.~\eqref{eq:sol3f50g50} into~\eqref{eq:red_shift_eccentric}   and comparing with  Eq.~\eqref{eq:up15o2e2}  for the given order in the PN and eccentricity expansions is
\begin{align}
\Delta U|_{u^{15/2}_p e^2}&=
\left[-\frac{20772764}{225}+\frac{576 \gamma_E }{5}+\frac{37445 \pi ^2}{768}\right] \chi^3 \nonumber\\
&\quad 
+\!\left[\frac{86233969 \pi ^2}{6144}-\frac{39716 \gamma_E }{35}-\frac{10851864049}{44100}\right. \nonumber\\
&\qquad
\left.-\frac{185660}{21} \log (2)-\frac{2430 \log (3)}{7}\right]\chi \nonumber\\
&\quad -\frac{2687231 \pi }{4410}\,,
\end{align}
which are purely S$^{0,1,3}$  contributions in the SME. In particular,  these are    N$^3$LO pieces  for the S$^3$ term, and N$^5$LO for the S$^1$ term, which are 3-loop and 5-loop scattering angle computations, respectively,  which are beyond the scope of this work. 

Finally, one can check that the solutions for the $f_{50},\, g_{50}$ coefficients given  in  Eq.~\eqref{eq:sol3f50g50} satisfy the 0SF constraint in Eq.~\eqref{eq:0SFspin5constrain}, as expected since the  Compton amplitude  has the correct  factorization properties into the Kerr 3-point spin exponential. 

Furthermore, notice that although we previously assumed that the unknowns in the scattering angle ansatz are independent of the masses and spins, having the $f_{50}$ and $g_{50}$ coefficients depend on the spins as in the above solution is not a problem when computing the redshift or spin-precession frequency using Eq.~\eqref{zOmegaOrbAvg}. 
This is because the derivatives are taken with respect to the mass or spin of the secondary object, which are not affected by the dependence of the coefficients of $a_1^5$ on the mass and spin of the primary (Kerr BH). 
However, because of the symmetry under the exchange of the two bodies' labels, the coefficients of $a_2^5$ in the scattering angle ansatz~\eqref{deltatheta56} would be given by the solution~\eqref{eq:sol3f50g50} after replacing the Kerr spin by the spin of the lighter object.
Thus, having the $f_{50}$ and $g_{50}$ coefficients be functions of spin would not change the calculations in this paper.

To summarize,  the non-perturbative in $\chi$ matching has been possible thanks to 
1) the SME of the Compton amplitude, which is inherited by the redshift observable, 
and 2) the inclusion of both near and far-zone contributions in the  solution of the Teukolsky equation.

\begin{table*}[ht]
\setlength\extrarowheight{4pt}
\caption{Summary of the matched scattering angle free coefficients at each order in spin, as determined by GSF and Compton/Teukolsky information. The explicit expression for   $c_{3}^{\text{NZ},(0)}$ is given in  Eq.~\eqref{eq:coeffsnear}. The $h_{22}$ coefficient is unconstrained as it requires currently unavailable 1SF results.  }
\label{tab:summary_coeffs}
\begin{ruledtabular}
\begin{tabular}{ccccccccccccc} 
Matching &  &  & \multirow{2}{*}{Spin order} & \multicolumn{2}{c}{} & \multirow{2}{*}{$f$-coefficients} &  &  & \multirow{2}{*}{$g$-coefficients} &  &  & \multirow{2}{*}{$h$-coefficients}\tabularnewline
prescription &  &  &  &  &  &  &  &  &  &  &  & \tabularnewline
\midrule
\multirow{2}{*}{Polynomial} &  &  & \multirow{2}{*}{NNLO S$^{3}$} &  &  & $f_{30}=-180$ &  &  & $g_{30}=-100$ &  &  & $h_{30}=\frac{451}{3}$\tabularnewline
 &  &  &  &  &  & $f_{21}=-540$ &  &  & $g_{21}=-100$ &  &  & $h_{21}=371$\tabularnewline
\midrule
\multirow{3}{*}{Polynomial} &  &  & \multirow{3}{*}{NNLO S$^{4}$} &  &  & $f_{40}=650$ &  &  & $g_{40}=400$ &  &  & $h_{40}=-\frac{631}{2}$\tabularnewline
 &  &  &  &  &  & $f_{31}=2600$ &  &  & $g_{31}=800$ &  &  & $h_{31}=-446$\tabularnewline
 &  &  &  &  &  & $f_{22}=3900$ &  &  &  &  &  & $h_{22}$\tabularnewline
\midrule
Polynomial &  &  & \multirow{3}{*}{NLO S$^{5}$} &  &  &  $f_{50}=-\frac{519}{32}$ &  &  & $g_{50}=-\frac{321}{32}$ &  &  & \tabularnewline
Small $\chi$ expansion &  &  &  &  &  & $f_{50}=48\zeta(3)-48\zeta(5)-\frac{687}{32}$ &  &  & $g_{50}=-48\zeta(3)+48\zeta(5)-\frac{153}{32}$ &  &  & \tabularnewline
Non-Pert. $\chi$ &  &  &  &  &  & $f_{50}=\frac{15}{128}(9c_{3}^{\text{NZ},(0)}-100)$ &  &  & $g_{50}=-\frac{15}{128}(9c_{3}^{\text{NZ},(0)}+124)$ &  &  & \tabularnewline
\midrule
\multirow{2}{*}{Polynomial / Non-Pert. $\chi$} &  &  & \multirow{2}{*}{NLO S$^{6}$} &  &  & $f_{60}=\frac{2205}{64}$ &  &  & $g_{60}=\frac{735}{64}$ &  &  & \tabularnewline
 &  &  &  &  &  & $f_{51}=210$ &  &  & $g_{51}=\frac{735}{16}$ &  &  & \tabularnewline
\end{tabular}
\end{ruledtabular}
\end{table*}

\subsection{NLO \texorpdfstring{S$^6$}{S6} Results}
In Sec.~\ref{sec:zPsiCirc}, we discussed how to fully fix the  $f_{60},\, g_{60},\, f_{51}$ and $g_{51}$ coefficients directly from GSF observables, as transcendental functions do not appear in the redshift at NLO S$^6$.  
The absence of irrational functions  at this order  is however only  an accident for the aligned-spin scenario as, in general, the S$^6$ Compton coefficients depend on such irrational functions.  
To illustrate this point, we carry the analogous S$^6$ coefficient fixing using non-perturbative in $\chi$ Compton results only. 

We  proceed as follows: replacing the BHPT solutions for the near-zone \eqref{eq:coeffsnear} + far-zone \eqref{eq:far_Zone_solutions} via the splitting \eqref{eq:coeffs_decomp}  into Eq.~\eqref{eq:angle_compton}, we obtain the following expressions for the unknowns:
\begin{equation}\label{S6Soln_compton}
\begin{alignedat}{2}
f_{60} &= \frac{2205}{64}, &\qquad g_{60} &= \frac{735}{64}, \\
f_{51} &= 210, &\qquad g_{51} &= \frac{735}{16}\,.
\end{alignedat}
\end{equation}
These are in perfect agreement with the ones reported in Eq.~\eqref{S6Soln}. However, the  solutions as computed from the Compton approach  expose  a very non-trivial cancellation of the irrational-in-spin  contributions for the $f_{60}$ and $g_{60}$ coefficients. That is, notice that the non-trivial Teukolsky  terms contributing  to $f_{60}$ and $g_{60}$  in Eq.~\eqref{eq:angle_compton}  come from the $c_{6}^{(1)}$  and $c_{6}^{(2)}$ coefficients, which individually contain polygamma and $\log\kappa$ dependence as seen from Eq.~\eqref{eq:coeffsnear}.  This cancellation signals that the non-perturbative in $\chi$ matching is a good prescription to capture interesting features of Kerr BHs, which we further study in the next section.   
As a final remark, one can check   that replacing the dictionary from Eqs.~\eqref{eq:angle_compton} into the eccentric redshift in Eq.~\eqref{eq:red_shift_eccentric} and precession frequency~\eqref{psi1SFeccen},  and taking  the test-mass  limit ($\delta\to1$), the $c_i^{(j)}$ contact deformation  drop out and one recovers the test-mass results given by Eqs.~\eqref{U0test} and \eqref{psi0Test} respectively, at the given S$^{5,6}$ orders.

For the reader's convenience, in Table~\ref{tab:summary_coeffs} we collect the results for the matched scattering angle coefficients for the different spin orders considered in this work. The polynomial and the small-$\chi$ matching prescriptions can be obtained as  special cases of the  non-perturbative in $\chi$ prescription.

\section{Worldline description for Kerr}
\label{sec:worldlineKerr}
\subsection{Worldline Compton amplitude vs on-shell  Compton ansatz vs BHPT solutions }
\label{sec:worldlineCompton}
\begin{table*}[ht]
\setlength\extrarowheight{4pt}
\caption{Free coefficients for the Compton amplitude derived in Ref.~\cite{Scheopner:2023rzp} at each order in the spin-multipole expansion.  }
\label{tab:tableWorldline}
\begin{ruledtabular}
\begin{tabular}{c|c|c|c} 
Spin & Free coeff. helicity-preserving & Free coeff. helicity-reversing & Coeff. helicity-preserving ansatz \eqref{eq:ansatzspin} \tabularnewline 
\hline 
S$^2$ & $C_{2}$ & $C_{2}$ & \tabularnewline 
S$^3$ & $C_{2},C_{3}$ & $C_{2},C_{3}$ & \tabularnewline 
S$^4$ & $C_{2},C_{3,}C_{4}\,\,\,\,D_{4a,...,f}$ & $C_{2},C_{3,}C_{4}\,\,\,\,D_{4a,...,f}\,\,\,\,E_{4b}$ & $c_{1}^{(i)},\,\,\,\,i=0,1,2$\tabularnewline 
S$^5$ & $C_{2},C_{3,}C_{4},C_{5}\,\,\,\,D_{5a,...j}\,\,\,\,E_{4a},\,E_{5a,c,d}$ & $C_{2},C_{3,}C_{4},C_{5}\,\,\,\,D_{5a,...j}\,\,\,\,E_{4a},\,E_{5a,...,e}$ & $c_{2}^{(i)},\,\,\,\,i=0,1,2$, $c_{3}^{(i)},\,\,\,\,i=0,1$\tabularnewline 
\end{tabular}
\end{ruledtabular}
\end{table*}

In the previous sections, we discussed a Compton amplitude with free coefficients that were matched  to the UV BHPT solutions. In the spirit of having an EFT description for the  Kerr BH, we would like to understand the nature of these free coefficients and classify the kind of BH finite-size effects they parametrize. 
Recall that BH finite-size effects are divided in two groups: tidal effects (present for both spinless and spinning BHs), and spin-induced multipole moments  (present for spinning BHs only)~\cite{Saketh:2023bul}. 
The latter are characterized by their  linear-in-curvature dependence whereas the former are quadratic (and higher) in curvature.
In addition, these quadratic in curvature contribution can be static (time independent) or dynamic (time dependent),  giving birth to the static and dynamical Love numbers for BHs, respectively.

From a worldline perspective, the dynamics of  rotating objects can be studied via the action~\cite{Vines:2016unv,Levi:2015msa}
\begin{equation}\label{lagrw}
S= \!\int \di \tau \!\left[
p{\cdot} \dot{z}+\frac{1}{2}S^{\mu\nu}\Omega_{\mu\nu}+\beta_\mu p_\nu S^{\mu\nu}+\frac{\alpha}{2}\left(
p^2+\mathcal{M}^2
\right)
\right],
\end{equation}
where $\dot{z}^\mu$ is the tangent to the worldline $z(\tau)$.
In this action,  $\alpha$ and $\beta_\mu$ are Lagrange multipliers imposing, respectively, the  on-shell condition for the ``dynamical mass function'' $\mathcal{M}$, with respect to the   magnitude of the momentum $p^\mu$,  and the covariant spin-supplementary condition $p_\mu S^{\mu\nu}=0$.  
In a recent work \cite{Scheopner:2023rzp}, the dynamical mass function was presented to fifth order in spin and including quadratic-in-curvature contributions. 
It takes the form 
\begin{equation}
\mathcal{M}^2 = m^2+\delta \mathcal{M}_1^2+\delta \mathcal{M}_2^2+\mathcal{O}(R^3)\,,
\end{equation}
where $\delta\mathcal{M}_{1,2}^2$ are linear- and quadratic-in-Riemann contributions, respectively\footnote{Quadratic-in-curvature contributions from a QFT Lagrangian approach were presented in Ref.~\cite{Bern:2022kto}, where the free coefficients were related to the Compton coefficients from Eq.~\eqref{eq:ansatzspin} in Table III of Ref.~\cite{Bautista:2023szu}. }.  

The explicit expressions for the dynamical mass function to fifth order in spin are given in Ref.~\cite{Scheopner:2023rzp} by Eqs.~(8.54)--(8.58) for the linear-in-Riemann terms and by Eqs.~(8.65)--(8.67) for the quadratic-in-Riemann pieces. 
For clarity, we write here the quadratic-in-Riemann part of the  S$^4$ and S$^5$ contributions to the dynamical mass
\begin{subequations}
\begin{align}
&\delta\mathcal{M}_{2 S^4}^2 = \frac{D_{4a}}{m^2} (E_{SS})^2 +\! \frac{D_{4b}}{m^2} S^2 E^\mu{}_S E_{\mu S} +\! \frac{D_{4c}}{m^2} S^4 E^{\mu\nu} E_{\mu\nu} \nonumber \\
&\quad + \frac{D_{4d}}{m^2} (B_{SS})^2 + \frac{D_{4e}}{m^2} S^2 B^\mu{}_S B_{\mu S} + \frac{D_{4f}}{m^2} S^4 B^{\mu\nu}B_{\mu\nu},        \\
&\delta\mathcal{M}_{2 S^5}^2 = \nonumber\\
&\quad \frac{D_{5a}}{m^3} B_{SS} E_{SSS} 
+ \frac{D_{5b}}{m^3} S^2 {B^\mu}_S E_{\mu SS}+ \frac{D_{5c}}{m^3} S^4 B^{\mu\nu}E_{\mu\nu S}   \nonumber \\
&\quad 
+\! \frac{D_{5d}}{m^3} E_{SS}B_{SSS} 
+\! \frac{D_{5e}}{m^3} S^2 E^\mu{}_S B_{\mu SS} 
+\! \frac{D_{5f}}{m^3} S^4 E^{\mu\nu}B_{\mu\nu S} \nonumber\\ 
&\quad + \!\bigg(
\frac{D_{5g}}{m^3}E_{S\mu}\dot E_{S\nu}
+\frac{D_{5h}}{m^3}S^2E^{\lambda}{}_{\mu}\dot E_{\lambda\nu}
+\frac{D_{5i}}{m^3}B_{S\mu}\dot B_{S\nu} \nonumber \\
&\qquad
+\frac{D_{5j}}{m^3}S^2B^{\lambda}{}_{\mu}\dot B_{\lambda\nu} \bigg)  S^2 \epsilon^{\mu\nu}{}_{\rho\sigma}u^\rho S^\sigma .
\end{align}
\end{subequations}
where notation such as $E_{SS}$ is a shorthand for $E_{\mu\nu}S^\mu S^\nu$.
The coefficients $D_{4a,\dots,f}$ are often denoted as static tidal numbers, while the $D_{5a,\dots,f}$ are tidal mixing numbers, and $D_{5g,\dots,j}$ are dynamical numbers.
In Table~\ref{tab:tableWorldline}, we list and classify the Wilson coefficients at each order in spin.

The Compton amplitudes presented in Ref.~\cite{Scheopner:2023rzp} sets $C_i=1$ in order to match the Kerr 3-point exponential. 
One  can in addition  map the free coefficients in the Compton ansatz~\eqref{eq:ansatzspin}, imposing  the exponential \eqref{eq:helicityreversing},   to the worldline coefficients. Requiring spin-exponentiation for the helicity-reversing amplitude---as dictated by the non-interfering Teukolsky solutions---ensures that the helicity-reversing part of the worldline Compton amplitude provides zero contribution to the scattering angle.  
The map is the following \footnote{Note that the map in Eqs.~\eqref{eq:map_4} and~\eqref{eq:map_5} for the Compton coefficients is a bit different from the one presented in Eq.~(8.79) of the second version of Ref.~\cite{Scheopner:2023rzp} on \texttt{arXiv}. At $\Order(S^4)$, it is due to dropping the $c_1^{(i)}$ coefficients in Ref.~\cite{Scheopner:2023rzp}, whereas at $\Order(S^5)$, the difference  is due to  some typos in that reference. 
}
\begin{equation}\label{eq:map_4}
\begin{split}
D_{4a} &= c_1^{(0)} - c_1^{(1)} + c_1^{(2)},\\
D_{4b} &=
\frac{ 1}{2} \big(E_{4b} - 2 c_1^{(0)} + c_1^{(1)}\big),\\
 D_{4c} &= -\frac{1}{6}E_{4b} + \frac{1}{4}c_1^{(0)}, \\
   D_{4d} &= 
 c_1^{(0)} - c_1^{(1)} + c_1^{(2)}, \\
 D_{4e} &=
 \frac{1}{2} \big({-}E_{4b} - 2 c_1^{(0)} + c_1^{(1)}\big),\\
 D_{4f} &= 
 \frac{1}{6}E_{4b} +\frac{1}{4} c_1^{(0)}\,,
 \end{split}
\end{equation}
at $\Order(S^4)$, and
\begin{align}\label{eq:map_5}
 D_{5a} &= -\frac{1}{10} - 2 c_2^{(0)} + 
  2 c_2^{(1)} - 2 c_2^{(2)}, \nonumber\\
  D_{5b} &=
\frac{1}{60}  \big(1 - 3 E_{4a} - 16 E_{5b} - 5 E_{5c} + 120 c_2^{(0)} - 
    60 c_2^{(1)}\big), \nonumber\\
    D_{5c} &=
 \frac{1}{180} \big(18 E_{5b} + 5 E_{5c} + 3 E_{5e} - 90 c_2^{(0)}\big), \nonumber\\ 
 D_{5d} &=
 \frac{4}{15} + 2 c_2^{(0)} - 2 c_2^{(1)} + 
  2 c_2^{(2)}, \nonumber\\
  D_{5e} &=
 \frac{1}{180} \big(E_{5a}\! -15  - 48 E_{5b}\! - 17 E_{5c}\! - 360 c_2^{(0)}\! + 
    180 c_2^{(1)}\big), \nonumber\\
    D_{5f} &= 
\frac{1}{180} \big(18 E_{5b} + 6 E_{5c} - E_{5d} + 3 E_{5e} + 90 c_2^{(0)}\big), \nonumber\\
 D_{5g} &= 
\frac{1}{180} \big(-19 - 18 E_{4a}\! + 3 E_{5a}\! - E_{5c}\! + 180 c_3^{(0)}\! - 
    90 c_3^{(1)}\big), \nonumber\\
    D_{5h} &= 
\frac{1}{180} \big(6 E_{5b} + E_{5c} - 2 E_{5d} + 9 E_{5e} - 45 c_3^{(0)}\big), \nonumber\\
 D_{5i} &=
 \frac{1}{180} \big(21 + 27 E_{4a} - 2 E_{5a} - E_{5c} + 180 c_3^{(0)} - 
    90 c_3^{(1)}\big), \nonumber\\
    D_{5j} &= 
\frac{1}{180} \big(-6 E_{5b} + E_{5d} - 9 E_{5e} - 45 c_3^{(0)}\big)
\end{align}
at $\Order(S^5)$.  We stress that the map in Eqs.~\eqref{eq:map_4} and~\eqref{eq:map_5} holds on the support of the Teukolsky solutions, as we have used it to set to zero any contact deformation of the helicity-reversing exponential.

A natural choice for the coefficient  $E_{4,b}$  is either  to set it to zero or to reabsorb it into the $D_{4}$ coefficients via a field redefinition. From an on-shell approach, this is expected to be doable as    there are only 6 free coefficients at this spin order that parametrize a crossing-symmetric Compton ansatz  corresponding to the  6-contact deformations of the BCFW exponentials: 3 for the helicity-preserving amplitude and 3 for the helicity-reversing one (see Tables 1 and 2 of Ref.~\cite{Haddad:2023ylx}). 
Therefore, this choice  results  in setting $D_{4,a,...,f}=0$ when evaluating the near+far-zone Teukolsky solutions,  which is nothing but the on-shell manifestation of the vanishing of the static tidal Love numbers for Kerr BHs~\cite{Pani:2015hfa,Landry:2015zfa,Chia:2020yla,Charalambous:2021mea,Hui:2022vbh,Ivanov:2022qqt,Saketh:2023bul},  since  such operators are of the form $(E_{SS})^2$ or $(B_{SS})^2$, which are the electric/magnetic components of the Weyl tensor $E_{\mu\nu}$/$B_{\mu\nu}$ contracted with the spin vector.

At $\Order(S^5)$---discarding the $|a|$ operators---there are 10 crossing-symmetric contact deformations of the BCFW exponentials on-shell: 5 for the helicity-preserving amplitude and 5 for the helicity-reversing one \cite{Haddad:2023ylx}. This counting agrees with the 10 $D_{5a,...,j}$ coefficients in Eqs.~\eqref{eq:map_5}; therefore, one expects to be able to reabsorb the 5 free $E_{5a,...e}$ coefficients via field redefinitions.
The operators with coefficients $D_{5g,h,i,j}$ are in general not vanishing when evaluating the near+far-zone Teukolsky solutions. Such operators are of dynamical type  and as is well known, the dynamical Love numbers of Kerr BH are in general non-vanishing~\cite{Saketh:2023bul}. 
Hence, one expects such coefficients to be non-vanishing even after field redefinitions.
It is interesting to note that the $D_{5g,h,i,j}$  coefficients are proportional to $c_{3}^{(i)}$, which are also the coefficients with non-trivial polygamma and $\log \kappa\epsilon$ dependence, capturing dynamical finite-size effects.

More intriguing are the coefficients $D_{5a,...f}$, as they are proportional to the $c_2^{(i)}$ coefficients, the latter vanishing on the Teukolsky solutions. These operators appear because of the breaking of the spherical symmetry by the Kerr spin, and feature a quadrupole-octupole mixing of (derivatives of) the tidal fields $E_{\mu\nu}$ and $B_{\mu\nu}$. 
In Ref.~\cite{Saketh:2023bul}, it was shown that this type of mixing produces coefficients proportional to those of the static tidal numbers, thus vanishing on the support of the  Teukolsky solutions, while the even-in-spin contributions vanish by parity arguments.
However, in Subsection \ref{subsec:constrains_wilson_coeffs}, we find that starting at fifth order in  spin, this mixing in general produces non-vanishing coefficients, such as $D_{5a}=-1/10$ and $D_{5d}=4/15$, which are independent of any value of the linear-in-Riemann $E_{4,5,a...}$ coefficients, thereby remain non-vanishing even after field redefinitions. 

To understand the meaning of the resulting $D_{5a}=-1/10$ and $D_{5d}=4/15$ values, one needs to  trace back their appearance  when evaluating the Teukolsky amplitude. 
The Compton ansatz in Eq.~\eqref{eq:ansatz} is constrained by the 3-point factorization, and on top of this, it needs to be supplemented with constraints from the cancellation of spurious poles. 
Indeed, at fifth order in the SME,  these two constraints  result into the requirement  $c_{3}^{(2)}=4/15-c_{3}^{(0)}+c_{3}^{(1)}$, as given in Eq. \eqref{eq:constrainspurious5}. 
The rational number $4/15$ appears from  3-point factorization arguments of  the Compton ansatz which is captured by the exponential $e^{-(2w-k_2-k_3)\cdot a}$ in  Eq.~\eqref{eq:ansatzspin}. 
This  number can be traced to a pure far-zone contribution in the Teukolsky amplitude as shown in Ref.~\cite{Bautista:2023szu}. However, far-zone contributions were discarded in Ref.~\cite{Saketh:2023bul}. 
Notice from the spurious pole constraint, this rational number can be understood as  the result of a combination of coefficients that also appear in the dynamical coefficients $D_{5g,h,i,j}$, which receive contributions from both far-zone and near-zone physics, as shown in Eqs.~\eqref{eq:far_Zone_solutions} and \eqref{eq:coeffsnear}. This does not come as a surprise since starting at fifth order in spin, the near-zone and far-zone solutions get mixed in the contact terms.

\subsection{Scattering angle for a spinning test body in Kerr background}
\label{sec:testScatter}
In this subsection, we compute the aligned-spin scattering angle for a spinning test body moving in a Kerr background, starting from the action in Eq.~\eqref{lagrw}.
We derive the integrand of the scattering angle without a PM expansion, to all orders in the Kerr spin, and to fifth order in the spin of the test body.
Then, we compute the scattering angle in a PM expansion to 6PM order and spin expansion to sixth order in the Kerr spin, though we only write in Eq.~\eqref{testAngle} the 3PM angle, which was needed in Sec.~\ref{sec:boundObserv} to obtain the NNLO S$^3$ and S$^4$ test-body coefficients.
The angle we derive also provides an independent check of the mapping in Eqs.~\eqref{eq:map_4} and~\eqref{eq:map_5}, and we use it in the following subsection to obtain additional constraints on the Wilson coefficients from GSF results.

The Mathisson-Papapetrou-Dixon (MPD) equations  describe the motion of a multipolar test body in an arbitrary (vacuum) curved background~\cite{mathisson2010republication,Papapetrou:1951pa,Dixon:1970zza,Harte:2014wya}, and they can be written as 
\begin{subequations}
\label{MPDeqns}
\begin{align}
&\frac{\mr D p_\mu}{\di\tau} + \frac{1}{2} R_{\mu\nu\rho\sigma} \dot{z}^\nu S^{\rho\sigma} = F_\mu , \\
&\frac{\mr D S^{\mu\nu}}{\di\tau} - 2 p^{[\mu}u^{\nu]} = N^{\mu\nu},
\end{align}
\end{subequations}
where $F_\mu$ and $N^{\mu\nu}$ are the `force' and `torque', along with the covariant spin-supplementary condition (SSC)
\begin{equation}
\label{covSSC}
p_\mu S^{\mu\nu} = 0,
\end{equation}
which determines the tangent $\dot{z}^\mu$ to the worldline $z(\tau)$.

The 4-velocity $u^\mu$ is not in the same direction as $\dot{z}^\mu$, but is related to the 4-momentum $p^\mu$ via
\begin{equation}
u^\mu = \frac{p^\mu}{\sqrt{-p^2}},
\end{equation}
where $p^2$ is related to the dynamical mass $\mathcal{M}$ through the mass-shell constraint
\begin{equation}
p^2 = -\mathcal{M}^2(z,u,S).
\end{equation}
The dynamical mass in terms of generic Wilson coefficients was obtained in Ref.~\cite{Scheopner:2023rzp} to fifth order in the test body's spin.

One can verify that the same equations of motion (in a vacuum background) are produced by the action~\eqref{lagrw} with the Lagrange multiplier~\cite{Vines:2017hyw,Vines:2016unv}
\begin{equation}
\alpha = \frac{\dot{z}\cdot p}{p^2} = - \frac{\dot{z} \cdot u}{\mathcal{M}} = \frac{1}{\mathcal{M}},
\end{equation}
where we use the normalization condition $\dot{z}\cdot u =-1$.
The force $F_\mu$ and torque $N^{\mu\nu}$ are given by
\begin{subequations}
\begin{align}
F_\mu &= - \frac{\alpha}{2} \frac{\mr D}{\mr D z^\mu} \mathcal{M}^2, \\
N^{\mu\nu} &= -\alpha \left(u^{[\mu} \frac{\mr D}{\mr D u_{\nu]}} + 2 {S^{[\mu}}_\lambda \frac{\mr D}{\mr D S_{\nu]\lambda}}\right) \mathcal{M}^2 \nonumber\\
&= -\alpha \left(u^{[\mu} \frac{\mr D}{\mr D u_{\nu]}} + \sigma^{[\mu} \frac{\mr D}{\mr D \sigma_{\nu]}}\right) \mathcal{M}^2,
\end{align}
\end{subequations}
where the spin vector $\sigma^\mu$ is related to the spin tensor $S^{\mu\nu}$ via
\begin{equation}
\sigma^\mu = -\frac{1}{2} {\eta^\mu}_{\nu\rho\lambda} u^\nu S^{\rho\lambda}, \qquad
S^{\mu\nu} = {\eta^{\mu\nu}}_{\rho\lambda} u^\rho \sigma^\lambda,
\end{equation}
with $\eta_{\mu\nu\rho\lambda}$ being the volume form.
We denote the magnitude of the spin vector by $\sigma = \sqrt{\sigma^\mu \sigma_\mu}$, and define $s := \sigma / m$, where $m$ is the mass of the test body, so that $s$ has units of length like the Kerr spin $a$.

To solve for $\dot{z}^\mu$, we first solve for the components of $u_\mu = (u_t, u_r, 0, u_\phi)$.
We use the conserved quantities from the Killing vectors $t^\mu := (\partial_t)^\mu$ and $\phi^\mu := (\partial_\phi)^\mu$, which are the energy and angular momentum~\cite{ehlers1977dynamics}
\begin{subequations}
\begin{align}
E &= - p_\mu t^\mu - \frac{1}{2} S^{\mu\nu} \nabla_\mu t_\nu, \nonumber\\
&= -\mathcal{M} u_t + \frac{M \sigma (a u_t + u_\phi)}{r^3}, \\
J &= p_\mu \phi^\mu + \frac{1}{2} S^{\mu\nu} \nabla_\mu \phi_\nu, \nonumber\\
&= \mathcal{M} u_\phi + \frac{\sigma}{r^3} (a M u_\phi + a^2 M u_t - u_t r^3).
\end{align}
\end{subequations}
These conserved quantities, together with the normalization condition $u^2 = -1$, can be solved for the three components of $u^\mu$.

After that, we apply $\mr D/\di\tau$ on the SSC condition~\eqref{covSSC} and use the MPD equations~\eqref{MPDeqns}, yielding 
\begin{align}
S^{\mu\nu} F_\nu + N^{\mu\nu} p_\nu - \frac{1}{2} S^{\mu\nu} R_{\nu\lambda\rho\sigma} \dot{z}^\lambda S^{\rho\sigma}
+ 2 p^{[\mu} \dot{z}^{\nu]} p_\nu = 0.
\end{align}
The last two terms can be written as follows:
\begin{subequations}
\begin{align}
2 p^{[\mu} \dot{z}^{\nu]} p_\nu &= \mathcal{M}^2 \left(\dot{z}^\mu - u^\mu\right), \\
- \frac{1}{2} S^{\mu\nu} R_{\nu\lambda\rho\sigma} \dot{z}^\lambda S^{\rho\sigma} &= 
\dot{z}^\mu \, \tensor[^*]{R}{^*_{usus}} + \tensor[^*]{R}{^{*\mu}_{sus}} \nonumber\\
&\quad + \dot{z}\cdot \sigma \, \tensor[^*]{R}{^{*\mu}_{uus}},
\end{align}
\end{subequations}
where we used the double (left and right) dual of the Riemann tensor $\tensor[^*]{R}{^*_{\mu\nu\rho\sigma}} = \frac{1}{2} {\eta_{\mu\nu}}^{\kappa\lambda} R_{\kappa\lambda\tau\chi} {\eta^{\tau\chi}}_{\rho\sigma} \frac{1}{2}$.

For aligned spins, $\dot{z}\cdot \sigma = 0$, and the above equations can be solved for $\dot{z}^\mu$, leading to
\begin{align}
\dot{z}^\mu = \frac{\mathcal{M}^2 u^\mu - \tensor[^*]{R}{^{*\mu}_{sus}} - S^{\mu\nu} F_\nu - N^{\mu\nu} p_\nu}{\mathcal{M}^2 + \tensor[^*]{R}{^*_{usus}}}.
\end{align}
This relation can then be evaluated for the components of $\dot{z}^\mu = (\dot{z}^t, \dot{z}^r, 0, \dot{z}^\phi)$, yielding the orbital equation $\di\phi / \di r = \dot{z}^\phi /\dot{z}^r$.
Our result for $\di\phi / \di r$ agrees with the $\Order(s^2)$ result given by Eq.~(65) of Ref.~\cite{Bini:2017pee}.
In the Supplemental Material, we provide $\di\phi / \di r$ to all PM orders, all orders in the Kerr spin, and fifth order in the spin of the test body. 

The scattering angle can be computed from the integral
\begin{equation}
\theta = \int \di \phi = \int_{0}^{1/b} 2 \frac{\di\phi}{\di u} \di u \,,
\end{equation}
where $b$ is the impact parameter and we defined $u := M/r$. 
The integral can be evaluated in a PM expansion while taking the \emph{partie-finie}, i.e., neglecting the divergent terms~\cite{Damour:2016gwp}.
Performing the integration to 3PM order and expanding to sixth order in both spins, we obtain 
\begin{widetext}
\begin{subequations}\label{testAngle}
\begin{align}
\theta_\text{test} &:= \frac{G M}{b} \theta_\text{test}^{(1)} + \pi \frac{G^2 M^2}{b^2} \theta_\text{test}^{(2)} + \frac{G^3 M^3}{b^3} \theta_\text{test}^{(3)} + \Order(G^4), \\
\theta_\text{test}^{(1)} &= \frac{2 \left(v^2+1\right)}{v^2} - \frac{4 (a+s)}{b v} + \frac{2 \left(v^2+1\right) (a+s)^2}{b^2 v^2} 
- \frac{4 (a+s)^3}{b^3 v} 
+ \frac{2 \left(v^2+1\right) (a+s)^4}{b^4 v^2} 
- \frac{4 (a+s)^5}{b^5 v}  \nonumber\\
&\quad+ \frac{2 a  \left(v^2+1\right) \left(a^5+6 a^4 s+15 a^3 s^2+20 a^2 s^3+15 a s^4+6 s^5\right)}{b^6 v^2}
+ \Order(a^7,s^6), \\
\theta_\text{test}^{(2)} &= \frac{3}{v^2}+\frac{3}{4} 
-\frac{\left(3 v^2+2\right) (4 a+3 s)}{2 b v^3}
+ \frac{3}{16 b^2 v^4} \left[a^2 \left(15 v^4+72 v^2+8\right)+8 a s \left(3 v^4+15 v^2+2\right)+4 s^2 \left(2 v^4+11 v^2+2\right)\right] \nonumber \\
&\quad
- \frac{3}{4 b^3 v^3} \left[4 a^3 \left(5 v^2+4\right)+a^2 s \left(51 v^2+44\right)+40 a s^2 \left(v^2+1\right)+4 s^3 \left(2 v^2+3\right)\right] \nonumber \\
&\quad
+ \frac{15}{b^4 v^4 \left(1-v^2\right)} \bigg\{\!
a^3 s \!\left(1-\frac{5 v^6}{4}-\frac{11 v^4}{2}+\frac{23 v^2}{4}\right)\!
+a^2 s^2 \!\left(\frac{3}{2}-\frac{3 v^6}{2}-\frac{59 v^4}{8}+\frac{59 v^2}{8}\right)\!
+a s^3 \!\left(\! 1-\frac{2 v^6}{3}-\frac{13 v^4}{3}+4 v^2 \!\right)\! \nonumber\\
&\qquad
+ a^4 \left(\frac{1}{4}-\frac{35 v^6}{96}-\frac{145 v^4}{96}+\frac{13 v^2}{8}\right)
+ s^4 \bigg[\frac{1}{4} + v^2 \left(\frac{3}{4}-\frac{D_4^\text{NLO}}{32}\right)
+v^4 \left(-\frac{3D_4^\text{NLO}}{32} +\frac{5 D_4^\text{NNLO}}{64}-1\right) \nonumber\\
&\quad\qquad
+v^6 \left(\frac{105 D_{4 c}}{128}-\frac{41 D_4^\text{NLO}}{512}-\frac{5 D_4^\text{NNLO}}{512}+\frac{35 E_{4b}}{256}\right)
\bigg]\bigg\} \nonumber\\
&\quad
+ \frac{15}{2 b^5 v^3 \left(1-v^2\right)} \bigg\{
a^5 \!\left(\frac{7 v^4}{2}-\frac{v^2}{2}-3 \!\right)\! +a^4 s \!\left(\frac{125 v^4}{8}-\frac{11 v^2}{8}-\frac{57}{4}\right)\!
+a^3 s^2 \left(27 v^4-27\right)
+a^2 s^3 \!\left(\! 22 v^4+\frac{7 v^2}{2}-\frac{51}{2} \!\right) \nonumber\\
&\qquad
+ a s^4 \left[-12 +v^2 \left(\frac{3 D_4^\text{NLO}}{4}-\frac{3 D_4^\text{NNLO}}{8}+4\right) +v^4 \left(-\frac{63 D_{4 c}}{8}+\frac{39 D_4^\text{NLO}}{32}-\frac{9 D_4^\text{NNLO}}{32}-\frac{21 E_{4b}}{16}+8\right)\right] \nonumber\\
&\qquad
+ s^5 \left[-\frac{9}{4} +v^2 \left(\frac{3 D_5^\text{NLO}}{8}+\frac{247}{160}\right) + v^4 \left(\frac{25 D_5^\text{NLO}}{64}-\frac{7 D_5^\text{NNLO}}{64}+\frac{1639}{1280}\right)\right]
\bigg\} \nonumber\\
&\quad
+ \frac{15}{4 b^6 v^4 \left(1-v^2\right)^2} \bigg\{
a^6 \left(\frac{147 v^8}{64}+\frac{245 v^6}{32}-\frac{1309 v^4}{64}+\frac{35 v^2}{4}+\frac{7}{4}\right)
+a^5 s \left(\frac{49 v^8}{4}+\frac{175 v^6}{4}-\frac{455 v^4}{4}+\frac{189 v^2}{4}+\frac{21}{2}\right) \nonumber\\
&\qquad
+a^4 s^2 \left(\frac{105 v^8}{4}+\frac{833 v^6}{8}-\frac{1043 v^4}{4}+\frac{833 v^2}{8}+\frac{105}{4}\right)
+a^3 s^3 \left(28 v^8+133 v^6-315 v^4+119 v^2+35\right) \nonumber\\
&\qquad
+ a^2 s^4 \bigg[\frac{105}{4}
+v^2 \!\left(\frac{147}{2}-\frac{21 D_4^\text{NLO}}{16} +\frac{7 D_4^\text{NNLO}}{16}\right)
+v^4 \!\left(\frac{441 D_{4 c}}{16}-\frac{441 D_4^\text{NLO}}{64}+\frac{189 D_4^\text{NNLO}}{64}+\frac{147 E_{4b}}{32}-\frac{847}{4}\right) \nonumber\\
&\qquad\quad
+v^6 \left(98-\frac{441 D_{4 c}}{64}+\frac{1449 D_4^\text{NLO}}{256}-\frac{819 D_4^\text{NNLO}}{256}-\frac{147 E_{4b}}{128}\right)
+v^8 \bigg(14-\frac{1323 D_{4 c}}{64}+\frac{651 D_4^\text{NLO}}{256}\nonumber\\
&\qquad\qquad
-\frac{49 D_4^\text{NNLO}}{256}-\frac{441 E_{4b}}{128}\bigg)
\bigg] 
+ a s^5 \bigg[\frac{21}{2}
+v^2 \bigg(-\frac{7 D_5^\text{NLO}}{8}+\frac{7 D_5^{\text{NLO\,S}^6}}{40}+\frac{357}{16}\bigg) \nonumber\\
&\qquad\quad
+v^4 \bigg(-\frac{77 D_5^\text{NLO}}{12}+\frac{49 D_5^\text{NNLO}}{96}+\frac{63 D_5^{\text{NLO\,S}^6}}{160}-\frac{49 D_5^{\text{NNLO\,S}^6}}{80}-\frac{50211}{640}\bigg)\nonumber\\
&\qquad\quad
+v^6 \bigg(\frac{2555 D_5^\text{NLO}}{384}-\frac{49 D_5^\text{NNLO}}{192}+\frac{147 D_5^{\text{N$^3$LO\,S}^6}}{320}-\frac{399 D_5^{\text{NLO\,S}^6}}{640}+\frac{49 D_5^{\text{NNLO\,S}^6}}{64}+\frac{27027}{640}\bigg)\nonumber\\
&\qquad\quad
+v^8 \bigg(\frac{245 D_5^\text{NLO}}{384}-\frac{49 D_5^\text{NNLO}}{192}-\frac{147 D_5^{\text{N$^3$LO\,S}^6}}{320}+\frac{7 D_5^{\text{NLO\,S}^6}}{128}-\frac{49 D_5^{\text{NNLO\,S}^6}}{320}+\frac{273}{80}\bigg)\bigg]
\bigg\} + \Order(a^7, s^6), \\
\theta_\text{test}^{(3)} &= \frac{2 \left(5 v^6+45 v^4+15 v^2-1\right)}{3 v^6}
-\frac{4(3 a+2 s)\! \left(5 v^4+10 v^2+1\right) }{b v^5} \nonumber\\
&\quad
+ \frac{4 }{b^2 v^6} \Big[a^2 \left(7 v^6+75 v^4+45 v^2+1\right)+2 a s \left(5 v^6+55 v^4+35 v^2+1\right)+s^2 \left(3 v^6+35 v^4+25 v^2+1\right)\Big] \nonumber\\
&\quad
- \frac{40 (a+s)^2}{3 b^3 v^5} \left[a \left(21 v^4+50 v^2+9\right)+2 s \left(3 v^4+10 v^2+3\right)\right]\nonumber\\
&\quad
+ \frac{15}{b^4 v^4 \left(1-v^2\right)} \bigg\{
a^4 \left(48-6 v^6-64 v^4+20 v^2+\frac{2}{v^2}\right)+a^3 s \left(\frac{496}{3}-\frac{56 v^6}{3}-208 v^4+\frac{160 v^2}{3}+\frac{8}{v^2}\right)\nonumber\\
&\qquad
+a^2 s^2 \left(208-20 v^6-240 v^4+40 v^2+\frac{12}{v^2}\right)+a s^3 \left(112-8 v^6-112 v^4+\frac{8}{v^2}\right) \nonumber\\
&\qquad
+ s^4 \bigg[
\frac{2}{v^2} -\frac{8}{15}  D_4^\text{NLO}+\frac{64}{3}
+v^2 \left(-\frac{72}{25}  D_4^\text{NLO}+\frac{48 D_4^\text{NNLO}}{25}-\frac{20}{3}\right)
+v^4 \bigg(\frac{4608 D_{4 c}}{175}-\frac{632 D_4^\text{NLO}}{175}+\frac{96 D_4^\text{NNLO}}{175}\nonumber\\
&\qquad\qquad
+\frac{768 E_{4b}}{175}-16\bigg) 
+v^6 \left(\frac{512 D_{4 c}}{175}-\frac{152 D_4^\text{NLO}}{525}-\frac{16 D_4^\text{NNLO}}{525}+\frac{256 E_{4b}}{525}-\frac{2}{3}\right)
\bigg]\bigg\} 
+ \Order(a^5,s^5),
\end{align}
\end{subequations}
\end{widetext}
To simplify the above expressions, and since we are working with BHs, we set the coefficients $C_2 = C_3 = C_4 = C_5 = 1$, and truncated the spin expansions to the orders needed below, but our result in the Supplemental Material retains the dependence on all Wilson coefficients, in addition to including the 6PM expansion of the scattering angle.

In Eq.~\eqref{testAngle}, we defined the following combinations of Wilson coefficients:
\begin{subequations}
\begin{align}
D_4^\text{NLO} &:= 2 D_{4a}+2 D_{4b}+12 D_{4c}+E_{4b}, \\
D_4^\text{NNLO} &:= 12 D_{4c}-6 D_{4e}-12 D_{4f}+E_{4b}, \\
D_5^\text{NLO} &:= 3 D_{5b}+6 D_{5c}-2 D_{5d}-2 D_{5e}-8 D_{5f} \nonumber\\
&\qquad
-3 D_{5h}+\frac{3 }{20} E_{4a}+\frac{1}{5} E_{5b}, \\
D_5^\text{NNLO} &:= 4 D_{5d}+4 D_{5e}-14 D_{5f}-12 D_{5h}+9 D_{5i} \nonumber\\
&\qquad
+9 D_{5j}-\frac{6 E_{4a}}{5}+\frac{7 E_{5b}}{5}, \\
D_5^{\text{NLO\,S}^6} &:= 5 D_{5d}+5 D_{5e}+20 D_{5f}-15 D_{5h}-E_{5b}, \\
D_5^{\text{NNLO\,S}^6} &:= 5 D_{5d}+5 D_{5e}+45 D_{5f}-2 E_{5b}, \\
D_5^{\text{N$^3$LO\,S}^6} &:= 5 D_{5d}+5 D_{5e}+E_{5b},
\end{align}
\end{subequations}
such that in a PN expansion, the NLO S$^4$ scattering angle (or Hamiltonian) only depends on $D_4^\text{NLO}$ while the NNLO only depends on $D_4^\text{NLO}$ and $D_4^\text{NNLO}$, and similarly for the dependence of the S$^5$ contributions on $D_5^\text{NLO}$ and $D_5^\text{NNLO}$.
The contributions at sixth order in spin depend on different combinations of the $D_5$ and $E_5$ coefficients, denoted by $D_5^{\text{NLO\,S}^6}$, $D_5^{\text{NNLO\,S}^6}$, and $D_5^{\text{N$^3$LO\,S}^6}$.

The 2PM part of our scattering angle, to fourth order in the test body's spin, agrees with Ref.~\cite{Siemonsen:2019dsu}, and agrees after setting the Wilson coefficients to $D_{4a,\dots,f}=E_{4b} = 0$ with Ref.~\cite{Guevara:2018wpp}. 
We also checked that our angle, at quadratic order in the test body's spin and up to 5PM order, agrees with the results of Ref.~\cite{Damgaard:2022jem}, which are valid to all orders in the Kerr spin.
To our knowledge, our result at fifth order in the test body's spin is new, and so are the higher PM contributions at lower orders in spin.
A generalization of the scattering angle for generic spin orientations, through the impulse and spin kick, was obtained at lower orders in spin in, e.g., Refs.~\cite{Bern:2020buy,Kosmopoulos:2021zoq,Jakobsen:2022zsx,Gonzo:2024zxo}.

\subsection{Constraints on the Wilson coefficients}\label{subsec:constrains_wilson_coeffs}

\subsubsection{Worldline test-body scattering angle vs Compton angle}
We can compare the scattering angle derived from the Compton ansatz in Eqs.~\eqref{eq:angle_compton} to the 2PM part of the test-body scattering angle in Eq.~\eqref{testAngle}, which leads to the following relations:
\begin{subequations}
\label{eq:test_angle_const}
\begin{align}
D_4^\text{NLO} &= 3 c_1^{(0)}-c_1^{(1)}+2 c_1^{(2)} = 0, \\
D_4^\text{NNLO} &= 6 c_1^{(0)}-3 c_1^{(1)} = 0, \\
D_{4c} &= -\frac{E_{4b}}{6} + \frac{c_1^{(0)}}{4} = -\frac{E_{4b}}{6}, \\
D_5^\text{NLO} &= -\frac{19}{60}+\frac{3 c_3^{(0)}}{4}-c_2^{(1)}-4 c_2^{(2)} \nonumber\\
&= -\frac{19}{60}+\frac{3 c_3^{(0)}}{4}, \\
D_5^\text{NNLO} &= \frac{11}{6} +\frac{39 c_3^{(0)}}{4}-\frac{9 c_3^{(1)}}{2}-7 c_2^{(1)}+8 c_2^{(2)} \nonumber\\
&= \frac{11}{6} +\frac{39 c_3^{(0)}}{4}-\frac{9 c_3^{(1)}}{2}, \\
D_5^{\text{NLO\,S}^6} &= \frac{11}{12}+15 c_2^0+\frac{15 c_3^0}{4}-5 c_2^1+10 c_2^2 \nonumber\\
&= \frac{11}{12}+\frac{15 c_3^0}{4}, \\
D_5^{\text{NNLO\,S}^6} &= \frac{11}{12}+\frac{45 c_2^0}{2}-5 c_2^1+10 c_2^2=\frac{11}{12}, \\
D_5^{\text{N$^3$LO\,S}^6} &= \frac{11}{12}-5 c_2^1+10 c_2^2=\frac{11}{12}\,,
\end{align}
\end{subequations}
where the latter equality follows on the support of the Teukolsky solutions, obtained via the splitting Eq. \eqref{eq:coeffs_decomp} and the explicit solutions Eqs. \eqref{eq:far_Zone_solutions} and \eqref{eq:coeffsnear}.

If we combine the above solution with the maps obtained in Eqs.~\eqref{eq:map_4} and \eqref{eq:map_5}, then the $D$ coefficients drop and we get that the $E$ coefficients satisfy
\begin{subequations}\label{eq:econstrains}
\begin{align}
&2 E_{5a}+66 E_{5b}+32 E_{5c}-14 E_{5d}+33 E_{5e}=0,\\
&27 E_{4a}+47 E_{5d}-14 E_{5a}-318 E_{5b}-173 E_{5c}\nonumber\\
&\qquad\qquad -231 E_{5e}=9, \\
&30 E_{5b}-E_{5a}-4 E_{5c}-2 E_{5d}+15 E_{5e} = 0, \\
&E_{5a}+42 E_{5b}+37 E_{5c}-9 E_{5d}+27 E_{5e} = 0, \\
&E_{5a}-12 E_{5b}-17 E_{5c} = 0,
\end{align}
\end{subequations}
which can be solved for the $E_5$ coefficients in terms of $E_{4a}$ leading to
\begin{alignat}{2}
E_{5a}&= -\frac{4}{7} + \frac{12 E_{4a}}{7},\qquad &E_{5b}&= -\frac{1}{28} + \frac{3 E_{4a}}{28},\nonumber\\
E_{5d}&= -\frac{2}{7} + \frac{6 E_{4a}}{7}, \qquad &E_{5c}&= -\frac{1}{119} + \frac{3 E_{4a}}{119},\nonumber\\
E_{5e}&= -\frac{5}{714} + \frac{5 E_{4a}}{238}. &&
\end{alignat}
Substituting this solution into the map~\eqref{eq:map_5} leads to
\begin{subequations}
\begin{align}
D_{5a} &= -\frac{1}{10}, \\
D_{5b} &= \frac{16}{595}-\frac{48 E_{4a}}{595}, \\
D_{5c} &= -\frac{1}{255}+\frac{E_{4a}}{85}, \\
D_{5d} &= \frac{4}{15}, \\
D_{5e} &= -\frac{8}{105}-\frac{3 E_{4a}}{140}, \\
D_{5f} &= -\frac{1}{420}+\frac{E_{4a}}{140}, \\
D_{5g} &= -\frac{88}{765}+ c_3^{(0)} -\frac{c_3^{(1)}}{2} -\frac{73 E_{4a}}{1020},  \\
D_{5h} &= \frac{1}{630} -\frac{c_3^{(0)}}{4} -\frac{E_{4a}}{210},  \\
D_{5i} &= \frac{659}{5355} + c_3^{(0)} - \frac{c_3^{(1)}}{2} +\frac{467 E_{4a}}{3570},  \\
D_{5j} &= -\frac{1}{21420} -\frac{c_3^{(0)}}{4} + \frac{E_{4a}}{7140}.
\end{align}
\end{subequations}
Thus, we obtain nonzero values for all the $D_{5a,\dots,j}$ Wilson coefficients, but while the two coefficients $D_{5a}$ and $D_{5d}$ are uniquely determined, the others are given in terms of $E_{4a}$, which remains an unknown.
These values are in tension with some of the results of Ref.~\cite{Saketh:2023bul}, which showed that the coefficients of tidal mixing terms, such as $E_{SS}B_{SSS}$, are zero.
This discrepancy could be due to an ambiguity in defining the tidal terms in the action when spin is included, or due to a missing far-zone contribution in Ref.~\cite{Saketh:2023bul}, as discussed at the end of Subsection~\ref{sec:worldlineCompton}.

\subsubsection{Constraints from GSF invariants}
In Sec.~\ref{sec:boundObserv}, we started from a scattering angle ansatz with unknowns coefficients, related it to a Hamiltonian, then computed the redshift and spin-precession invariants and compared them to GSF results to solve for the unknowns.
The same steps can be performed but using the test-body scattering we derived in the previous subsection, which includes Wilson coefficients. 

Specifically, we take the scattering angle ansatz from Sec.~\ref{sec:angleParam} and use the angle from Eq.~\eqref{testAngle} to determine the coefficients in the test-mass limit, at NLO S$^4$, NNLO S$^4$, NLO S$^5$, and (partially) NLO S$^6$.
The result of the matching for the $a_1^4$ part of the scattering angle in terms of Wilson coefficients reads
\begin{align}
\frac{\theta_{a_1^4}}{\Gamma} &= \frac{a_1^4}{b^4} \Bigg\{
\frac{G M\left(\frac{2}{v^2}+2\right)}{b}
+ \frac{\pi G^2 M^2}{b^2v^4} \bigg\{\frac{15}{4} \nonumber\\
&\qquad
+ v^2 \left[\frac{105 \delta }{16}+\left(\frac{15 \delta }{64}-\frac{15}{64}\right) D_4^\text{NLO}+\frac{345}{16}\right] \nonumber\\
&\qquad
+ v^4 \bigg[\frac{175}{64}+\frac{175 \delta }{64}+\left(\frac{15 \delta }{16}-\frac{15}{16}\right) D_4^\text{NLO} \nonumber\\
&\quad\qquad
+\left(\frac{75}{128}-\frac{75 \delta }{128}\right) D_4^\text{NNLO}\bigg]\bigg\} 
+ \frac{G^3M^3}{b^3v^6} \bigg\{
30 \nonumber\\
&\qquad
+v^2 [200 \delta +(4 \delta -4) D_4^\text{NLO}-50 \nu +550] \nonumber\\
&\qquad
+ v^4 \bigg[650+400 \delta +\left(\frac{128 \delta }{5}-\frac{128}{5}\right) D_4^\text{NLO} \nonumber\\
&\quad\qquad
+\left(\frac{72}{5}-\frac{72 \delta }{5}\right) D_4^\text{NNLO}+h_{40} \nu\bigg]
\bigg\}
\Bigg\},
\end{align}
and for the $a_1^5$ contribution
\begin{align}
&\frac{\theta_{a_1^5}}{\Gamma} = \frac{v a_1^5}{b^5} \Bigg\{
{-}\frac{4 G M}{b v} + \frac{\pi G^2 M^2}{b^2 v^3} \bigg\{
-\frac{45 \delta }{16}-\frac{315}{16} \nonumber\\
&\quad
+ v^2 \!\left[\frac{45 D_5^\text{NLO}}{32}-\frac{2019}{128}
- \delta  \left(\frac{45 D_5^\text{NLO}}{32}+\frac{1341}{128}\right)\right]\!
\bigg\}\!
\Bigg\},
\end{align}
where instead of the two unknowns $f_{50}$ and $g_{50}$ in Eq.~\eqref{deltatheta56}, we now have one unknown, which is the combination $D_5^\text{NLO}$ of Wilson coefficients. 
At the NLO S$^6$, we get the following contribution:
\begin{align}
&\frac{\theta_{a_1^5a_2}}{\Gamma} = \frac{a_1^5a_2}{b^6} \bigg\{
\frac{12G M}{b v^2} \left(1+v^2\right) 
+ \frac{\pi G^2 M^2}{b^2 v^4} \bigg\{
\frac{315}{8} \nonumber\\
&\quad
+ v^2 \bigg[\frac{26775}{128}-\frac{105 D_5^\text{NLO}}{64}+\frac{21 D_5^{\text{NLO\,S}^6}}{64}\nonumber\\
&\qquad
+\delta  \bigg(\frac{105 D_5^\text{NLO}}{64}-\frac{21 D_5^{\text{NLO\,S}^6}}{64}+\frac{5985}{128}\bigg) \bigg]
\bigg\}\bigg\},
\end{align}
which replaces the $f_{51}$ and $g_{51}$ unknowns from Eq.~\eqref{deltatheta56} by $D_5^\text{NLO}$ and $D_5^{\text{NLO\,S}^6}$.

Next, we obtain a Hamiltonian in terms of the scattering angle coefficients and compute from it the redshift and spin-precession invariants, expanded to first order in the mass ratio.
Comparison with circular-orbit GSF results allows us to obtain the following solution at $\Order(S^4)$:
\begin{equation}
\begin{alignedat}{2}
&h_{40} = - \frac{631}{2}, &\qquad &h_{31} = -446, \\
&D_4^\text{NLO} = 0, &\qquad &D_4^\text{NNLO} = 0,
\end{alignedat}
\end{equation}
which is the same solution obtained in Sec.~\ref{sec:boundObserv} for $h_{40}$ and $h_{31}$, in addition to confirming that the $\Order(S^4)$ Wilson coefficients are consistent with being zero for BHs~\cite{Siemonsen:2019dsu,Saketh:2023bul}.
As for the $D_5^\text{NLO}$ coefficient at $\Order(S^5)$, we get 
\begin{subequations}
\begin{alignat}{2}
&\text{polynomial:} \quad  &D_5^\text{NLO} &= -\frac{19}{60}, \\
&\text{spin expansion:} \quad &D_5^\text{NLO} &= -\frac{81}{20} + \frac{512 \zeta (3)}{15}-\frac{512 \zeta (5)}{15},
\end{alignat}
\end{subequations}
where the first solution is obtained by matching our result to the polynomial piece of the 1SF redshift, while the second solution is obtained by first expanding the irrational functions in the 1SF redshift in a small-spin expansion, as explained in Sec.~\ref{sec:NLOS5soln}.
Similarly, from the NLO S$^6$ part, we get
\begin{subequations}
\begin{alignat}{2}
&\text{polynomial:} \quad  &D_5^{\text{NLO\,S}^6} &= \frac{11}{12}, \\
&\text{spin expansion:} \quad &D_5^{\text{NLO\,S}^6} &= -\frac{71}{4}+\frac{512 \zeta (3)}{3}-\frac{512 \zeta (5)}{3},
\end{alignat}
\end{subequations}
in addition to the same solution for $f_{60}$ and $g_{60}$ as in Eq.~\eqref{S6Soln}.
These solutions are all in agreement with Eqs.~\eqref{eq:angle_compton} when combined with the constraints~\eqref{eq:test_angle_const}.

\section{Conclusions}
\label{sec:conclusions}
Improving the accuracy of gravitational waveform models is crucial for the increasing sensitivity of current and future gravitational-wave detectors~\cite{Purrer:2019jcp}.
Since spin plays an important role in compact binary inspirals, in this paper, we looked into improving the spin description in binary dynamics, by deriving new PN results at high orders in spin. 

In particular, in this work we studied the conservative, aligned-spin binary dynamics at NNLO S$^3$ and S$^4$, in addition to the NLO S$^5$ and S$^6$ PN contributions for Kerr BHs, using a synergy of methods including worldline field theory, the Tutti-Frutti method, on-shell Compton amplitudes, and BHPT.  We obtained several observables for binary systems, namely the scattering angle for hyperbolic orbits, and the redshift and spin-precession frequency for bound eccentric orbits without an eccentricity expansion.

Our result for the NNLO S$^4$ dynamics contains one unknown coefficient of order $\Order(\nu a_1^2 a_2^2)$. 
Solving for that unknown using the Tutti-Frutti method requires 1SF results for the redshift at quadratic order in both the secondary and Kerr spins, which are currently unavailable.
We leave such a calculation for future investigation.
The inclusion of higher-spin, and higher $\ell$ near-zone BHPT contributions is also left for future work. 

At NLO S$^5$, we encountered an ambiguity while solving for the unknown coefficients, due to transcendental functions of the Kerr spin appearing in the redshift at that order. 
We suggested three possibilities to determine the unknowns: matching the polynomial S$^5$ piece only, expanding in small spin, or a non-perturbative matching as a function of the Kerr spin; with  the first two  prescriptions obtained as special cases of the latter. 
While the polynomial and the small spin matching  prescriptions are of theoretical interest, for instance for obtaining an all-orders-in-spin expression for the tree-level gravitational Compton amplitude, the non-perturbative matching is relevant for  phenomenological applications as it  has the advantage of encapsulating, via simple contact term in the Compton amplitude, Kerr BH finite-size effects that are present for   spinning observables in  binary systems, as derived directly in the alternative GSF computations.

Furthermore, from a parametrized worldline action, we derived a scattering angle for a spinning test body, to fifth order in its spin, in a Kerr background, and used it to obtain constraints on the Wilson coefficients through comparisons with both GSF and Compton amplitudes results.
Our solutions for the Wilson coefficients are consistent between the two approaches, but they are in tension with recent EFT results~\cite{Saketh:2023bul}.
Therefore, fully resolving the issues with the spin$^5$ dynamics probably requires the derivation of the NLO S$^5$ PN contribution from first principles, and finding additional ways to fix the value of Wilson coefficients.

\section*{Acknowledgments}  
We would like to thank Lucille Cangemi, Giorgio Di Russo, Diego Gallego, David Kosower,  Jose Francisco Morales, Donal O'Connell, M. V. S. Saketh, Trevor Scheopner, and Zihan Zhou for useful discussions.  We are grateful to Trevor Scheopner for rechecking the Compton dictionary presented in  Ref.~\cite{Scheopner:2023rzp}, which he confirmed agrees with ours given in  Eq.~\eqref{eq:map_5}.
We also thank the organizers of the workshop ``Amplifying Gravity at All Scales'', held in the summer of 2023 in  Nordita, where the idea of constraining the Compton amplitude from eccentric gravitational self-force computations was born. 
Y.F.B. thanks  the hospitality of the physics department of the    Universidad Pedagógica y Tecnológica de Colombia  where  part of the writing in the final stage of this work was done. 
This work made use of the tensor algebra package \texttt{xAct}~\cite{xAct,martin2008xperm} in Mathematica, and the Black Hole Perturbation Toolkit \cite{BHPToolkit}.

The  work of Y.F.B. and M.S.  has been supported by the European Research Council under Advanced Investigator Grant ERC–AdG–885414. 
M.K.'s work is supported by Perimeter Institute for Theoretical Physics. Research at Perimeter Institute is supported in part by the Government of Canada through the Department of Innovation, Science and Economic Development and by the Province of Ontario through the Ministry of Colleges and Universities.
 C.K. thanks David Kosower and  the financial support from  the Grant ERC–AdG–885414 during his visit to the IPhT during the spring of 2024. C.K. acknowledges support from Science Foundation Ireland under Grant number 21/PATH-S/9610.
%%%%%%%%%%%%%%%%%%%%%%%%%%

\appendix

\section{Test-body redshift and spin-precession invariants for eccentric orbits}
\label{app:zPsiTest}
The equations of motion for a test mass on an equatorial orbit ($\theta = \pi/2$) in a Kerr background are given by (see, e.g., Ref.~\cite{Bini:2016iym})
\begin{subequations}
\begin{align}
\Delta\, r^2 \frac{\di t}{\di \tau} &= (r^2 E - a \sqrt{K}) (r^2 + a^2) + \Delta a \sqrt{K}, \\
r^4 \left(\frac{\di r}{\di \tau}\right)^2 &= (r^2 E - a \sqrt{K})^2 - \Delta (r^2 + K), \\
\Delta\, r  \frac{\di \phi}{\di \tau} &= r L - 2 m_1 \sqrt{K},
\end{align}
\end{subequations}
where $K = (L - a E)^2$ is Carter's constant and $\Delta := r^2 - 2 m_1 r + a^2$, with $m_1$ and $a$ being the mass and spin of the Kerr BH.

Using the Keplerian parametrization $r = m_1/\left[u_p (1 + e \cos\zeta)\right]$, we can obtain four equations of motion for $(\di \tau/\di \zeta,\, \di t/\di \zeta,\, \di r/\di \zeta,\, \di \phi/\di \zeta)$.
Then, we solve $\di r/\di \zeta = 0$ at the turning points $r = \left[u_p (1 \pm e)\right]^{-1}$ for $E(u_p,e)$ and $L(u_p,e)$, in a spin expansion to $\Order(a^6)$.

The periods of an eccentric orbit can be evaluated through the following integrals:
\begin{subequations}
\begin{align}
T_\tau^{(0)} &:= \oint \di \tau = \int_{0}^{2\pi} \frac{\di \tau}{\di \zeta} \di\zeta, \\
T_r^{(0)} &:= \oint \di t = \int_{0}^{2\pi} \frac{\di t}{\di \zeta} \di\zeta, \\
T_\phi^{(0)} &:= \oint \di \phi = \int_{0}^{2\pi} \frac{\di \phi}{\di \zeta} \di\zeta.
\end{align}
\end{subequations}
The orbital frequency is $\Omega^{(0)} = T_\phi^{(0)}/T_r^{(0)}$, and we define the gauge invariant variables
\begin{subequations}
\begin{align}
y(u_p,e) &:= (m_1 \Omega^{(0)})^{3/2}, \\
\lambda(u_p,e) &:= \frac{3 y}{T_\phi^{(0)} / (2\pi) - 1}.
\end{align}
\end{subequations}
We computed these periods, in addition to $y$ and $\lambda$, in a PN and spin expansion to $\Order(a^6 u_p^8)$, but without an eccentricity expansion. 
The results are lengthy, so we provide them in the Supplemental Material.

The redshift can be computed from the periods via $z^{(0)} :=  T_\tau^{(0)} / T_r^{(0)}$, with $U^{(0)} := 1/z^{(0)}$, leading to
\begin{widetext}
\begin{align}
\label{U0test}
U^{(0)} &= 1 +\frac{3 \varepsilon ^2 u_p}{2}
+\left(6 \varepsilon ^3-\frac{21 \varepsilon ^4}{8}\right) u_p^2
+\left(\frac{55 \varepsilon ^6}{16}-12 \varepsilon ^5-6 \varepsilon ^4+23 \varepsilon ^3\right) u_p^3
+ \bigg(-\frac{525 \varepsilon ^8}{128}+\frac{75 \varepsilon ^7}{4}+12 \varepsilon ^6-105 \varepsilon ^5-24 \varepsilon ^4 \nonumber\\
&\qquad
+\frac{249 \varepsilon ^3}{2}\bigg) u_p^4 
+ \left(\frac{1197 \varepsilon ^{10}}{256}-\frac{105 \varepsilon ^9}{4}-\frac{57 \varepsilon ^8}{4}+\frac{969 \varepsilon ^7}{4}-\frac{27 \varepsilon ^6}{2}-780 \varepsilon ^5-96 \varepsilon ^4+\frac{5943 \varepsilon ^3}{8}\right) u_p^5 + \Order(u_p^6) \nonumber\\
&\quad
+ \chi \, \varepsilon ^3 \bigg[
(-6+3 \varepsilon ) u_p^{5/2}
+ \left(36 \varepsilon ^2-\frac{21 \varepsilon ^3}{2}+18 \varepsilon -57\right) u_p^{7/2}
+ \left(\frac{165 \varepsilon ^5}{8}-\frac{375 \varepsilon ^4}{4}-36 \varepsilon ^3+441 \varepsilon ^2+90 \varepsilon -\frac{945}{2}\right) u_p^{9/2} \nonumber\\
&\qquad
+ \left(-\frac{525 \varepsilon ^7}{16}+\frac{735 \varepsilon ^6}{4}+\frac{159 \varepsilon ^5}{4}-\frac{5745 \varepsilon ^4}{4}+\frac{531 \varepsilon ^3}{2}+4176 \varepsilon ^2+420 \varepsilon -\frac{30345}{8}\right) u_p^{11/2} + \Order(u_p^{13/2})
\bigg] \nonumber\\
&\quad
+ \chi^2 \, \varepsilon ^3 \bigg[\!
\left(2-\frac{3 \varepsilon }{2}\right)\! u_p^3
+\left(\frac{39 \varepsilon ^3}{4}-39 \varepsilon ^2-18 \varepsilon +54\right)\! u_p^4
+\left(\frac{711 \varepsilon ^4}{4}-\frac{585 \varepsilon ^5}{16}+18 \varepsilon ^3-\frac{1479 \varepsilon ^2}{2}-120 \varepsilon +\frac{1485}{2}\right)\! u_p^5 \nonumber\\
&\qquad
+ \left(\frac{2835 \varepsilon ^7}{32}-\frac{4115 \varepsilon ^6}{8}+\frac{327 \varepsilon ^5}{4}+\frac{6891 \varepsilon ^4}{2}-1226 \varepsilon ^3-\frac{37713 \varepsilon ^2}{4}-672 \varepsilon +\frac{16863}{2}\right) u_p^6 + \Order(u_p^7)
\bigg] \nonumber\\
&\quad
+ \chi^3 \, \varepsilon ^3 \bigg[
\left(-3 \varepsilon ^3+18 \varepsilon ^2+6 \varepsilon -22\right) u_p^{9/2}
+\left(27 \varepsilon ^5-\frac{645 \varepsilon ^4}{4}+45 \varepsilon ^3+\frac{1251 \varepsilon ^2}{2}+60 \varepsilon -\frac{2445}{4}\right) u_p^{11/2} \nonumber\\
&\qquad
+ \left(-\frac{915 \varepsilon ^7}{8}+\frac{5901 \varepsilon ^6}{8}-\frac{1815 \varepsilon ^5}{4}-4293 \varepsilon ^4+2583 \varepsilon ^3+\frac{93279 \varepsilon ^2}{8}+420 \varepsilon -\frac{21315}{2}\right) u_p^{13/2} + \Order(u_p^{15/2})
\bigg] \nonumber\\
&\quad
+ \chi^4 \, \varepsilon ^3 \bigg[
\left(3-3 \varepsilon ^2\right) u_p^5
+\left(-\frac{57 \varepsilon ^5}{8}+\frac{141 \varepsilon ^4}{2}-63 \varepsilon ^3-276 \varepsilon ^2+\frac{555}{2}\right) u_p^6 \nonumber\\
&\qquad
+\left(\frac{1125 \varepsilon ^7}{16}-\frac{4587 \varepsilon ^6}{8}+\frac{1413 \varepsilon ^5}{2}+\frac{23421 \varepsilon ^4}{8}-2919 \varepsilon ^3-\frac{17211 \varepsilon ^2}{2}+\frac{33705}{4}\right) u_p^7 + \Order(u_p^8)
\bigg] \nonumber\\
&\quad
+ \chi^5 \, \varepsilon ^3 \bigg[
\left(-12 \varepsilon ^4+27 \varepsilon ^3+\frac{117 \varepsilon ^2}{2}-6 \varepsilon -\frac{135}{2}\right) u_p^{13/2} \nonumber\\
&\qquad
+\left(-\frac{33 \varepsilon ^7}{2}+\frac{1841 \varepsilon ^6}{8}-513 \varepsilon ^5-\frac{4089 \varepsilon ^4}{4}+\frac{3659 \varepsilon ^3}{2}+\frac{31119 \varepsilon ^2}{8}-84 \varepsilon -\frac{17269}{4}\right) u_p^{15/2} + \Order(u_p^{17/2})
\bigg] \nonumber\\
&\quad
+ \chi^6 \, \varepsilon ^3 \bigg[
\left(-3 \varepsilon ^3-\frac{9 \varepsilon ^2}{2}+\frac{15}{2}\right) u_p^7
+\left(-\frac{75 \varepsilon ^6}{2}+\frac{351 \varepsilon ^5}{2}+\frac{501 \varepsilon ^4}{4}-612 \varepsilon ^3-\frac{4275 \varepsilon ^2}{4}+\frac{2835}{2}\right) u_p^8 + \Order(u_p^9)
\bigg],
\end{align}
where we recall that $\varepsilon := \sqrt{1-e^2}$ and $\chi := a / m_1$.

The spin-precession frequency can be derived as in Ref.~\cite{Bini:2016iym}, utilizing Marck's parallel propagated frame~\cite{marck1983parallel}. 
The angular velocity of the gyro-fixed axes in the Marck frame is given by
\begin{equation}
\mathcal{T} = \frac{a + E \sqrt{K}}{r^2 + K}.
\end{equation}
The instantaneous spin-precession frequency and its orbit average can then be computed using
\begin{equation}
\Omega_S^\text{inst} = \frac{\di\tau}{\di t} \left(\frac{\di\phi}{\di\tau} - \mathcal{T}\right), \qquad
\Omega_S^{(0)} = \frac{1}{T_r^{(0)}} \int_{0}^{T_r^{(0)}} \Omega_S^\text{inst} \di t,
\end{equation}
from which the spin-precession invariant is defined as $\psi^{(0)} := \Omega_S^{(0)} / \Omega^{(0)}$.
Expanding to $\Order(\chi^5 u_p^{13/2})$, but without an eccentricity expansion, we obtain
\begin{align}
\label{psi0Test}
\psi^{(0)} &= \frac{3 u_p}{2}
+ \left(\frac{21}{8}-\frac{3 \varepsilon ^2}{2}\right) u_p^2
+\left(\frac{3 \varepsilon ^4}{16}-\frac{69 \varepsilon ^2}{8}+\frac{81}{8}\right) u_p^3
+\left(-\frac{3 \varepsilon ^6}{32}+\frac{33 \varepsilon ^4}{8}-\frac{1665 \varepsilon ^2}{32}+\frac{6549}{128}\right) u_p^4 + \Order(u_p^5) \nonumber\\
&\quad
+ \chi \left[
-u_p^{3/2}
+\!\left(3 \varepsilon ^2-\frac{9}{2}\right)\! u_p^{5/2}
+\!\left(-\frac{9 \varepsilon ^4}{8}+\frac{117 \varepsilon ^2}{4}-\frac{63}{2}\right)\! u_p^{7/2}
+\!\left(\frac{3 \varepsilon ^6}{4}-\frac{75 \varepsilon ^4}{4}+\frac{1965 \varepsilon ^2}{8}-\frac{3777}{16}\right)\! u_p^{9/2} + \Order(u_p^{11/2})
\right]\nonumber\\
&\quad
+ \chi^2 \left[
\left(2-\frac{3 \varepsilon ^2}{2}\right) u_p^3
+\left(\frac{3 \varepsilon ^4}{8}-\frac{75 \varepsilon ^2}{2}+\frac{315}{8}\right) u_p^4
+\left(-\frac{33 \varepsilon ^6}{16}+\frac{285 \varepsilon ^4}{16}-\frac{969 \varepsilon ^2}{2}+\frac{7635}{16}\right) u_p^5 + \Order(u_p^6)
\right]\nonumber\\
&\quad
+ \chi^3 \left[
\left(\frac{9 \varepsilon ^4}{4}+24 \varepsilon ^2-\frac{107}{4}\right) u_p^{9/2}
+\left(\frac{15 \varepsilon ^6}{8}+\frac{201 \varepsilon ^4}{8}+\frac{4203 \varepsilon ^2}{8}-\frac{4449}{8}\right) u_p^{11/2} + \Order(u_p^{13/2})
\right]\nonumber\\
&\quad
+ \chi^4 \left[
\left(-\frac{33 \varepsilon ^4}{16}-\frac{75 \varepsilon ^2}{8}+\frac{183}{16}\right) u_p^5
+\left(-\frac{47 \varepsilon ^6}{32}-60 \varepsilon ^4-\frac{10899 \varepsilon ^2}{32}+\frac{6443}{16}\right) u_p^6 + \Order(u_p^7)
\right]\nonumber\\
&\quad
+ \chi^5 \left[
\left(\frac{3 \varepsilon ^4}{8}+\frac{9 \varepsilon ^2}{4}-\frac{21}{8}\right) u_p^{11/2}
+\left(3 \varepsilon ^6+\frac{93 \varepsilon ^4}{2}+123 \varepsilon ^2-\frac{345}{2}\right) u_p^{13/2} + \Order(u_p^{15/2})
\right].
\end{align}

In both $U^{(0)}$ and $\psi^{(0)}$, we truncated the PN expansion at each order in spin to the orders needed for the calculations in Sec.~\ref{sec:boundObserv}.
In the Supplemental Material, we include the full 8PN expressions for $y(u_p,e)$, $\lambda(u_p,e)$, $z^{(0)}$ and $\psi^{(0)}$, to complement the 1SF 8PN results derived in Refs.~\cite{Munna:2023wce,Kavanagh:2016idg,Bini:2018ylh,Bini:2019lkm} for circular orbits or in an eccentricity expansion.

\section{Full Compton Coefficients}
\label{app_B}
In this appendix we present relevant expressions for the Compton amplitude derivation using Teukolsky solutions, presented in Sec. \ref{sec:compton}. We start by presenting the extended form of the near-zone phase-shift given in Eq. \eqref{eq:deltaNZl2_squem}. It reads
\begin{align}\label{eq:deltaNZl2}
\blue{{}_{-2}\delta^{\text{NZ}}_{\ell m}|_{\ell\to2}} =&
 \Big[\frac{m \chi }{18000} \Big(10 \left(\left(m^2-4\right) \chi ^2+4\right) \left(\left(m^2-1\right) \chi ^2+1\right) \psi ^{(0,m)}(\chi)
+ \left(-605 m^2+40 \gamma_E  \left(5 m^2-8\right)+824\right) \chi ^2\nonumber
\\
&
+\left(-133 m^4+605 m^2+40 \gamma_E  \left(m^4-5 m^2+4\right)-412\right) \chi ^4+160 \gamma_E -412
\Big)\nonumber
\\
&
+\frac{1}{900} m \chi  \left(\left(5 m^2-8\right) \chi ^2+\left(m^4-5 m^2+4\right) \chi ^4+4\right) \log (2 \kappa  \epsilon )\Big]\epsilon ^6\nonumber\\
& +\Big[\frac{1}{3402000}\Big(
- \frac{9 m \chi }{\kappa} \left(\left(5 m^2-8\right) \chi ^2+\left(m^4-5 m^2+4\right) \chi ^4+4\right) (107 \kappa  \tilde{\psi}^{(0,m)}(\chi)+210 \tilde{\psi}^{(1,m)}(\chi))\nonumber\\
&
+210 \left(9 \pi  m \left(5 m^2-8\right) \chi ^3+\left(72-115 m^2\right) \chi ^2+5 m^2 \left(m^4-5 m^2+4\right) \chi ^6+9 \pi  m \left(m^4-5 m^2+4\right) \chi ^5\right.\nonumber\\
&\left.
+\left(-20 m^4+95 m^2-36\right) \chi ^4+36 \pi  m \chi -36\right) \psi ^{(0,m)}(\chi)+114345 \pi  \kappa ^2 m^3 \chi ^3\nonumber\\
&
+963 \kappa  \left(\left(5 m^2-8\right) \chi ^2+\left(m^4-5 m^2+4\right) \chi ^4+4\right)+840 \gamma_E  \left(9 \pi  m \left(5 m^2-8\right) \chi ^3+\left(72-115 m^2\right) \chi ^2\right.\nonumber\\
&\left.
+5 m^2 \left(m^4-5 m^2+4\right) \chi ^6+9 \pi  m \left(m^4-5 m^2+4\right) \chi ^5+\left(-20 m^4+95 m^2-36\right) \chi ^4+36 \pi  m \chi -36\right)\nonumber\\
&+7 m^2 \chi ^2 \left(\chi ^2 \left(-3591 \pi  m^3 \chi +7930 m^2-5 \left(437 m^4-2005 m^2+1388\right) \chi ^2-35125\right)+42065\right)\nonumber\\
&-77868 \kappa ^4 (\pi  m \chi -1)\Big)+
\frac{1}{8100}
\Big(
\left(9 \pi  m \left(5 m^2-8\right) \chi ^3+\left(72-115 m^2\right) \chi ^2+5 m^2 \left(m^4-5 m^2+4\right) \chi ^6\right.\nonumber\\
&\left.
+9 \pi  m \left(m^4-5 m^2+4\right) \chi ^5+\left(-20 m^4+95 m^2-36\right) \chi ^4+36 \pi  m \chi -36\right) \log (2 \kappa  \epsilon )\Big)\Big]\epsilon^7
+\nonumber\\
&\Big[
\frac{\chi m \epsilon ^6}{1800} \left(\chi^2 \left(m^2-4\right)+4\right) \left(\chi^2 \left(m^2-1\right)+1\right)+\frac{ \epsilon ^7}{16200}\left(\chi \left(9 \pi  m \left(\chi^2 \left(m^2-4\right)+4\right) \left(\chi^2 \left(m^2-1\right)+1\right)\right.\right.\nonumber
\\
&\left.\left.+\chi \left(\chi^2 \left(5 m^2 \left(\chi^2 \left(m^4-5 m^2+4\right)-4 m^2+19\right)
-36\right)-115 m^2+72\right)\right)-36\right)\Big]\frac{1}{\ell-2}+ \mathcal{O}(\epsilon^8)\,.
\end{align}

We also present the explicit matched coefficients in the Compton ansatz, as obtained from the gravitational wave scattering Teukolsky solutions. The spurious pole cancellation conditions read:
\begin{align}
c_{3}^{(2)}&=4/15-c_{3}^{(0)}+c_{3}^{(1)}\label{eq:constrainspurious5}\\
c_{10}^{(2)} & =c_{10}^{(1)}-c_{10}^{(0)}\\
d_{1}^{(0)}=&-\frac{8}{45}+4 \sum_{i=0}^{2}(-1)^i c_7^{(i)}+\sum_{j=5}^6\sum_{i=0}^{2}(-1)^i c_j^{(i)}\\
f_{1}^{(0)}  =&\frac{4}{45} +\sum_{i=0}^2\sum_{j\in\{6,8\}}(-1)^i c_j^{(i)}\,,
\end{align}
with the c-coefficients spitted as in Eq. \eqref{eq:coeffs_decomp}. The far-zone solutions are 
\begin{equation}\label{eq:far_Zone_solutions}
\begin{split}
c_j^{\text{FZ},(i)}&=0\,,\quad i=0,1,2, \quad j =1,2,4,5,7,9,10\,,\\
d_1^{\text{FZ},(0)}&=0\,,\\
c_3^{\text{FZ},(0)}&=\frac{64}{15}\,,\quad c_3^{\text{FZ},(1)}=\frac{16}{3}\,, \quad
c_3^{\text{FZ},(2)}=\frac{4}{3}\,\\
c_6^{\text{FZ},(0)}&=\frac{128}{45}\,,\quad c_6^{\text{FZ},(1)}=\frac{32}{9}\,,\quad c_6^{\text{FZ},(2)} = \frac{8}{9}\,,\\
c_8^{\text{FZ},(0)}&=-\frac{512}{45}\,,\quad c_8^{\text{FZ},(1)}=-\frac{160}{9}\,, \quad c_8^{\text{FZ},(2)} = -\frac{64}{9}\,,\\
f_1^{\text{FZ},(0)}&=-\frac{4}
{9}\,,
\end{split}
\end{equation}

whereas the non-perturbative in $\chi$, $\ell=2$ near-zone solutions are 

\begin{align}\label{eq:coeffsnear}
c_j^{\text{NZ},(i)}&=0\,,\quad c_j^{\log,\text{NZ},(i)} \,,\quad i=0,1,2, \quad j =2,4,5,7,9\,,\nonumber\\
 c_3^{\text{NZ},(0)}&=\frac{32 \left(10 \left(3 \chi ^2+1\right) \psi^{ (0,2)}(\chi)+20 \left(3 \chi ^2+1\right) \log (2 \kappa )-30 \chi ^4+3 (40 \gamma_E -133) \chi ^2+40 \gamma_E -103\right)}{225 \chi ^4}\nonumber\\
  c_3^{\text{NZ},(1)}&=\frac{1}{225 \chi ^4}\Big(
4 \left(-3 \chi ^2 \left(10 \psi^{ (0,1)}(\chi)-120 \psi^{ (0,2)}(\chi)+100 \chi ^2+1523\right)+8 (5 \psi^{ (0,1)}(\chi)+15 \psi^{ (0,2)}(\chi)-206)\right.\nonumber
&\left.\qquad\qquad\quad
+20 \left(33 \chi ^2+16\right) \log (2 \kappa )+40 \gamma_E  \left(33 \chi ^2+16\right)\right)\Big)\nonumber\\
\\
c_{3}^{\text{NZ},(2)}&=4/15-c_{3}^{\text{NZ},(0)}+c_{3}^{\text{NZ},(1)}\nonumber\\
 c_3^{\log,\text{NZ},(0)} &= \frac{128 \left(3 \chi ^2+1\right)}{45 \chi ^4}\,,\quad c_3^{\log,\text{NZ},(1)}  =\frac{16 \left(33 \chi ^2+16\right)}{45 \chi ^4}\,,\quad
 c_{3}^{\log,\text{NZ},(2)}=4/15-c_{3}^{\log,\text{NZ},(0)}+c_{3}^{\log,\text{NZ},(1)}\nonumber\\
 %d_1^{\text{NZ},(0)}&=0\,,\quad d_1^{\log,\text{NZ},(0)}=0 \,,\quad
  %f_1^{\text{NZ},(0)}=0\,,\quad f_1^{\log,\text{NZ},(0)}=0 \,,\nonumber\\
%
 c_6^{\log,\text{NZ},(0)} &=\frac{2}{3}c_3^{\log,\text{NZ},(0)}\,,\quad c_6^{\log,\text{NZ},(1)}  = \frac{2}{3}c_3^{\log,\text{NZ},(1)}\,,\quad c_6^{\log,\text{NZ},(2)}  =\frac{32 \left(9 \chi ^2+8\right)}{135 \chi ^4}\,,
 \nonumber\\
  c_6^{\text{NZ},(0)}&= \frac{2}{3}c_3^{\text{NZ},(0)}\,,\quad
c_6^{\text{NZ},(1)}= \frac{2}{3} c_3^{\text{NZ},(1)}\nonumber\\
 c_6^{\text{NZ},(2)}&=\frac{1}{675 \chi ^4} \Big(8 \left(-3 \chi ^2 \left(10 \psi^{ (0,1)}(\chi)-40 \psi^{ (0,2)}(\chi)+20 \chi ^2+459\right)+8 (5 \psi^{ (0,1)}(\chi)+5 \psi^{ (0,2)}(\chi)-103)\right.\nonumber\\
 &\left.\qquad\qquad\qquad+20 \left(9 \chi ^2+8\right) \log (2 \kappa )+40 \gamma_E  \left(9 \chi ^2+8\right)\right)\Big)\nonumber\\
 c_8^{\log,\text{NZ},(0)}
 &= -\frac{128 \left(4  \left(3 \chi ^2+1\right) \chi ^2+\left(6 \chi ^4-97 \chi ^2-9\right)\right)}{405 \chi ^6}\,,\quad
 c_8^{\log,\text{NZ},(1)}=
 -\frac{32 \left(4  \left(\chi ^2+2\right) \chi ^2+\left(29 \chi ^4-273 \chi ^2-36\right)\right)}{135 \chi ^6}\nonumber\\
 c_8^{\log,\text{NZ},(2)}&=\frac{32 \left(15  \chi ^4+\left(-24 \chi ^4+158 \chi ^2+36\right)\right)}{135 \chi ^6}\nonumber\\
c_8^{\text{NZ},(0)}&=-\frac{64  \left(3 \chi ^2+1\right) \left(210 \kappa  \tilde{\psi }^{(1,2)}(\chi)+107 \left(\chi ^2-1\right) \tilde{\psi }^{(0,2)}(\chi)\right)}{4725 \chi ^5 \left(\chi ^2-1\right)}\nonumber\\
&
+\frac{1}{42525 \chi ^6}\Big(32 \left(-210 \left(6 \chi ^4 (2+1)+\chi ^2 (4 -97 )-9 \right) (\psi^{(0,2)}(\chi)+2 \log (2 \kappa ))\right.\nonumber\\
&\left.
+\chi ^2 (84 (103-40 \gamma_E ) +(-2889 \kappa +81480 \gamma_E -255521))+2520 \chi ^6 (1+5 )\right.\nonumber\\
&\left.-84 \chi ^4 (60 \gamma_E  (2 +1)-399 -52 )+9  (-107 \kappa +840 \gamma_E -2163)\right)\Big)\nonumber\\
c_8^{\text{NZ},(1)}& = \frac{8 } {4725 \chi ^5 \left(\chi ^2-1\right)}\left(210 \kappa  \left(\left(3 \chi ^2-4\right) \tilde{\psi }^{(1,1)}(\chi)-16 \left(3 \chi ^2+1\right) \tilde{\psi }^{(1,2)}(\chi)\right)\right.\nonumber\\
&\left.
+107 \left(3 \chi ^4-7 \chi ^2+4\right) \tilde{\psi }^{(0,1)}(\chi)-1712 \left(3 \chi ^4-2 \chi ^2-1\right) \tilde{\psi }^{(0,2)}(\chi)\right)\nonumber\\
& -\frac{1}{14175 \chi ^6}\Big(8 \left(70 \left(\left(39 \chi ^4+4 \left(4-3 \chi ^2\right) \chi ^2-43 \chi ^2-36\right) \psi^{(0,1)}(\chi)+8 \left(9 \chi ^4-96 \chi ^2-9\right) \psi^{(0,2)}(\chi)\right.\right.\nonumber\\
&\left.\left.+6 \left(29 \chi ^4+4 \left(\chi ^2+2\right) \chi ^2-273 \chi ^2-36\right) \log (2 \kappa )\right)+963 \kappa  \left(7 \chi ^2+4\right)+84 \left(-193 \chi ^2+40 \gamma_E  \left(\chi ^2+2\right)-206\right) \chi ^2\right.\nonumber\\
&\left.+727629 \chi ^2-7 \left(4500 \chi ^2+7021\right) \chi ^4+840 \gamma_E  \left(29 \chi ^4-273 \chi ^2-36\right)+77868\right)\Big)\nonumber\\
c_{8}^{\text{NZ},(2)}&= \frac{840 \kappa  \left(\left(3 \chi ^2-4\right) \tilde{\psi}^{(1,1)}(\chi)-6 \left(3 \chi ^2+1\right) \tilde{\psi}^{(1,2)}(\chi)\right)+428 \left(3 \chi ^4-7 \chi ^2+4\right) \tilde{\psi}^{(0,1)}(\chi)-2568 \left(3 \chi ^4-2 \chi ^2-1\right) \tilde{\psi}^{(0,2)}(\chi)}{1575 \chi ^5 \left(\chi ^2-1\right)}\nonumber\\
&+\frac{4}{14175 \chi ^6} \left(14 \chi ^2 \left(3 \chi ^2 \left(-195 \psi^{(0,1)}(\chi)-90 \psi^{(0,1)}(\chi)+750 \chi ^2+2527\right)+645 \psi^{(0,1)}(\chi)+4365 \psi^{(0,1)}(\chi)-61882\right)\right.\nonumber\\
&\left.+210 \chi ^2 \left(3 \chi ^2 (2 \psi^{(0,1)}(\chi)+8 \psi^{(0,1)}(\chi)+40 \gamma -121)-8 \psi^{(0,1)}(\chi)+8 \psi^{(0,1)}(\chi)-36 \chi ^4\right)\right.\nonumber\\
&\left.+378 (20 \psi^{(0,1)}(\chi)+15 \psi^{(0,1)}(\chi)-412)
+963 \kappa  \left(\chi ^4-8 \chi ^2-6\right)+840 \left(-9 \chi ^4+158 \chi ^2+36\right) \log (2 \kappa )\right.\nonumber\\
&\left.-2100 \gamma  \left(21 \chi ^4-130 \chi ^2-27\right)\right)\nonumber\\
 c_{10}^{\log,\text{NZ},(0)}
 &=\frac{64 (12 \pi +\green{ (24 \gamma_E -25) i}) \left(3 \chi ^2+1\right)}{135 \chi ^5}\,,\quad
  c_{10}^{\log,\text{NZ},(1)}
 =\frac{8 (12 \pi +\green{(24 \gamma_E -25) i }) \left(33 \chi ^2+16\right)}{135 \chi ^5}\,,\nonumber\\
 c_{10}^{\text{NZ},(0)}
 &=\frac{16  (12 \pi +\green{(24 \gamma_E -25) i }) \left(10 \left(3 \chi ^2+1\right) \psi^{ (0,2)}(\chi)+20 \left(3 \chi ^2+1\right) \log (2 \kappa )-30 \chi ^4-399 \chi ^2+40 \gamma_E  \left(3 \chi ^2+1\right)-103\right)}{675 \chi ^5}\nonumber\\
 c_{10}^{\text{NZ},(1)}
 &=\frac{2}{675 \chi ^5}
 (12 \pi +\green{(24 \gamma_E -25) i} ) \left(-3 \chi ^2 \left(10 \psi^{(0,1)}(\chi)-120 \psi^{(0,2)}(\chi)+100 \chi ^2+1523\right)\right.\nonumber\\
 &\qquad\left.+8 (5 \psi^{(0,1)}(\chi)+15 \psi^{(0,2)}(\chi)-206)
 +20 \left(33 \chi ^2+16\right) \log (2 \kappa )+40 \gamma_E  \left(33 \chi ^2+16\right)\right)\nonumber\\
 c_{10}^{\text{NZ},(2)} & =c_{10}^{\text{NZ},(1)}-c_{10}^{\text{NZ}(0)}\,,\qquad c_{10}^{\log,\text{NZ},(2)} =c_{10}^{\log,\text{NZ},(1)}-c_{10}^{\log,\text{NZ}(0)}\nonumber\\
d_{1}^{\text{NZ},(0)}=&-\frac{8}{45}+4 \sum_{i=0}^{2}(-1)^i c_7^{\text{NZ},(i)}+\sum_{j=5}^6\sum_{i=0}^{2}(-1)^i c_j^{\text{NZ},(i)}\,,\quad d_{1}^{\log,\text{NZ},(0)}=-\frac{8}{45}+4 \sum_{i=0}^{2}(-1)^i c_7^{\log,\text{NZ},(i)}+\sum_{j=5}^6\sum_{i=0}^{2}(-1)^i c_j^{\log,\text{NZ},(i)} \nonumber\\
f_{1}^{\text{NZ},(0)}  =&\frac{4}{45} +\sum_{i=0}^2\sum_{j\in\{6,8\}}(-1)^i c_j^{\text{NZ},(i)}\,,\qquad f_{1}^{\log,\text{NZ},(0)}  =\frac{4}{45} +\sum_{i=0}^2\sum_{j\in\{6,8\}}(-1)^i c_j^{\log,\text{NZ},(i)}\,,
\end{align}
For the reader's convenience, we include 
the  expressions for these coefficients in the supplementary Material. 
\end{widetext}

\bibliography{bibl}

\end{document}